\documentclass[twocolumn,trackchanges]{aastex701} 

\usepackage{graphicx}	
\usepackage{amsmath}	
\usepackage{braket}
\usepackage{lipsum}
\usepackage{float}
\usepackage{xcolor}
\usepackage{epsfig}
\usepackage{float}
\usepackage{placeins}
\usepackage{mwe}
\usepackage{ulem}
\usepackage{caption}
\usepackage{threeparttable}

\begin{document}

\title[TXS 0518+211]{Broad-band temporal and spectral study of TeV blazar TXS 0518+211}

\author[orcid=0000-0002-9526-0870,gname=Avik Kumar, sname='Das']{Avik Kumar Das} 
\affiliation{Astronomy and Astrophysics division, Physical Research Laboratory, Ahmedabad, 380009, Gujarat, India}
\affiliation{Department of Physical Sciences, Indian Institute of Science Education and Research Mohali, Knowledge City, Sector 81, SAS Nagar, Punjab 140306, India}
\email[show]{avikdas@prl.res.in}

\author[orcid=0000-0001-6890-2236,gname=Pankaj, sname='Kushwaha']{Pankaj Kushwaha} 
\affiliation{Department of Physical Sciences, Indian Institute of Science Education and Research Mohali, Knowledge City, Sector 81, SAS Nagar, Punjab 140306, India}
\email{}

\author[orcid=0000-0002-6040-4993,gname=Veeresh, sname='Singh']{Veeresh Singh} 
\affiliation{Astronomy and Astrophysics division, Physical Research Laboratory, Ahmedabad, 380009, Gujarat, India}
\email{}

\author[orcid=0000-0003-2445-9935,gname=Sandeep Kumar, sname='Mondal']{Sandeep Kumar Mondal} 
\affiliation{Aryabhatta Research Institute of Observational Sciences (ARIES), Manora Peak, Nainital 263001, India}
\altaffiliation{Tsung-Dao Lee Institute, Shanghai Jiao Tong University, 1 Lisuo Road, Shanghai, 201210, People’s Republic of China}
\email{sandeep@sjtu.edu.cn}
 
\author[orcid=0009-0008-1809-3256,gname=Goldy, sname='Ahuja']{Goldy Ahuja} 
\affiliation{Astronomy and Astrophysics division, Physical Research Laboratory, Ahmedabad, 380009, Gujarat, India}
\affiliation{Indian Institute of Technology Gandhinagar, Palaj, Gujarat-382355, India}
\email{}
 
\author[gname=Deekshya R., sname='Sarkar']{Deekshya R. Sarkar} 
\affiliation{Astronomy and Astrophysics division, Physical Research Laboratory, Ahmedabad, 380009, Gujarat, India}
\email{}

\begin{abstract}
We present a long-term broad-band temporal and spectral study of a TeV BL Lac source TXS 0518+211 by analyzing nearly 16 years (MJD 54682 -- 60670) of simultaneous optical, UV and X-ray light curves from \textit{Swift}-XRT/UVOT and gamma-ray light curves from \textit{Fermi}-LAT. Based on the availability of simultaneous multi-wavelength data and considering flux level as the depiction of AGN-jet activity we identified 11 epochs (named as Epoch-A to Epoch-K) and investigated  temporal as well as spectral variability during these epochs to understand the emission properties in this source. The fractional variability analysis reveals that, in all epochs, X-ray light curve exhibits relatively high degree of variability in compared to the optical, UV and gamma-ray light curves. The flux-flux plots among different bands, in general, show weak to moderate correlation with Spearman correlation coefficient ranging from 0.29 to 0.58. Notably, during Epoch-I, we detect a possible orphan flare exhibiting increase in the X-ray flux level ($\sim$ 2.4 times of the total average flux) but with no corresponding counterpart seen in the optical, UV bands. In contrast, during Epoch-K, we detect a significant decrease in the X-ray flux but no corresponding decrease in optical, UV and gamma-ray bands. Overall, our study reveals several changes in the flux states and complex nature of jet dominated emission processes. We tested one-zone and two-zone leptonic scenarios and for most of the epochs, the latter one provides a better description of the broad-band emission in this TeV BL Lac source.
\end{abstract}


\section{Introduction}
Blazars are the most luminous and extreme subclass of active galactic nuclei (AGNs). They are accepted to harbor highly collimated relativistic jets aligned closely to the line-of-sight of observers \citep{Urry1995Sep}, leading to a Doppler beaming of their emission, thereby enhancing apparent changes in observed spectral and temporal behaviour. Emission in these sources is dominated by non-thermal continuum originating predominantly within the jet and characterized by rapid variability, with time scales ranging from a few minutes to years across the entire accessible electromagnetic (EM) spectrum \citep{Blandford2019Aug}. Their continuum emission exhibits a characteristic broad double-hump spectral energy distribution (SED): the ``low-energy hump'' extends from radio up to X-rays, and the ``high-energy hump'' extends from X-ray to GeV-TeV gamma-ray. The ``low-energy hump'' is widely accepted to be predominantly synchrotron emission from relativistic electrons in the jet. This is evident by the polarization fraction variability observed in the radio and optical bands (e.g., \citealt{D'arcangelo2009May}). Conversely, origin of high-energy hump in blazars SEDs is still a subject of debate within the blazar community and can be explained by leptonic and/or hadronic emission models. In the leptonic scenario, this hump is attributed to the inverse Compton (IC) scattering of photons originated from the synchrotron emission within the jet itself (known as synchrotron self-Compton or SSC), and/or external to the emission region (i.e., external Compton or EC) such as the accretion disk, broad-line region (BLR), dusty torus (DT). On the other hand, the hadronic origin of emission processes involves phenomena such as proton synchrotron emission \citep{2001APh....15..121M} and/or photo-hadronic interactions (\citealt{1992A&A...253L..21M, 1999MNRAS.302..373B}). The latter can lead to the production of high-energy photons from $\pi^{0}$ decay and synchrotron emission of charged pions/secondary leptons, which can initiate cascades through $\gamma$-$\gamma$ pair productions, i.e., synchrotron-supported pair cascades \citep{Bottcher2013Apr}. \\
\\
Traditionally, depending on the strength of optical emission lines \citep{1991ApJ...374..431S}, blazars have been categorized into two sub-classes: Flat spectrum radio quasars (FSRQs) with broad strong emission lines (EW $>$ 5A$^\circ$) and BL Lacertae (BL Lacs) objects with weak or no emission lines (EW $<$ 5A$^\circ$). In the extended SED-based classification scheme (refer to \citealt{Abdo2010May}, for details), all non-thermal radiation domination or jet-dominated AGNs have been categorized into three sub-classes based on the peak frequency ($\nu_{peak}$) of the synchrotron emission: Low-synchrotron-peaked (LSP; $\nu_{peak} \leq$ 10$^{14}$ Hz), Intermediate-synchrotron-peaked (ISP; 10$^{14}$ $\leq \nu_{peak} \leq$ 10$^{15}$ Hz), and High-synchrotron-peaked (HSP; $\nu_{peak} \geq$ 10$^{15}$ Hz) sources. Almost all the known FSRQ-type blazars belong to the LSP subclass. On the other hand, BL Lacs can be found in all three sub-classes, also known as LBL, IBL, and HBL. A remarkable property of a particular spectral subclass is the stability of the location of the peaks (i.e., $\nu_{peak}$ and high-energy peak) despite significant flux variations, except for a few special cases where shift has been observed for either (e.g., December 2015 - May 2016 activity of OJ 287; \citealt{2018MNRAS.473.1145K}) or both (e.g., June 2012 outburst of Mrk 501; \citealt{Ahnen2018Dec}) of the peaks. Such a shift or change in the broadband SED provides valuable insights about the high-energy emission mechanism. A shift in the location of both peaks may imply that the same particle distribution contributes to the overall emission, whereas a shift only in the high-energy peak can be explained by the external Compton emission (where seed photons originating from the BLR, \citep{2018MNRAS.473.1145K}), emission from different regions \citep{Sahakyan2022Jul}, or hadronic processes \citep{2019MNRAS.489.4347O}. There is a growing evidence in the literature which suggests that several TeV-detected BL Lac objects are classified as IBLs during low-flux states but as HBLs during high-flux states, e.g., PKS 1424+240 (\citealt{2014ApJ...785L..16A, 2014A&A...567A.135A}), S2 0109+22 \citep{2018MNRAS.480..879M}, 1ES 1011+496 \citep{2016MNRAS.459.2286A}, PKS 0301-243 \citep{2013A&A...559A.136H}. Correlation studies across multi-wavelength regime are crucial for understanding such emission properties. \\
\\
In general, the broadband SEDs of IBl- and HBL-type BL Lac objects can be explained by sum of a synchrotron component and an associated SSC component under leptonic scenario. However, there are several sources where occasionally peculiar behaviors have been observed in their multi-wavelength light curves and SEDs such as the observation of orphan flares across different wavebands (\citealt{2013A&A...552A..11R, 2013ApJ...763L..11C, 2015ApJ...807...79H}), and/or plateaus in the low-energy gamma-ray SED (\citealt{2018A&A...611A..44P, 2015ApJ...798....2S}), which cannot be explained by the simple one-zone leptonic model. In some cases during such exceptional behaviour, a significant shift in the location of the synchrotron peak ($\nu_{\text{peak}}$) and/or the high-energy peak has been observed (e.g., \cite{2025ApJ...984....5L}). Therefore, to provide a comprehensive explanation, more complex emission models are required. The availability of both multi-wavelength temporal and spectral data are essential for gaining insight into the physical processes occurring in the emission regions. \\
\\
TXS 0518+211 is an IBL-type blazar located at R.A. = 05h21m45.9658s, Dec. = +21d12m51.451s \citep{Shaw2013Jan}. The redshift ($z$) of this source is still uncertain. However, in earlier studies, a lower limit of $z$=0.18 \citep{2017ApJ...837..144P} and an upper limit of $z$ = 0.34 \citep{Archambault2013Sep} have been reported. A recent study by \cite{2025A&A...694A.308M} reported an improved statistical upper limit of $z \leq 0.244$ at the 95\% confidence level using simultaneous high-energy and very-high-energy (VHE) observations. This source was first identified as a VHE (E $>$ 100 GeV) gamma-ray emitter (also known as VER J0521+211) in October 2009 by the VERITAS collaboration \citep{Ong2009Oct}. It has also been continuously monitored by \textit{Fermi}-LAT (Large Area Telescope) since August 2008, and the first gamma-ray outburst (E $>$ 100 MeV) was reported on October 14, 2013, with a daily averaged flux of (1.0$\pm$0.3)$\times$10$^{-6}$ ph cm$^{-2}$ s$^{-1}$ \citep{Buson2013Oct}. Following the report of this gamma-ray outburst, near-infrared (NIR) photometry in the $J$ band was carried out, which revealed correlated flux variability \citep{Carrasco2013Oct}. However, on February 25, 2020, the VERITAS collaboration \citep{Quinn2020Feb} reported an observation of simultaneous flares in the VHE and X-ray (0.3 to 10.0 keV) bands, with no flaring signature in the optical band \citep{Bachev2020Feb}. The observations of these exceptional variability signatures in this source motivates us to perform a long-term comprehensive study of both temporal and spectral behavior.  \\
\\
In this paper, we have focused on the study of TeV BL Lac TXS 0518+211 using optical, X-ray, and gamma-ray data collected over a period of $\sim$ 16 years.  Analysis methods of optical to high-energy gamma-ray data have been described in section \S\ref{sec:2}. The identification of different epochs using long-term multi-band light curves and temporal study in these epochs are described in section \S\ref{sec:3}. The broad-band spectral properties and SEDs modeling are presented in section \S\ref{sec:4}. The detailed discussions and conclusions are provided in section \S\ref{sec:5} and section \S\ref{sec:6}, respectively.

\section{MULTI-WAVELENGTH OBSERVATION AND DATA ANALYSIS} \label{sec:2}

\subsection{Simultaneous optical, UV, and X-ray observations from \textit{Swift} observatory} 
We utilized simultaneous optical, UV, and X-ray observations from the Neil Gehrels \textit{Swift} Observatory \footnote{\url{https://swift.gsfc.nasa.gov/about_swift/}} (\textit{Swift}, hereafter). The \textit{Swift}, a multi-wavelength space-based observatory, comprises of three primary instruments: UV/Optical telescope (UVOT) with wavelength coverage of 170 - 600 nm, X-ray telescope (XRT) sensitive in 0.3 - 10 keV band, and Burst alert telescope (BAT) operating in hard X-ray energy band of 15 - 150 keV. In this work, we analyzed \textit{Swift}-UVOT and -XRT data of TXS 0518+211 collected over a period of approximately 15 years (MJD 55131 - 60670) to perform the temporal and spectral study. A total of 131 observations were available during this time period. 

\subsubsection{The \textit{Swift}-UVOT data analysis} \label{sec:2.1.1}
We analyzed all the image mode Optical-UV data from \textit{Swift}-UVOT \citep{Roming2005Oct} available in six filters: $V$ (546.8 nm), $B$ (439.2 nm), $U$ (346.5 nm), $UVW1$ (260.0 nm), $UVM2$ (224.6 nm), and $UVW2$ (192.8 nm). The source and background regions are chosen 
as circles with radii of 5\arcsec{} (centerd on to the sources) and 10\arcsec{} (in a nearby source free region), respectively. The `\textit{uvotsource}' task is used to compute the magnitudes in all filters. 
The magnitude is further corrected for galactic extinction using a reddening E(B-V) = 0.5875 as reported in \citealt{Schlafly2011Aug}. Finally, reddening corrected magnitude is converted into flux using the zero points \citep{Breeveld2011Aug} and appropriate conversion factors \citep{Poole2007Dec}. The task `\textit{uvotimsum}' is used to aggregate multiple observation IDs.
\subsubsection{The \textit{Swift}-XRT data analysis} \label{sec:2.1.2}
We used \textit{Swift}-XRT data taken only in photon counting (PC) and windowed timing (WT) mode. The {\tt xrtpipeline} (version: 0.13.7) is used to perform the XRT data reduction of each set of observations. 
The source spectra are extracted from a circular region of 47\arcsec radius centered on to the source, while 
background spectra are extracted from an annular region with inner and outer radii of 80\arcsec and 180\arcsec, respectively, in a source free region. For pile-up correction, we followed recommended steps\footnote{\url{https://www.swift.ac.uk/analysis/xrt/pileup.php}} that include discarding 2-3 pixels from the central region of the source whenever the source count rate exceeds 0.5 count sec$^{-1}$ in PC mode or 100 counts sec$^{-1}$ in WT mode. Subsequently, we employed {\tt xrtmkarf} and {\tt grppha} tasks to create the corresponding ancillary response files, and to group the source spectrum with 20 counts per bin, respectively. In a few data sets, we encountered low photon counts and, in such cases, we grouped their spectra with one count per bin and employed the Cash statistic. 
Finally, these grouped spectra were loaded into {\tt XSPEC} (version: 12.12.1) to perform the spectral fitting using two models: {\tt power law} and {\tt logparabola}. The relatively complex model {\tt logparabola} is used to achieve the best fit if the $F$-test probability is $\leq$0.05. During the spectral fitting we kept all the parameters of the models including $N_{\rm H}$ as free. However, if the fitted $N_{\rm H}$ value is found less than or inconsistent with the the Galactic column density value, we fixed it to the Galactic value (4.38$\times$10$^{21}$ cm$^{-2}$; \cite{Kalberla2005Sep}) in the direction of source. To add more than one obs-IDs we used {\tt addspec} task. 

\subsection{\textit{Fermi}-Large Area Telescope (\textit{Fermi}-LAT)} \label{sec:2.2}
\textit{Fermi}-LAT is a gamma-ray (20 MeV to 300 GeV) imaging pair-conversion telescope that scans the entire sky in every two orbits with an orbital period of $\sim$ 96 minutes. It has a peak effective area of nearly 8000 cm$^{2}$.  
The detailed characteristics and specifications of the LAT are provided on the \textit{Fermi}-LAT website\footnote{\url{https://fermi.gsfc.nasa.gov/science/instruments/table1-1.html}}.

We used the weekly binned light curve of this source (catalog name: 4FGL J0521.7+2112) from the \textit{Fermi}-LAT Light Curve Repository (LCR)\footnote{\url{https://fermi.gsfc.nasa.gov/ssc/data/access/lat/LightCurveRepository/}} for the temporal study \citep{Abdollahi2023Mar}. To analyze the gamma-ray SED for different time durations, we utilized the Fermi Science tools\footnote{\url{https://fermi.gsfc.nasa.gov/ssc/data/analysis/scitools/python_tutorial.html}} of version 2.2.0 and followed the standard analysis procedures provided therein. For our analysis we opted the recommended {\tt evclass=128} and {\tt evtype=3}, an energy cut ranging from 0.1 to 300 GeV and a zenith angle cut of 90$^{\circ}$. The galactic emission model (gll\_iem\_v07.fits) and isotropic emission model (iso\_P8R3\_SOURCE\_V3\_v1.tx) are used as diffuse background models. The latest 4FGL DR3 catalog is used to create the model XML file. We used unbinned likelihood analysis method for the analysis of LAT data. We note that two extended sources are detected at $\sim$ 7$^{\circ}$ (S 147) and $\sim$ 12$^{\circ}$ (IC 443) away from the center of the region of interest (ROI). During the spectral fitting, we kept free all parameters of the model components within the ROI, as well as those of the two extended and diffuse background sources. We also chose a 10$^{\circ}$ additional patch around the ROI to account for the contribution of other sources in the analysis but kept all parameters within this region frozen during the spectral fitting procedure. 

\section{TEMPORAL STUDY} \label{sec:3}
In this section we present long-term (16 years; MJD 54682 -- 60670) temporal study of TXS 0518+211 in optical, UV, X-ray and gamma-ray bands.  
\begin{figure*}
\centering
\includegraphics[height=5.2in,width=8.0in]{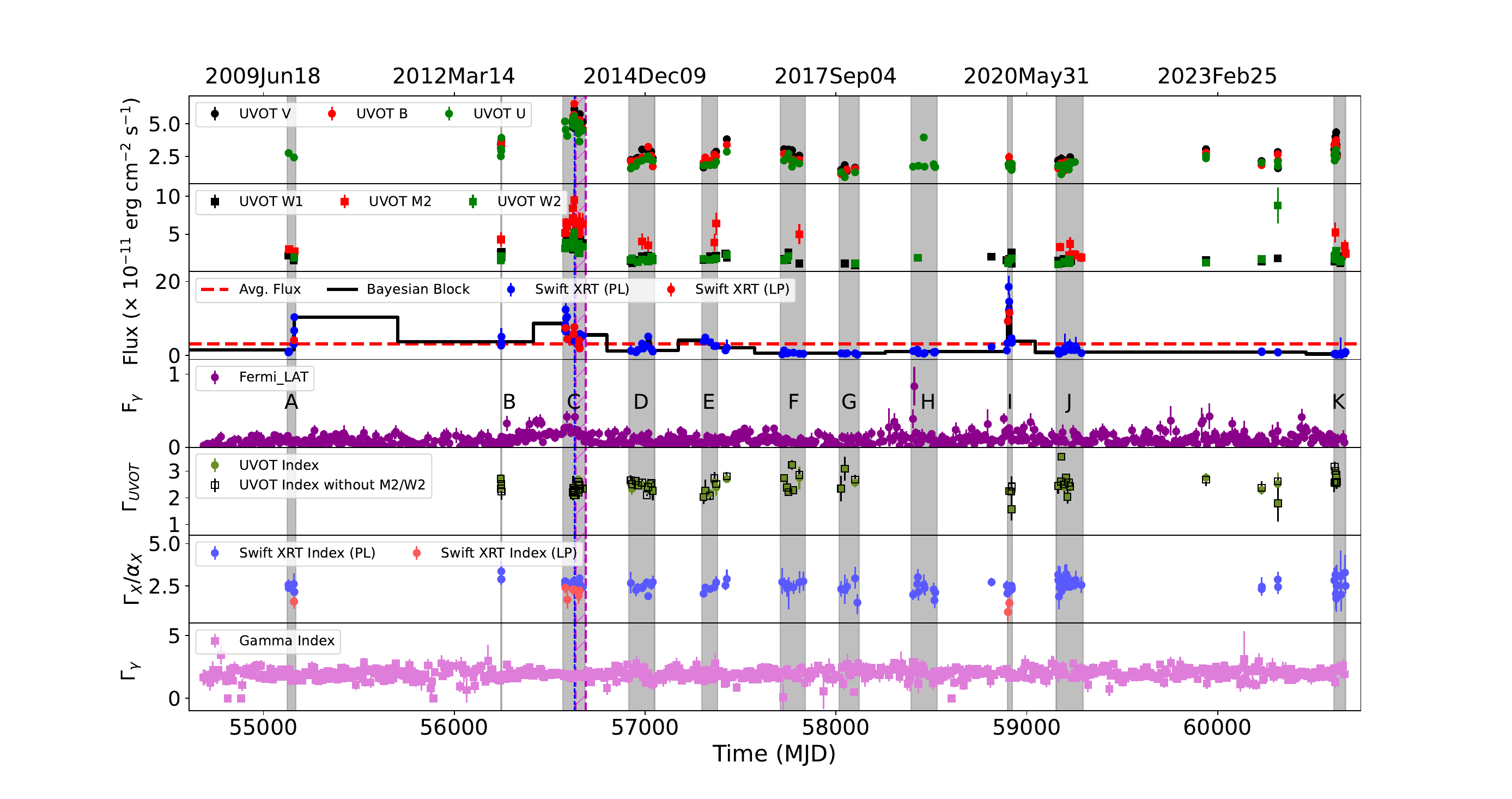}
\caption{Multi-wavelength light curve of TXS 0518+211 during MJD 54682 $-$ 60148. The eleven different epochs (Epoch-A to Epoch-K) identified based on the availability of simultaneous data across all bands are shown by grey shaded vertical strips. The first four panels from top to bottom represent optical ($V$, $B$, $U$ band fluxes plotted with different colours in the same panel), UV ($UVW1$, $UVM2$, $UVW2$ band fluxes plotted with different colours in the same panel), X-ray, and gamma-ray light curves. The last three panels from top to bottom represent optical-UV photon index (${\Gamma}_{\rm UVOT}$), X-ray photon (${\Gamma}_{\rm X}$), and gamma-ray photon index (${\Gamma}_{\gamma}$). The optical/UV, X-ray, and gamma-ray fluxes are in the unit of 10$^{-11}$ erg cm$^{-2}$ s$^{-1}$, 10$^{-11}$ erg cm$^{-2}$ s$^{-1}$, and 10$^{-6}$ ph cm$^{-2}$ s$^{-1}$, respectively. The blue (MJD 56628.5–56632.5; see also Figure-\ref{fig:B1}) and magenta (MJD 56632.5–56689.0) hatched regions indicate the time intervals of the VERITAS observations \citep{Adams2022Jun}.}
\label{fig:1}
\end{figure*}

\subsection{Optical, UV, X-ray, Gamma-ray light curves and spectral indices}
\label{sec:3.1}
Based on the availability of contemporaneous optical, UV, and X-ray observations, we divided the multi-wavelength light curve into 11 different epochs, labeled alphabetically from Epoch-A to Epoch-K, and displayed by grey shaded regions in Figure-\ref{fig:1}. We applied the Bayesian block method \footnote{\url{https://docs.astropy.org/en/stable/api/astropy.stats.bayesian_blocks.html}} \citep{2013ApJ...764..167S} to the long-term X-ray light curve (which exhibits higher variability compared to optical/UV and gamma-ray light curves) and verified that the flux level within each epoch was nearly the same. It is noted that all data points in the other wavebands (optical–UV and gamma-ray) within a given epoch are not strictly simultaneous to X-ray light curve. The corresponding observation times for each instrument for each epoch are discussed and listed in Appendix-\ref{app:A} and Table-\ref{tab:A1}, respectively. The first four panels (from top to bottom) in Figure-\ref{fig:1} represent the optical ($V$, $B$, $U$ band fluxes plotted with different colors in the same panel), UV ($UVW1$, $UVM2$, $UVW2$ band fluxes plotted with different colors in the same panel), X-ray ($F_{\rm 0.3-10~keV}$), and gamma-ray ($F_{\rm 20~MeV-300~GeV}$) bands light curves. The last three panels represent optical-UV (${\Gamma}_{\rm UVOT}$), X-ray (${\Gamma}_{\rm X}$) and gamma-ray (${\Gamma}_{\rm {\gamma}}$) spectral indices across time. To derive ${\Gamma}_{\rm UVOT}$ we fitted a power law to the joint optical-UV spectrum for each observation-ID. We employed this procedure only when there are at least three data points available for a given observation-ID. We also corrected optical UV fluxes for extinction by using E(V-B) = 0.5875 following \cite{Schlafly2011Aug}. We notice that \cite{Schlegel1998Jun} claimed E(V-B) = 0.703, a value nearly 16\% higher than the recent one. The discrepancy in reddening value can affect the extinction correction and consequently the derived fluxes from the magnitudes, specifically in the $M2$ and $W2$ band of UV filter by $\gtrsim$60\%. Therefore, we also computed ${\Gamma}_{\rm UVOT}$ using the same procedure as before, but excluding $M2$ and $W2$ bands. However, we found that this exclusion did not significantly change the indices as shown (by the open black square symbols) in Figure-\ref{fig:1}. We derived ${\Gamma}_{\rm X}$/$\alpha_{\rm X}$ by fitting the X-ray spectra with a power law/logparabola model (see section \ref{sec:2.1.2}). The ${\Gamma}_{\rm {\gamma}}$ are obtained from the \textit{Fermi}-LAT Light Curve Repository (LCR). 

\par
The average flux in all the bands during all the epochs is given Table-\ref{tab:1}. We note that Epoch-K represents the lowest X-ray flux epoch among eleven epochs (Average X-ray flux of 0.43$\pm$0.08). The gamma-ray flux in this epoch (average $\gamma$-ray flux of $0.10 \pm 0.01$) is similar to the average flux of the entire light curve. In optical-UV band, flux is low (i.e., below the average flux value of a particular band), but it does not correspond to the lowest flux value. Among all the Epochs, X-ray flux is highest during Epoch-I but optical-UV fluxes are very low i.e., even lower than the optical-UV flux level of Epoch-K, the low flux state. Our analysis of light curves find the detection of a possible orphan X-ray flare during Epoch-I. We note that, during Epoch-I, the average X-ray flux of 7.78$\pm$1.13, approximately two times higher than the overall average flux of 4.74$\pm$0.55. In the optical bands ($V$, $B$, $U$), UV ($W1$, $W2$), and gamma-ray bands, the average fluxes are 1.85$\pm$0.12, 1.85$\pm$0.15, 1.78$\pm$0.07, 1.71$\pm$0.18, 1.52$\pm$0.16, and 0.15$\pm$0.02, respectively. These values are either lower or similar to the overall average flux for their respective bands, and hence Epoch-I can be considered as a unique flux state, i.e., orphan X-ray flare (also reported in \cite{2025A&A...694A.308M}). 
In contrast, during Epoch-C, we observe simultaneous activity of the source in all three bands, with flux levels about $\sim$ 1.5 – 2.0 times higher than the overall average flux in each band. During Epoch-K, the optical/UV and gamma-ray light curves remained close to the overall average flux level (see Table-\ref{tab:1}), whereas the X-ray light curve showed a significant decrease. The average X-ray flux during this time period is $0.43 \pm 0.08$ (average flux of the entire X-ray light curve is 3.27$\pm$0.33). To the best of our knowledge, this is the first time such a flux state in this source is detected.  

\begin{table*}
\caption{The variability parameters} 
\label{tab:1}
\centering
\scalebox{0.87}{
\begin{tabular}{ccccccc}   
\hline                        
Waveband & Activity Epochs & $F_{var}$ & $\sigma^{2}_{NXS}$  &  Average flux & $\chi^{2}_{red}$ (df) & t$_{var}$ \\ [0.8ex] 
   (1)       &     (2)           &    (3)       &      (4)              &     (5)         &     (6)                   &    (7)        \\
\hline
 & Epoch-B & - & -0.003 & 3.55$\pm$0.08 & 0.51(3) & 21.72 \\
 & Epoch-C & 0.04$\pm$0.02 & 0.001$\pm$0.001 & 5.28$\pm$0.08 & 1.47(16) & 6.40 \\ 
 & Epoch-D & 0.11$\pm$0.04 & 0.012$\pm$0.01 & 2.50$\pm$0.12 & 2.54(8) & 38.11 \\
UVOT-$V$ & Epoch-E & 0.21$\pm$0.05 & 0.04$\pm$0.02 & 2.24$\pm$0.23 & 6.28(3) & 69.89 \\
(Optical) & Epoch-F & 0.05$\pm$0.07 & 0.002$\pm$0.006 & 2.75$\pm$0.12 & 1.34(5) & 31.79 \\
 & Epoch-G & - & -0.003 & 1.59$\pm$0.10 & 0.81(3) & 38.79 \\
 & Epoch-I & - & -0.01 & 1.85$\pm$0.12 & 0.63(2) & 5.44 \\
 & Epoch-J & - & -0.002 & 2.13$\pm$0.08 & 1.04(7) & 21.75 \\
 & Epoch-K & 0.16$\pm$0.04 & 0.02$\pm$0.01 & 3.52$\pm$0.25 & 5.06(5) & 8.08 \\
 & Full LC & 0.41$\pm$0.01 & 0.16$\pm$0.01 & 3.25$\pm$0.16 & 27.13(67) & 1.80 \\
\hline
 & Epoch-B &  0.013$\pm$0.12 & 0.0002$\pm$0.003 & 3.39$\pm$0.11 & 1.47(3) & 9.88 \\
 & Epoch-C & 0.07$\pm$0.01 & 0.004$\pm$0.002 & 5.24$\pm$0.11 & 2.52(16) & 4.74 \\
 & Epoch-D & 0.16$\pm$0.03 & 0.02$\pm$0.01 & 2.30$\pm$0.14 & 3.66(8) & 21.44 \\
UVOT-$B$ & Epoch-E & 0.05$\pm$0.07 & 0.03$\pm$0.01 & 2.36$\pm$0.11 & 1.82(3) & 51.56 \\
(Optical) & Epoch-F & 0.06$\pm$0.06 & 0.003$\pm$0.007 & 2.46$\pm$0.11 & 1.33(5) & 79.72 \\
 & Epoch-G & 0.03$\pm$0.23 & 0.001$\pm$0.01 & 1.38$\pm$0.10 & 1.05(3) & 63.09 \\
 & Epoch-I & 0.12$\pm$0.08 & 0.01$\pm$0.02 & 1.85$\pm$0.15 & 1.61(4) & 3.13 \\
 & Epoch-J & 0.03$\pm$0.12 & 0.001$\pm$0.008 & 1.85$\pm$0.08 & 1.20(8) & 13.09 \\
 & Epoch-K & 0.10$\pm$0.03 & 0.01$\pm$0.001 & 3.23$\pm$0.15 & 2.96(7) & 5.19 \\
 & Full LC & 0.44$\pm$0.01 & 0.19$\pm$0.01 & 3.05$\pm$0.16 & 32.49(72) & 3.13 \\
\hline
 & Epoch-A & 0.09$\pm$0.02 & 0.01$\pm$0.003 & 2.59$\pm$0.17 & 8.44(1) & 217.45 \\
 & Epoch-B & 0.17$\pm$0.04 & 0.03$\pm$0.01 & 3.11$\pm$0.29 & 6.39(3) & 3.42 \\
 & Epoch-C  & 0.09$\pm$0.01 & 0.01$\pm$0.002 & 4.81$\pm$0.11 & 5.07(19) & 5.73 \\
 & Epoch-D  & 0.10$\pm$0.05 & 0.01$\pm$0.009 & 2.06$\pm$0.10 & 1.90(8) & 64.00 \\
UVOT-$U$ & Epoch-E  & - & -0.01 & 1.89$\pm$0.05 & 0.34(3) & 75.26 \\
(Optical) & Epoch-F  & 0.10$\pm$0.06 & 0.01$\pm$0.01 & 2.19$\pm$0.14 & 1.57(5) & 29.37 \\
 & Epoch-G  & 0.04$\pm$0.34 & 0.001$\pm$0.03 & 1.17$\pm$0.12 & 1.10(2) & 57.83 \\
 & Epoch-H & 0.04$\pm$0.02 & 0.002$\pm$0.002 & 1.73$\pm$0.04 & 2.05(6) & 36.23 \\
 & Epoch-I & - & -0.01 & 1.78$\pm$0.07 & 0.86(5) & 3.64 \\
 & Epoch-J & 0.07$\pm$0.05 & 0.005$\pm$0.008 & 1.76$\pm$0.07 & 2.81(12) & 6.63 \\
 & Epoch-K & 0.04$\pm$0.08 & 0.002$\pm$0.007 & 2.59$\pm$0.11 & 1.12(6) & 5.73 \\
 & Full LC & 0.46$\pm$0.01 & 0.21$\pm$0.01 & 2.71$\pm$0.13 & 38.22(87) & 3.42 \\
\hline   
& Epoch-A & 0.20$\pm$0.06 & 0.04$\pm$0.03 & 1.88$\pm$0.29 & 6.54(1) & 90.14 \\
 & Epoch-B & - & -0.002 & 2.42$\pm$0.16 & 1.88(3) & 3.77 \\
 & Epoch-C  & 0.07$\pm$0.03 & 0.005$\pm$0.005 & 4.09$\pm$0.12 & 1.49(16) & 2.56 \\
 & Epoch-D  & 0.08$\pm$0.13 & 0.006$\pm$0.02 &  1.69$\pm$0.12 & 1.23(7) & 29.7 \\
UVOT-$W1$ & Epoch-E  & - & -0.01 & 1.83$\pm$0.12 & 0.81(3) & 28.29 \\
(UV) & Epoch-F  & 0.27$\pm$0.11 & 0.08$\pm$0.06 & 2.97$\pm$0.30 & 2.23(3) & 16.42 \\
 & Epoch-G  & - & -0.06 & 1.03$\pm$0.13 & 0.38(1) & 210.55 \\
 & Epoch-I & 0.20$\pm$0.12 & 0.04$\pm$0.04 & 1.66$\pm$0.21 & 1.98(5) & 1.27 \\
 & Epoch-J & - & -0.03 & 1.47$\pm$0.07 & 0.52(9) & 8.04 \\
 & Epoch-K & 0.14$\pm$0.08 & 0.02$\pm$0.02 & 1.90$\pm$0.14 & 4.05(8) & 4.60 \\
 & Full LC & 0.45$\pm$0.02 & 0.20$\pm$0.02 & 2.29$\pm$0.12 & 10.88(73) & 1.27 \\
\hline
& Epoch-A & - & -0.02 & 2.89$\pm$0.12 & 0.14(1) & 337.98 \\
 & Epoch-C  & - & -0.002 & 6.57$\pm$0.30 & 1.18(15) & 2.62 \\
UVOT-$M2$ & Epoch-D & - & -0.08 &  3.79$\pm$0.24 & 0.10(1) & 248.51 \\
(UV) & Epoch-E  & 0.23$\pm$0.23 & 0.05$\pm$0.11 & 5.15$\pm$1.25 & - & 18.47 \\
 & Epoch-J & 0.10$\pm$0.21 & 0.01$\pm$0.04 & 2.64$\pm$0.29 & 1.03(5) & 10.26 \\
 & Epoch-K & 0.28$\pm$0.18 & 0.08$\pm$0.10 & 3.70$\pm$0.82 & 2.52(2) & 11.55 \\
 & Full LC & 0.35$\pm$0.04 & 0.12$\pm$0.03 & 4.90$\pm$0.34 & 7.78(33) & 2.62 \\
\hline
 & Epoch-B & - & -0.02 & 1.87$\pm$0.12 & 0.89(3) & 3.22 \\
 & Epoch-C  & 0.15$\pm$0.03 & 0.02$\pm$0.009 & 3.59$\pm$0.16 & 2.59(18) & 2.62 \\
 & Epoch-D  & - & -0.04 & 1.63$\pm$0.08 & 0.32(7) & 30.13 \\
UVOT-$W2$ & Epoch-E  & - & -0.06 & 1.71$\pm$0.04 & 0.07(2) & 108.64 \\
(UV) & Epoch-F  & - & -0.04 & 1.81$\pm$0.16 & 0.39(2) & 42.32 \\
 & Epoch-I & - & -0.01 & 1.52$\pm$0.16 & 0.89(2) & 18.17 \\
 & Epoch-J & - & -0.03 & 1.32$\pm$0.10 & 0.79(4) & 9.90 \\
 & Epoch-K & - & -0.004 & 1.95$\pm$0.17 & 1.45(6) & 4.26 \\
 & Full LC & 0.50$\pm$0.03 & 0.25$\pm$0.03 & 2.41$\pm$0.17 & 7.34(59) & 2.62 \\
\hline   
\hline
\end{tabular}
}
\end{table*}

\addtocounter{table}{-1}

\begin{table*}[t!]
\caption{\textbf{Continued.}} 
\centering
\scalebox{0.87}{
\begin{tabular}{ccccccc}   
\hline
Waveband & Activity Epochs & $F_{var}$ & $\sigma^{2}_{NXS}$  &  Average flux & $\chi^{2}_{red}$ (df) & t$_{var}$ \\ [0.8ex] 
   (1)       &     (2)           &    (3)       &      (4)       &     (5)         &     (6)                   &    (7)        \\
\hline
 & Epoch-A & 0.96$\pm$0.04 & 0.92$\pm$0.08 & 4.02$\pm$1.58 & 202.45(5) & 2.07 \\
 & Epoch-B & 0.15$\pm$0.67 & 0.02$\pm$0.19 & 3.86$\pm$1.21 & 0.60(1) & 3.21 \\
 & Epoch-C & 0.37$\pm$0.05 & 0.13$\pm$0.03 & 5.79$\pm$0.48 & 8.34(26) & 0.99 \\ 
 & Epoch-D & 0.59$\pm$0.06 & 0.35$\pm$0.08 & 2.24$\pm$0.49 & 12.54(7) & 10.58 \\
\textit{Swift}-XRT & Epoch-E & 0.23$\pm$0.07 & 0.05$\pm$0.03 & 3.46$\pm$0.50 & 3.19(2) & 38.60 \\
(X-ray) & Epoch-F & 0.36$\pm$0.07 & 0.13$\pm$0.05 & 3.11$\pm$0.50 & 8.39(5) & 19.26 \\
 & Epoch-G & - & -0.22 & 0.47$\pm$0.07 & 1.46(4) & 13.06 \\
 & Epoch-H & 0.17$\pm$0.19 & 0.03$\pm$0.06 & 0.88$\pm$0.11 & 1.09(8) & 5.44 \\
 & Epoch-I & 0.71$\pm$0.07 & 0.50$\pm$0.09 & 7.78$\pm$1.73 & 83.75(10) & 2.96 \\
 & Epoch-J & 0.29$\pm$0.13 & 0.08$\pm$0.08 & 1.53$\pm$0.18 & 5.13(18) & 1.28 \\
 & Epoch-K & - & -0.20 & 0.43$\pm$0.08 & 0.91(7) & 3.21 \\
 &  Full LC & 1.10$\pm$0.03 & 1.02$\pm$0.05 & 3.27$\pm$0.33 & 72.82(109) & 0.99 \\
\hline
 & Epoch-A &  - & -0.09 & 0.09$\pm$0.01 & 1.42(6) & 7.86 \\
 & Epoch-C & 0.26$\pm$0.07 & 0.07$\pm$0.04 & 0.23$\pm$0.02 & 3.07(16) & 10.30 \\
 & Epoch-D & - & -0.02 & 0.09$\pm$0.01 & 8.63(17) & 4.70 \\
\textit{Fermi}-LAT & Epoch-E & - & -0.04 & 0.08$\pm$0.01 & 2.74(9) & 5.50 \\
(Gamma-ray) & Epoch-F & - & -0.17 & 0.08$\pm$0.01 & 3.65(12) & 8.70 \\
 & Epoch-G & - & -0.03 & 0.09$\pm$0.01 & 60.52(17) & 3.80 \\
 & Epoch-H & 0.89$\pm$0.15 & 0.79$\pm$0.26 & 0.20$\pm$0.06 & 3.71(12) & 5.70 \\
 & Epoch-I & 0.11$\pm$0.31 & 0.003$\pm$0.07 & 0.18$\pm$0.03 & 1.31(4) & 6.69 \\
 & Epoch-J & - & -0.09 & 0.12$\pm$0.01 & 1.17(12) & 4.76 \\
 & Epoch-K & - & -0.49 & 0.10$\pm$0.01 & 0.16(3) & 3.21 \\
 & Full LC & 0.45$\pm$0.03 & 0.20$\pm$0.02 & 0.11$\pm$0.01 & 14.77(674) & 3.71 \\
\hline   
\hline
\end{tabular}
}
\\
Notes. {Column 1 : Instrument/filter; Column 2 : Epoch; Column 3 : Fractional variability parameter ($F_{var}$); Column 4 : 
Normalized excess variance ($\sigma^{2}_{NXS}$); Column 5 : Average flux in the units of 10$^{-11}$ erg cm$^{-2}$ s$^{-1}$); Column 6 : Reduced chi-square values ($\chi^{2}_{red}$) of the fit obtained with a constant value and degree of freedom written within brackets; Column 7 : variability time-scale (t$_{var}$) in days units.}

\end{table*}

\begin{figure*}
\centering
\includegraphics[scale=0.16]{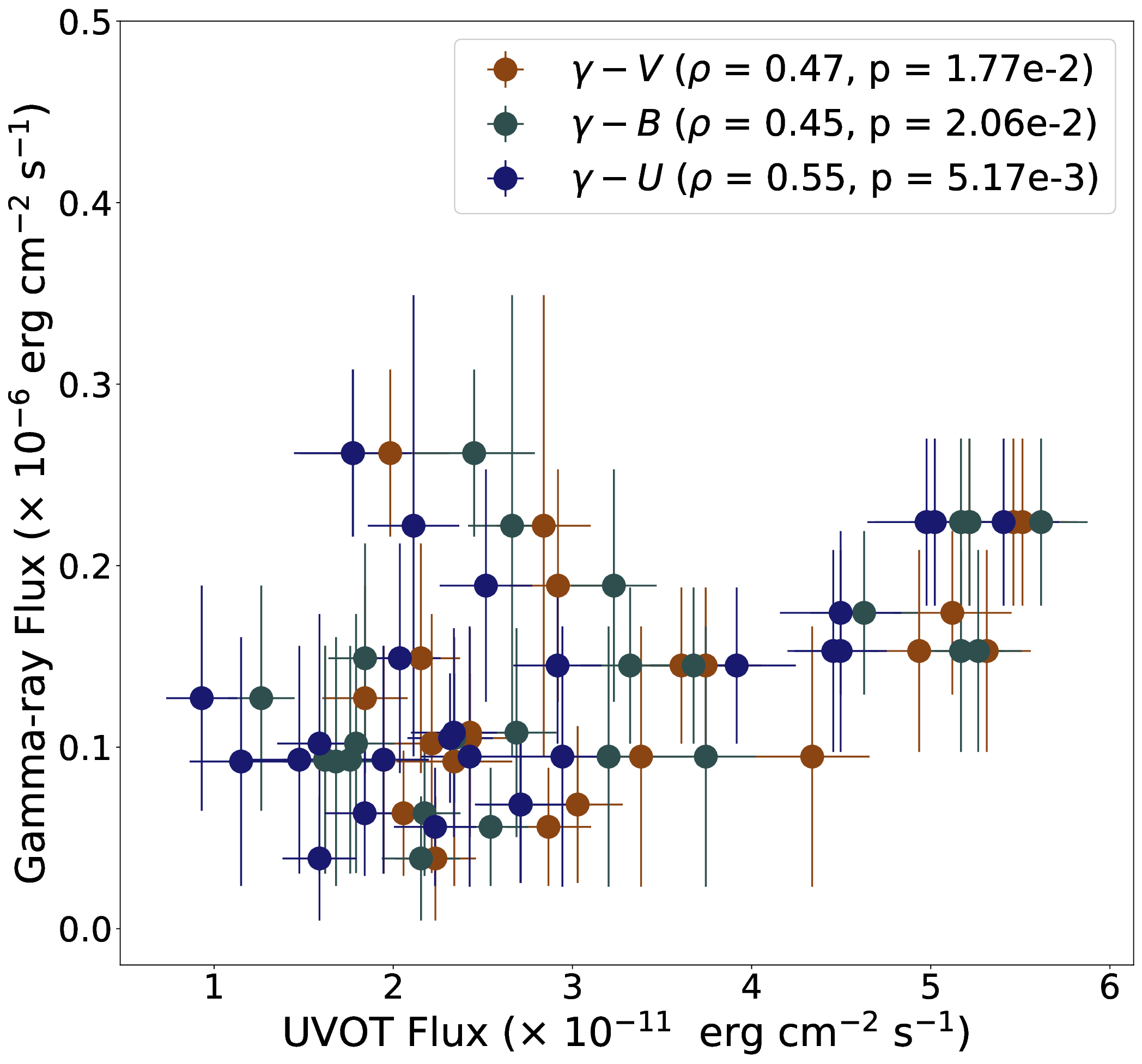}
\includegraphics[scale=0.16]{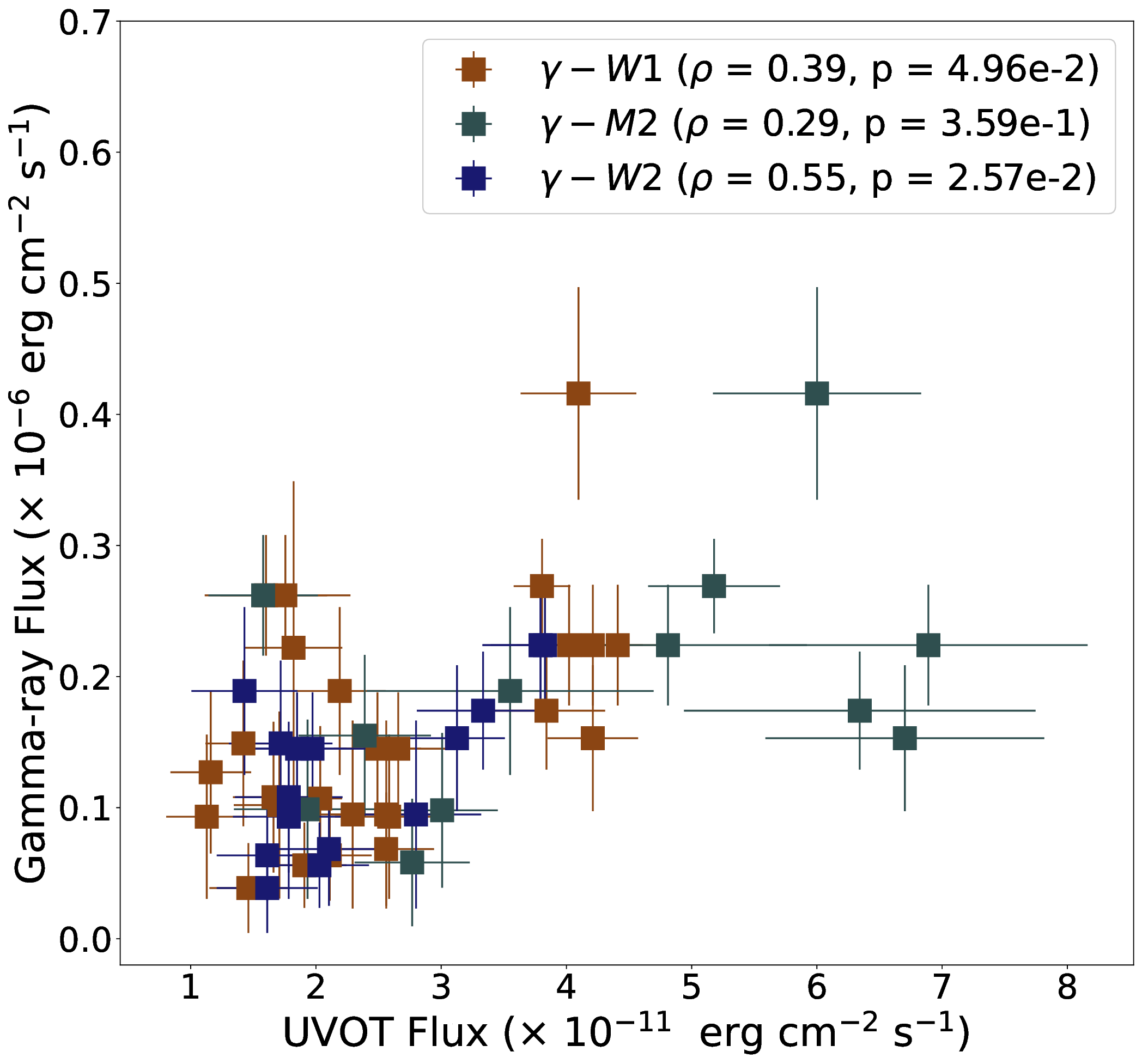}
\includegraphics[scale=0.16]{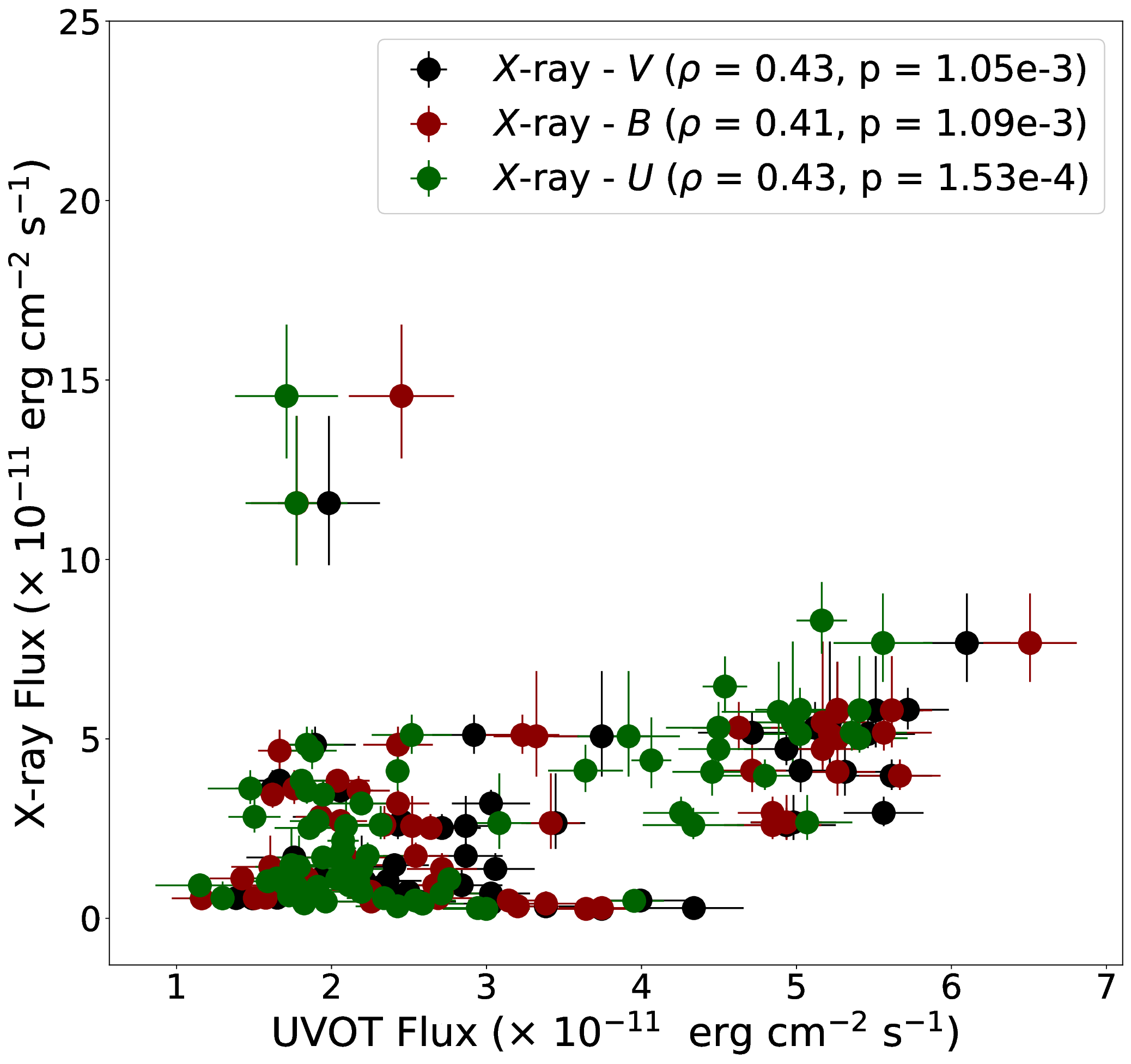}
\includegraphics[scale=0.16]{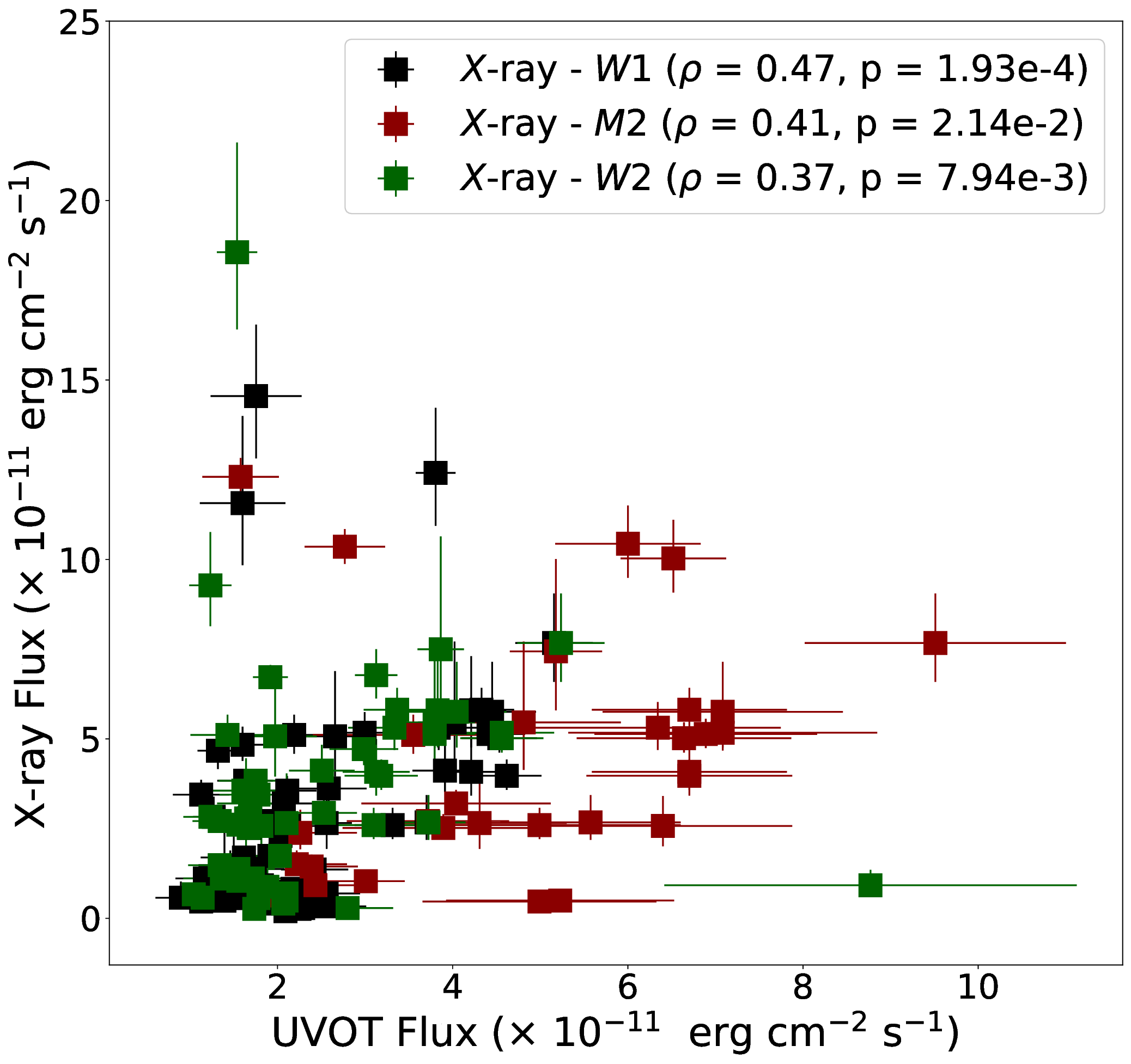}
\includegraphics[scale=0.16]{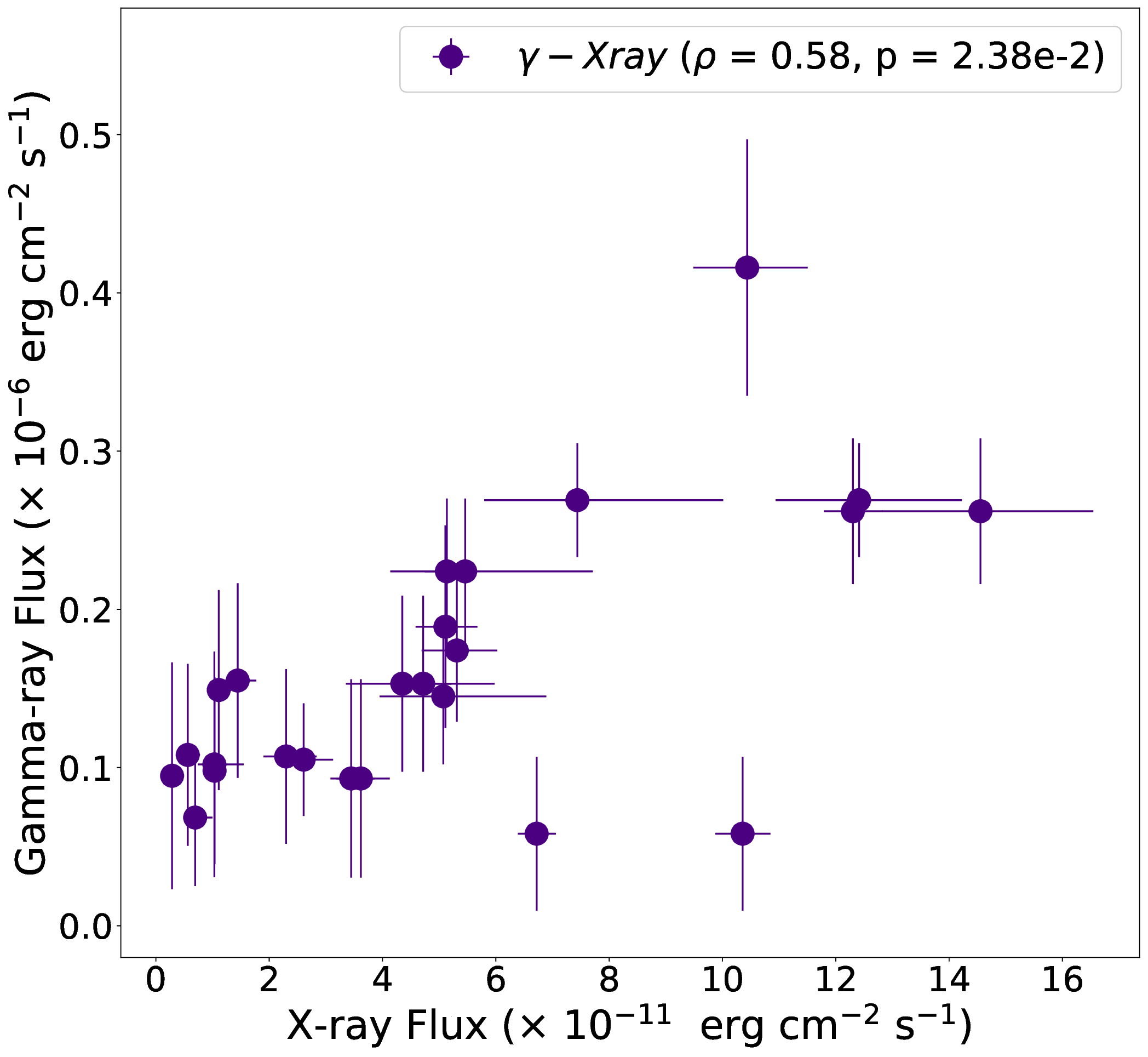}
\caption{Flux-flux correlation plots. }
\label{fig:2}
\end{figure*}

\subsection{Fractional variability}
To characterize the long-term variability in the light curve of each band, we used the well-known fractional variability amplitude. It is defined \citep{Vaughan2003Nov} as follows:

\begin{equation}
    F_{var} = \sqrt{\frac{S^{2}-\overline{\sigma_{err}^{2}}}{F^{2}}}
\end{equation}
where, $S^{2}$, $\overline{\sigma_{err}^{2}}$, and $F$ are the variance, mean square error ($\frac{1}{N} \sum_{i=1}^{N} \sigma_{err,i}^{2}$), and mean flux of the light curve segment, respectively. The uncertainty associated with the fractional variability is given by:

\begin{equation}
    \Delta F_{var} = \sqrt{\left(\frac{1}{\sqrt{2N}} \frac{\overline{\sigma_{err}^{2}}}{F^{2}F_{var}}\right)^{2} + \left(\frac{\overline{\sigma_{err}^{2}}}{N} \frac{1}{F} \right)^{2}}
\end{equation}

It is noted that when the signal-to-noise ratio is low or intrinsic variability is weak, then $S^{2}\lesssim\overline{\sigma_{err}^{2}}$, implying weak detection of variability. The values of fractional variability amplitudes ($F_{var}$), normalized excess variance ($\sigma_{NXS}^{2}$) estimated for multi-wavelength light curves in different epochs are given in Table-\ref{tab:1}. To compute the fractional variability error ($\Delta F_{var}$) in the X-ray band, where uncertainty on the flux is asymmetric, we used larger error value for each data point. Also, during a few epochs we lack some data points. For instance $V$ band optical data are missing in Epoch-A.
\par
We found that the X-ray flux variability across different epochs is relatively higher compared to that of the optical and gamma-ray bands, as indicated by the fractional variability analysis (see Table~\ref{tab:1}). More detailed results of the fractional variability study are discussed in Section~\ref{sec:5.1}.

\subsection{Variability timescales}
To characterize the fastest timescale of variability, we computed the `doubling time' using the procedure given in \cite{Zhang1999Dec}. It is defined by:
\begin{equation} \label{eq:3}
    \tau_{ij} = \left\lvert \frac{F\Delta t_{ij}}{\Delta F_{ij}} \right\rvert 
\end{equation}

where, $\Delta t_{ij} = t_{j} - t_{i}$, $\Delta F_{ij} = F_{j} - F_{i}$, and $F = \frac{F_{j} + F_{i}}{2}$. $F_{i}$ and $F_{j}$ are the flux values at time $t_{i}$ and $t_{j}$, respectively. Using equation \ref{eq:3}, the timescales for all pairs of two data points in a given light curve segment are computed, and the minimum value of timescale corresponds to the shortest variability time scale ($t_{var}$). This timescale is useful for constraining the size of the emission regions. We note that $t_{var}$ depends on the various quantities such as bin size, length, signal-to-noise ratio of the time series. 

For the entire light curve, the shortest $t_{var}$ is found in the X-ray band ($t_{var} \sim 1$ day), followed by the UV ($t_{var} \sim 2.2$ days, averaged over W1, M2, and W2 bands), optical ($t_{var} \sim 2.8$ days, averaged over V, B, and U bands), and gamma-ray ($t_{var} \sim 3.5$ days) bands. We list the variability timescales ($t_{var}$) for each light curve in different epochs in Table~\ref{tab:1}.

\begin{figure*}
\centering
\includegraphics[scale=0.16]{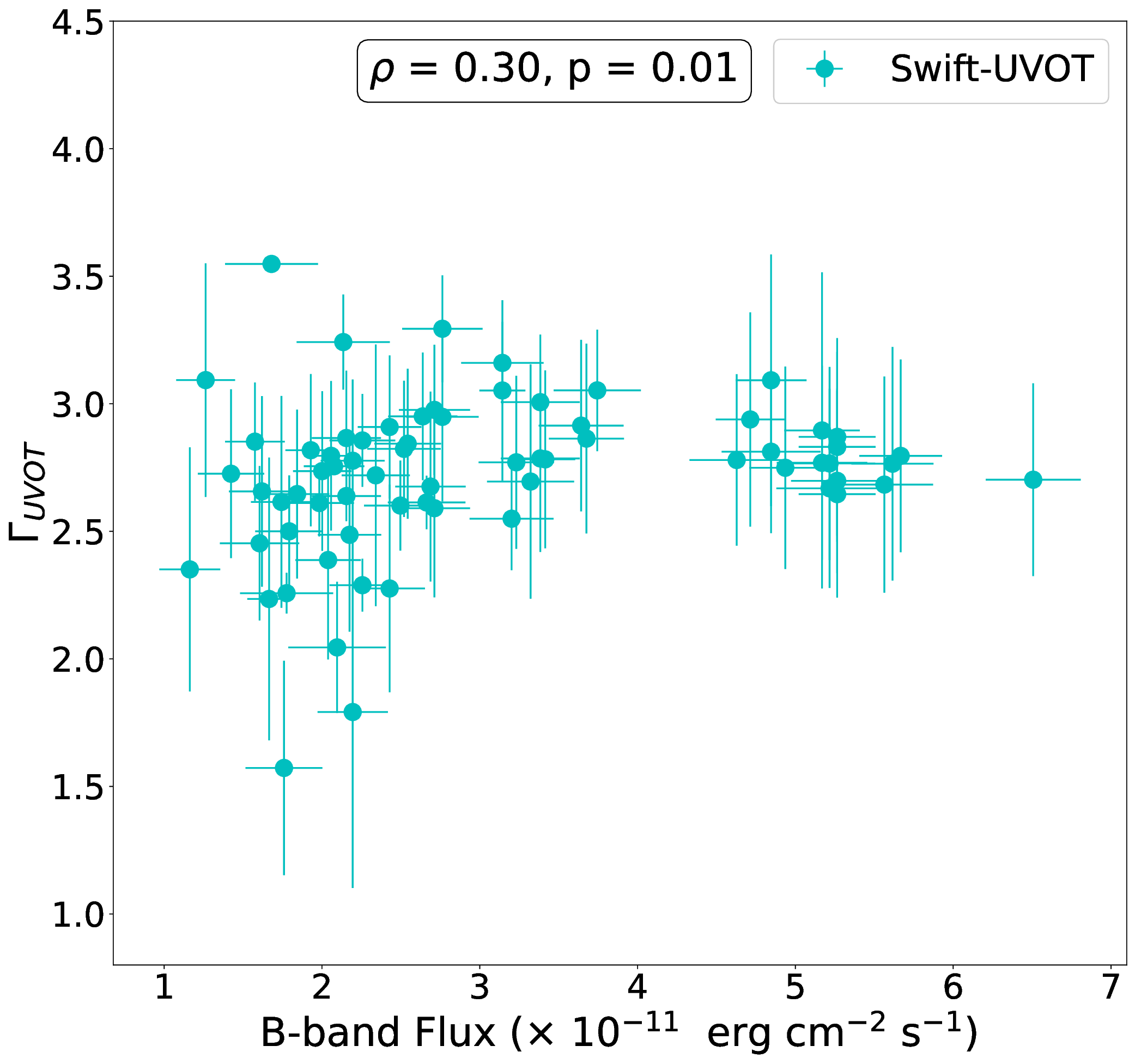}
\includegraphics[scale=0.16]{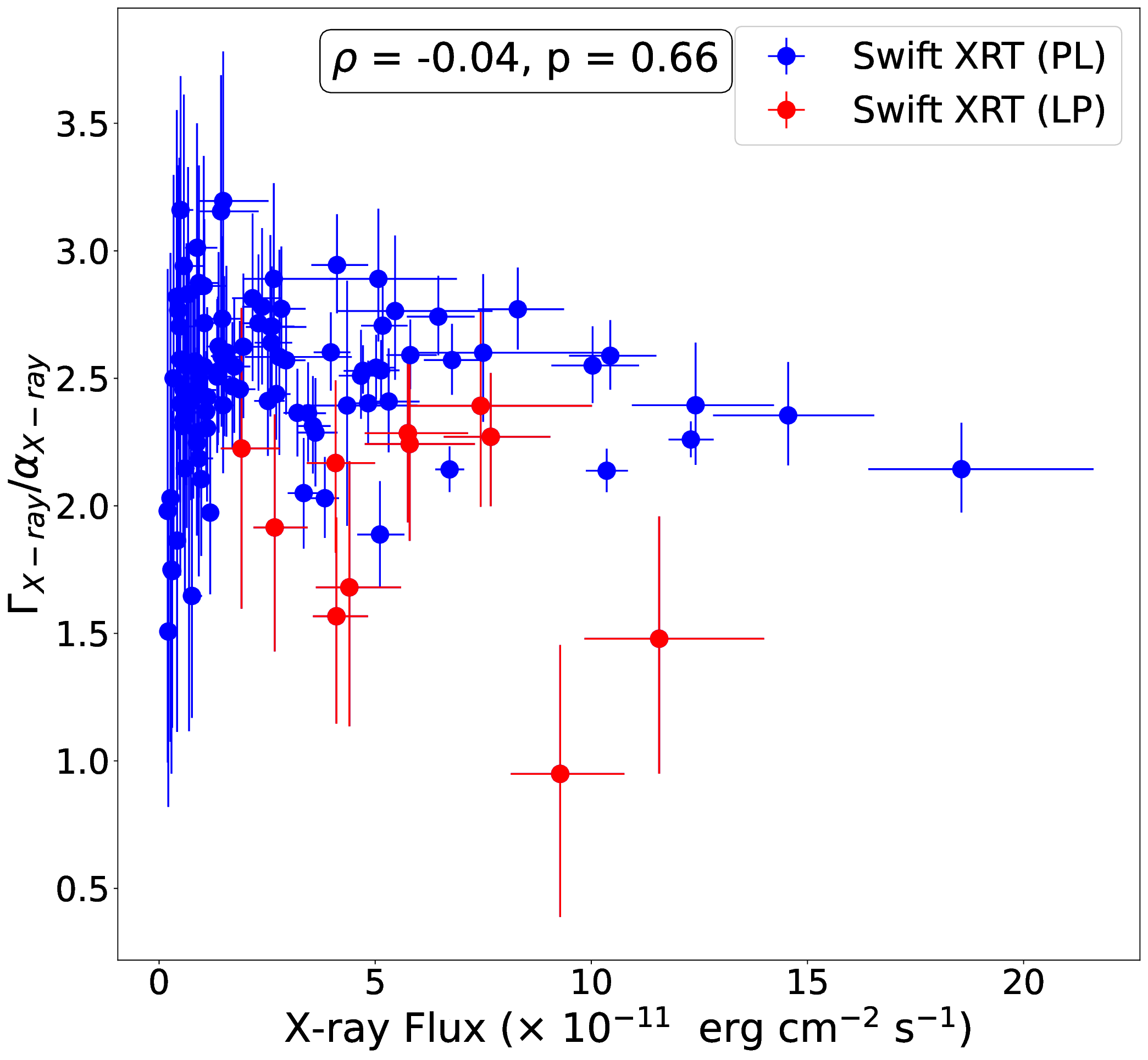}
\includegraphics[scale=0.16]{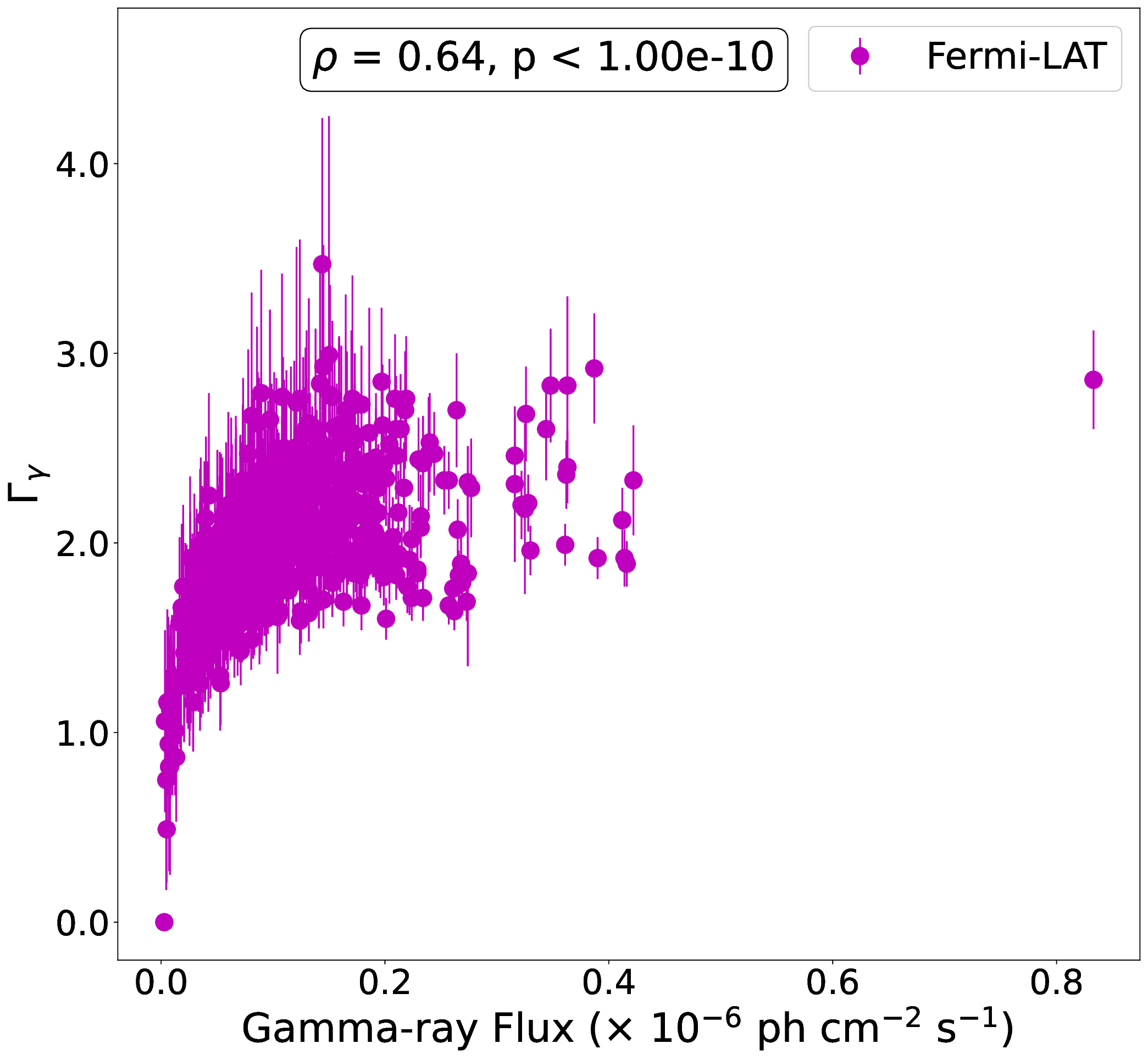}
\caption{Photon index/spectral index versus flux plots.}
\label{fig:3}
\end{figure*}

\subsection{The correlation between fluxes in different bands}
\label{sec:3.4}
The flux$-$flux correlations between different bands are useful to understand the location of the emission regions. We examined correlations between fluxes in optical-UV bands ($U$, $B$, $V$, $W1$, $W2$ and $M1$), X-ray and gamma-ray bands. The correlation plots are shown in Figure-\ref{fig:2}. For gamma -- optical/UV and gamma -- X-ray correlations where cadence of data is widely different in the two bands, we ensured simultaneity by considering only those data points in the two bands which lie within a 1-day window. To evaluate the strength of correlations we computed Spearman correlation coefficients ($\rho$) and $p$-values for each correlations. We find that, in general, there is a moderate trend of increase in flux in one band with the increase in flux in other band. The correlation coefficient ($\rho$) in flux-flux correlations among various bands are found to be on the range of 0.29 to 0.58 ($>$ 95\% significance level in all cases except one, gamma-ray vs. M2 band). X-ray and $\gamma$-ray, the two high energy bands, show relatively strong correlation with correlation coefficient ($\rho$) 0.58 and $p$-value = 0.024. From X-ray versus optical/UV correlation plot, a possible X-ray flare with no optical-UV counterpart is also evident, which is consistent from the result obtained from multi-wavelength light curves analysis i.e, an X-ray flare detected in Epoch-I but with no optical/UV counterpart (see Section~\ref{sec:3.1}).

\par 

Figure \ref{fig:3} shows correlation plots of spectral indices versus fluxes i.e., $F_{\rm optical}$ Vs ${\Gamma}_{\rm UVOT}$, $F_{\rm X-ray}$ Vs ${\Gamma}_{\rm X-ay}$ and $F_{\rm {\gamma}-ray}$ Vs ${\Gamma}_{\rm {\gamma}-ray}$. In ${\gamma}$-ray band, we find a clear `softer-when-brighter' trend with a high significance  (Spearman rank correlation coefficient ($\rho$) = 0.64 and $p$-value $<$ 1.00$\times$10$^{-10}$). In optical (B-band) too we observe a similar trend between flux and spectral index but with smaller correlation coefficient and lower significance level ($\rho$ = 0.30 and $p$ = 0.01). We note that the long-term observations do not show any clear correlation between X-ray flux and photon index with $\rho$ = -0.04 and $p$-value = 0.66. However, considering only index values with relatively lower error bars (corresponding to X-ray flux levels $\gtrsim$ 1.0 × 10$^{-11}$), we observe a very weak harder-when-brighter trend ($\rho$ = -0.32, p = 3.98$\times$10$^{-3}$). A similar trend was reported in \citet{Adams2022Jun}, but with a higher correlation coefficient (Pearson’s correlation coefficient, r = 0.74). Their broadband study covered a time span of $\sim$ 1 year (MJD $\sim$ 56244 - 56673). \\
\\

\subsection{Intra-Night Optical Variability (INOV)} 
\label{sec:3.5}
In addition to long-term variability study we also examined intra-night optical variability (INOV) in this source using $R$-band photometric monitoring observations from a 11K~$\times$~8.0K front illuminated CMOS imager installed on 1.2m telescope at the Mount Abu observatory\footnote{\url{https://www.prl.res.in/~miro/}}. 
Our observations include monitoring on three sessions, each lasting $\geq$ 3.0 hours, on three different nights. 
The imager is equipped with Johnson filter system ($U$, $B$, $V$, $R$, $I$ filters) and we chose to perform 
monitoring in $R$ filter owing to the fact that imager sensitivity is maximum in this filter.
The CMOS imager having 3.76 $\mu$m pixel size provides 9$^{\prime}$.0 $\times$ 8$^{\prime}$.0 field-of-view (FoV). 
It has readout noise of approximately 1e$^{-}$ at 0°C and gives a dark current of 0.003 e$^{-}$ pixel$^{-1}$ sec$^{-1}$  
During observations we set exposure time of each frame 3.0 to 5.0 minutes depending upon sky conditions and achieved signal-to-noise ratio (SNR) of 20$-$25. To perform flat fielding and bias subtraction we also acquired five flat frames taken at the dusk or dawn and 10 bias frames in each observing session.  
\par 
We reduced data using standard tasks in AstroImageJ\footnote{\url{https://www.astro.louisville.edu/software/astroimagej/}} software. We extracted instrumental magnitudes of both the target and comparison stars using standard aperture photometry \citealt{Collins2017Jan}. The comparison stars are chosen such that they are non-varying or least varying among a 
set chosen potential comparison stars, have counts/magnitudes similar to the target source. We note that aperture size is a crucial parameter while deducing the instrumental magnitude. To determine optimum aperture size we followed a method based on the stabilization of flux/magnitude with gradual increase in the aperture size (see \citealt{Ojha2021Mar}). We find optimum aperture size equal to 2$\times$FWHM. Further, to examine INOV in each monitoring session we plot Differential Light Curves (DLCs) of target source (say `T') {\it w.r.t} two comparison stars named as `C1' and `C2'. 
\par
The presence or absence of INOV is determined by performing two different versions of $F$-test: (i) standard $F$-test (hereafter $F^{\eta}$ test; \citealt{deDiego2010Feb, Goyal2012Aug}), and (ii) power-enhanced $F$-test (hereafter $F^{p-en}$ test; \citealt{deDiego2014Oct}). 
In case of $F^{\eta}$ test, we compute the $F$-statistic of each DLCs in a monitoring session using the following equation:

\begin{equation*}
    F^{\eta}_{1} = \frac{Var(T-C1)}{\eta^{2}\overline{\sigma_{err}^2(T-C1)}}, F^{\eta}_{2} = \frac{Var(T-C2)}{\eta^{2}\overline{\sigma_{err}^2(T-C2)}}, 
\end{equation*}
\begin{equation} \label{eq:4}
    F^{\eta}_{3} = \frac{Var(C1-C2)}{\eta^{2}\overline{\sigma_{err}^2(C1-C2)}}
\end{equation}

where, $Var(T-C1)$, $Var(T-C2)$, $Var(C1-C2)$ are the observed variances in the data and $\overline{\sigma_{err}^2(T-C1)}$, $\overline{\sigma_{err}^2(T-C2)}$, $\overline{\sigma_{err}^2(C1-C2)}$ are the expected variance (i.e., mean square rms errors of each DLCs, defined by $\frac{\sum_{i=1}^{N}\sigma^{2}_{err,i}}{N}$). The value of $\eta$ is set to 1.54, as described in \cite{2025ApJ...990...79S}. To asses the INOV we compared $F^{\eta}_{1}$ and  $F^{\eta}_{2}$ values with the critical value ($F_{crit}$), which corresponds to 95\% and 99\% confidence levels. A DLC is designated to be variable (V) if $F^{\eta}$ is higher than $F_{crit}(0.99)$ (i.e., 99\% significant), else non-variable (NV; $F^{\eta}$ < $F_{crit}(0.95)$) or probably variable (PV) if $F_{crit}(0.95)$ $\leq$ $F^{\eta}$ $\leq$ $F_{crit}(0.99)$). 
The target is considered variable if both DLCs i.e., T-C1 and T-C2 exhibit variability.
\par
The power enhanced $F$-test ($F^{p-en}$) is more robust than the standard one as it includes more than two comparison stars to compute the combined variance and hence enhance the power of the $F$-test (see \citealt{deDiego2014Oct} for more discussion). We used three steady stars (one Reference (R1) star and two comparison stars (C1, C2) for this test. 
The $F^{p-en}$ is defined as follows:

\begin{equation} \label{eq:5}
    F^{p-en} = \frac{Var(T-R1)}{\sigma^2_{comb}}  
\end{equation}

where, Var(T-R1) is the observed variance of the `T-R1' DLC and $\sigma^2_{comb}$ is the combined variance of `Companion - R1' DLCs.

\begin{equation}
    \sigma^{2}_{comb} = \frac{1}{(\sum_{j=1}^{p}N_{j}) - p} \sum_{j=1}^{p}\sum_{i=1}^{N} s^{2}_{ji}, \\ 
    s^{2}_{ji} = \omega_j(m_{ji} - \Bar{m}_{j})^{2}
\end{equation}
$N_{j}$ is the number of data points in $j$th `Companion-R1' DLC. In our case, j runs from 1 to 2. $\Bar{m}_{j}$ represents the mean magnitude of the $j$th `Companion-R1' DLC. $\omega_{j}$ is the scaling factor used to scale the variance of the $j$th `Companion-R1' DLC to the level of the `T-R1' DLC, and its value is taken as described in \citealt{Joshi2011Apr}. 
The INOV status of the target is derived by the same criteria as discussed in case of $F^{\eta}$ test.

The INOV status of our target sources TXS 0518+022 resulting from the two different versions of $F$-tests is listed 
in Table-\ref{tab:2}. The DLCs of all three monitoring sessions  are shown in Figure-\ref{fig:3}. Among three intra-night optical monitoring sessions carried out during January and March, 2025, none of the sessions showed strong INOV. During MJD 60755, DLC is identified as `variable'; however, after correcting for false discovery (i.e., the Bonferroni correction), it changes to `probable variable' case. 
It should be noted that the INOV results mentioned here are not absolute, and the source may have been in an INOV-quiescent phase during the total time period of observation \citep{1999A&AS..135..477R}.

\begin{figure*}
\centering
\includegraphics[scale=0.29]{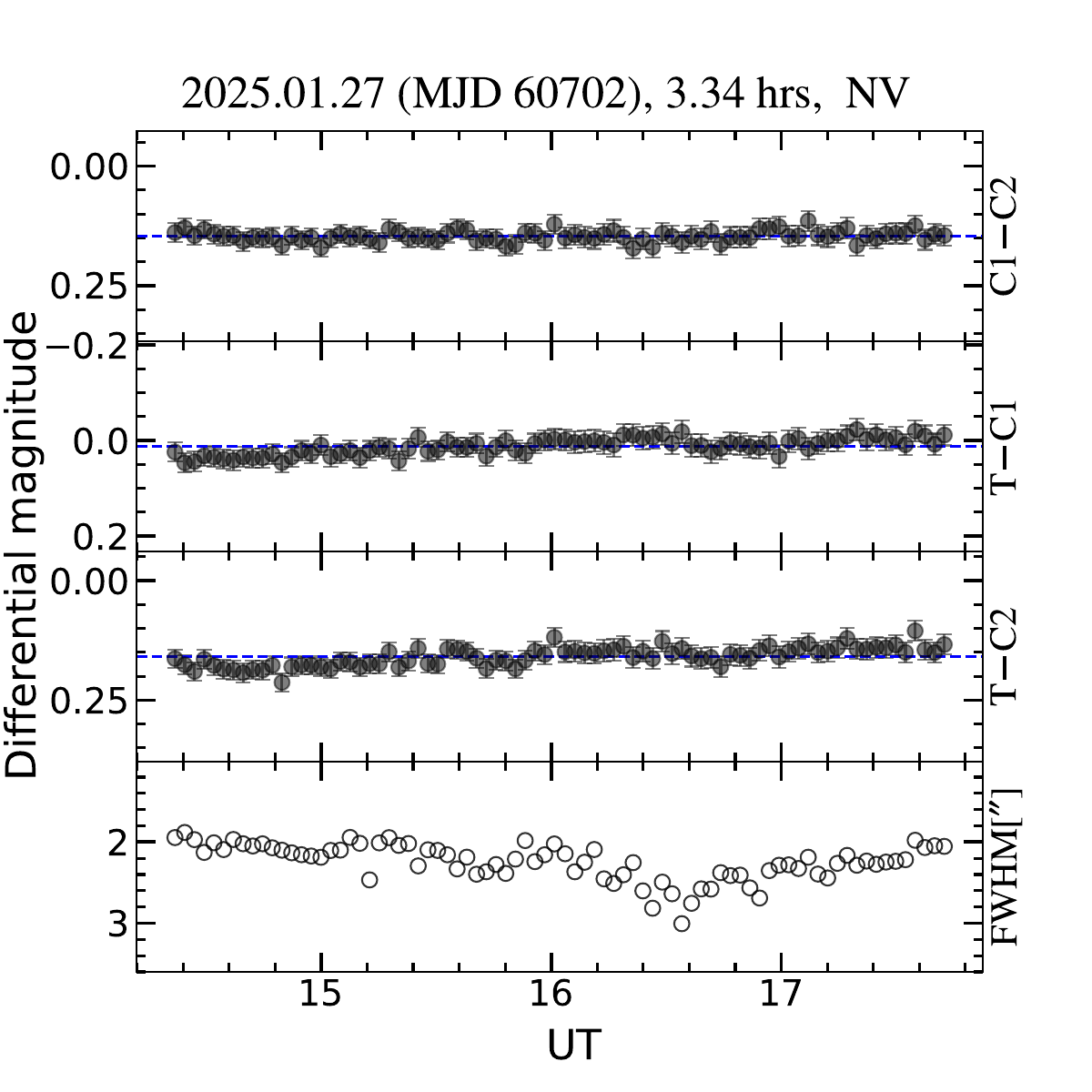}
\includegraphics[scale=0.29]{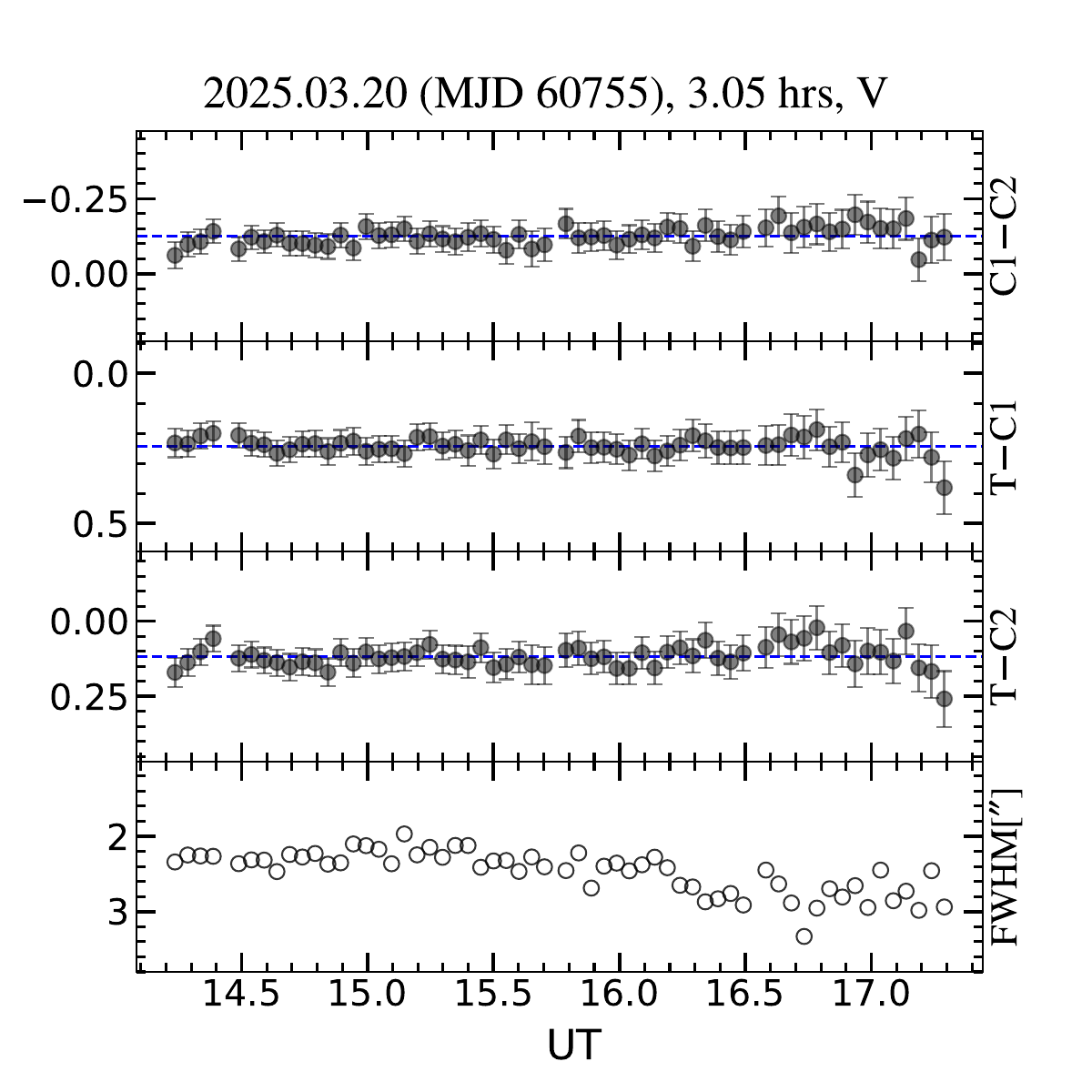}
\includegraphics[scale=0.29]{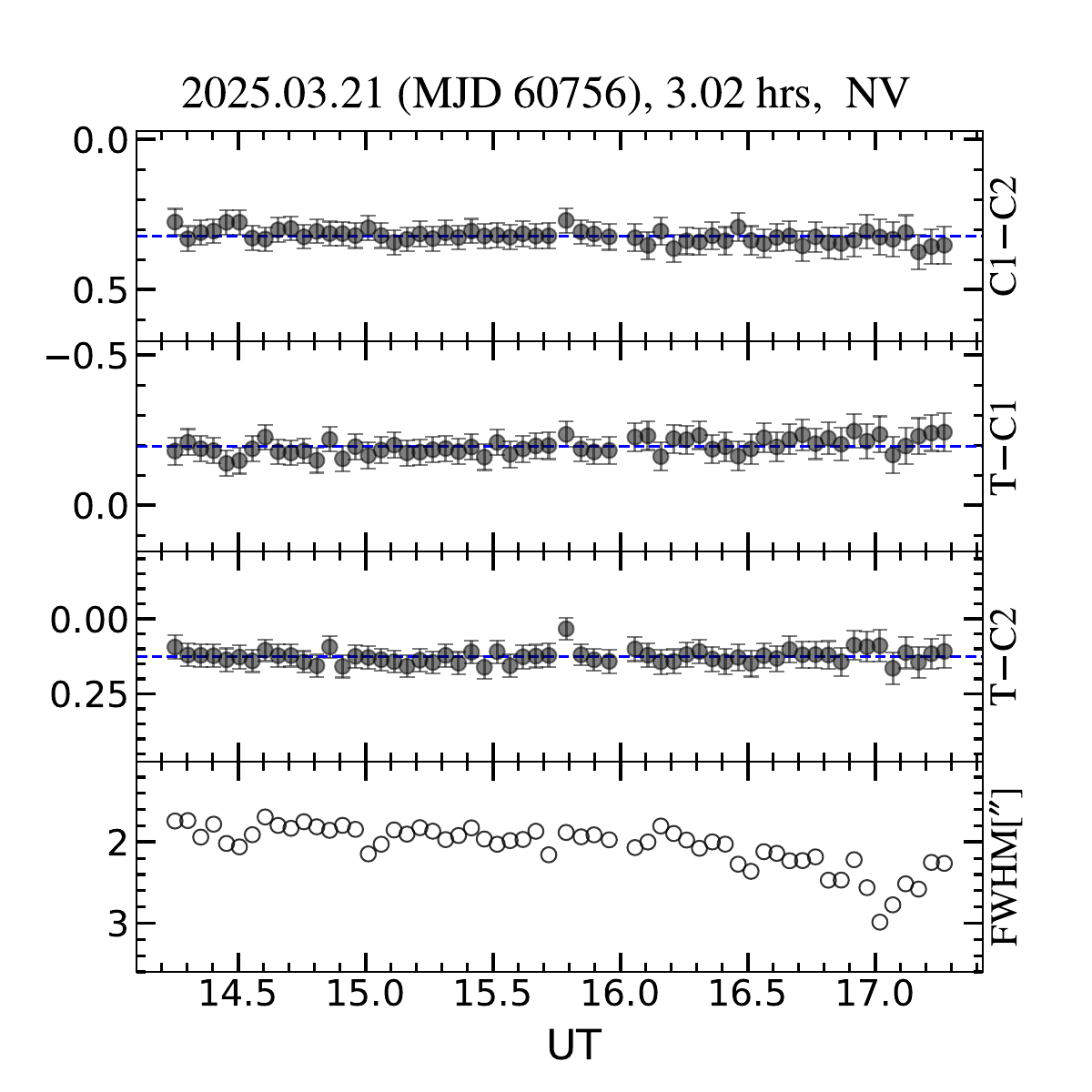}
\caption{The differential light curves (DLCs) of TXS 0518+211 in $R$-band for three intra-night monitoring sessions. 
The date, time duration, and variability status are mentioned in the top of each DLC plot. 
In each plot, the uppermost panel represents the DLC of two comparison sources (C1 and C2), while the other panels 
show DLCs of target {\it w.r.t.} to two comparison sources (T-C1, T-C2). 
The lowermost panel shows seeing (PSF) variations across the monitoring session.}
\label{fig:4}
\end{figure*}

\begin{table*}
\caption{INOV status based on the $F^{\eta}$ and $F^{p-en}$ statistical tests.} 
\label{tab:2}
\centering
\begin{tabular}{cccccc rrrr}   
\\
\hline\hline                        
Observation Date & N & Time Duration & Mean FWHM & $F^{\eta}_{1}$, $F^{\eta}_{2}$  & INOV status & $F^{\eta}_{3}$ & INOV status & $F^{p-ehn}$ & INOV status \\ [0.8ex] 
(yyyy.mm.dd) & & (Hours) & (arcseconds) & & & & & & \\
\hline
\hline
2025.01.27 & 80 & 3.34 & 2.24 & 0.28, 0.41 & NV, NV & 0.15 & NV & 0.96 & NV \\
2025.03.20 & 59 & 3.05 & 2.48 & 0.26, 0.25 & NV, NV & 0.13 & NV & 1.87 & V \\
2025.03.21 & 59 & 3.00 & 2.05 & 0.13, 0.18 & NV, NV & 0.18 & NV & 1.02 & NV \\
\hline   
\hline
\end{tabular}
\\ 
Notes -  NV : non-variable, V : variable. $N$ represents the number of data points in a DLC.
\end{table*}

\begin{figure*}
\centering
\includegraphics[scale=0.30]{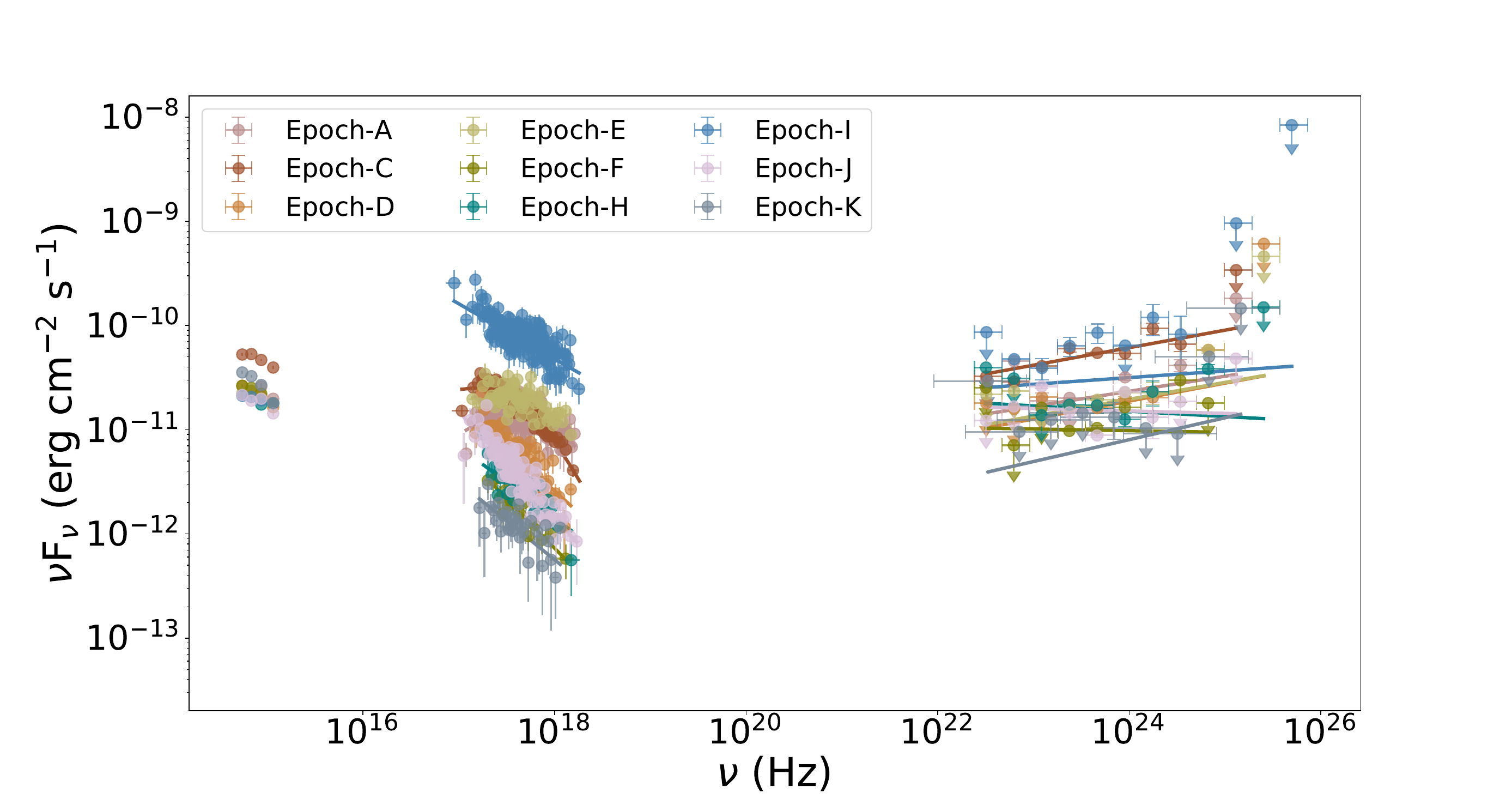}
\caption{Multi-wavelength SED for nine different epochs (including Epoch-K). }
\label{fig:5}
\end{figure*}

\section{BROADBAND SPECTRAL STUDY} \label{sec:4}
To understand change in flux state and emission mechanisms across various epochs we analyses broad-band SEDs in all epochs. 
In Figure-\ref{fig:5} we have plotted broad-band SEDs where each epoch is depicted by a different color. We note the exclusion of $UVM2$ and $UVW2$ bands data in each epoch due to issues in extinction correction (see Section \ref{sec:3}). 
In Epochs B, G, and K, gamma-ray fluxes are only upper limits, and hence, these epochs are not considered for SED modeling. As discussed earlier, the redshift of this source is still uncertain, 0.18 $\leq$ z $\leq$ 0.34. We assume z = 0.25 for the SED modeling. Also, we include X-ray fluxes only above 1 keV, as our source TXS 0518+211 is located near the Galactic plane and high absorption X-ray photons is expected at $\leq$ 1.0 keV.

\subsection{Broad-band SED modeling} 
To model the broad-band SED we used the publicly available code `GAMERA' \footnote{\url{http://libgamera.github.io/GAMERA/docs/main_page.html}} \citep{Hahn:2016CO} that solves time-dependent continuity equation for an input injected particle (only leptons are considered in our case) spectrum and compute the photon spectrum for different radiative losses of electrons (e.g., synchrotron, SSC) at different time stamps. This code also incorporates the Klein-Nishima cross section to compute the IC/SSC spectrum. To model the SED we have tried both one-zone and two-zone SSC models. More details on these models are given in the following sub-sections.

\subsubsection{One-zone SSC model:}
\begin{figure*}
\centering
\includegraphics[height=2.2in,width=3.0in]{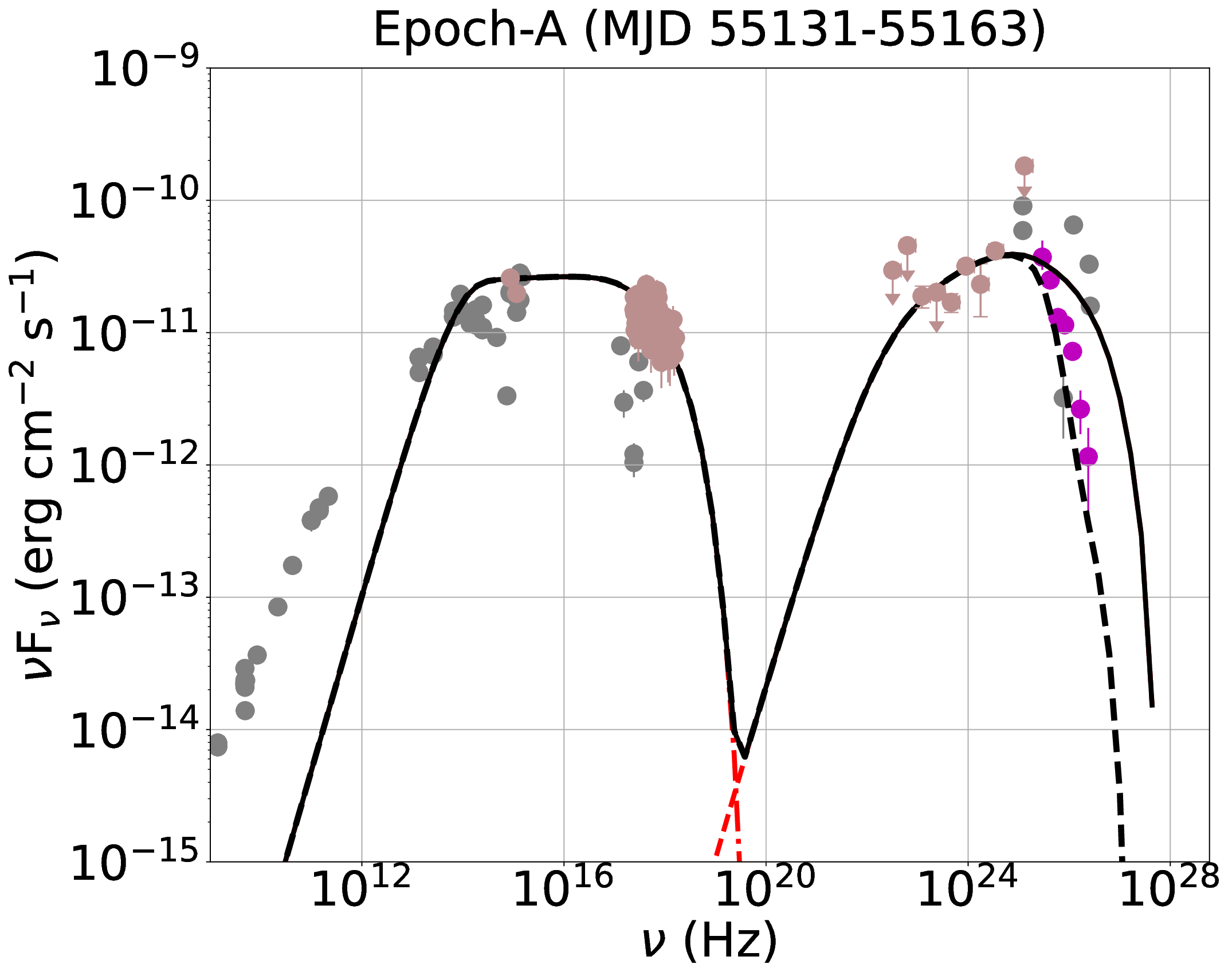}
\includegraphics[height=2.2in,width=3.0in]{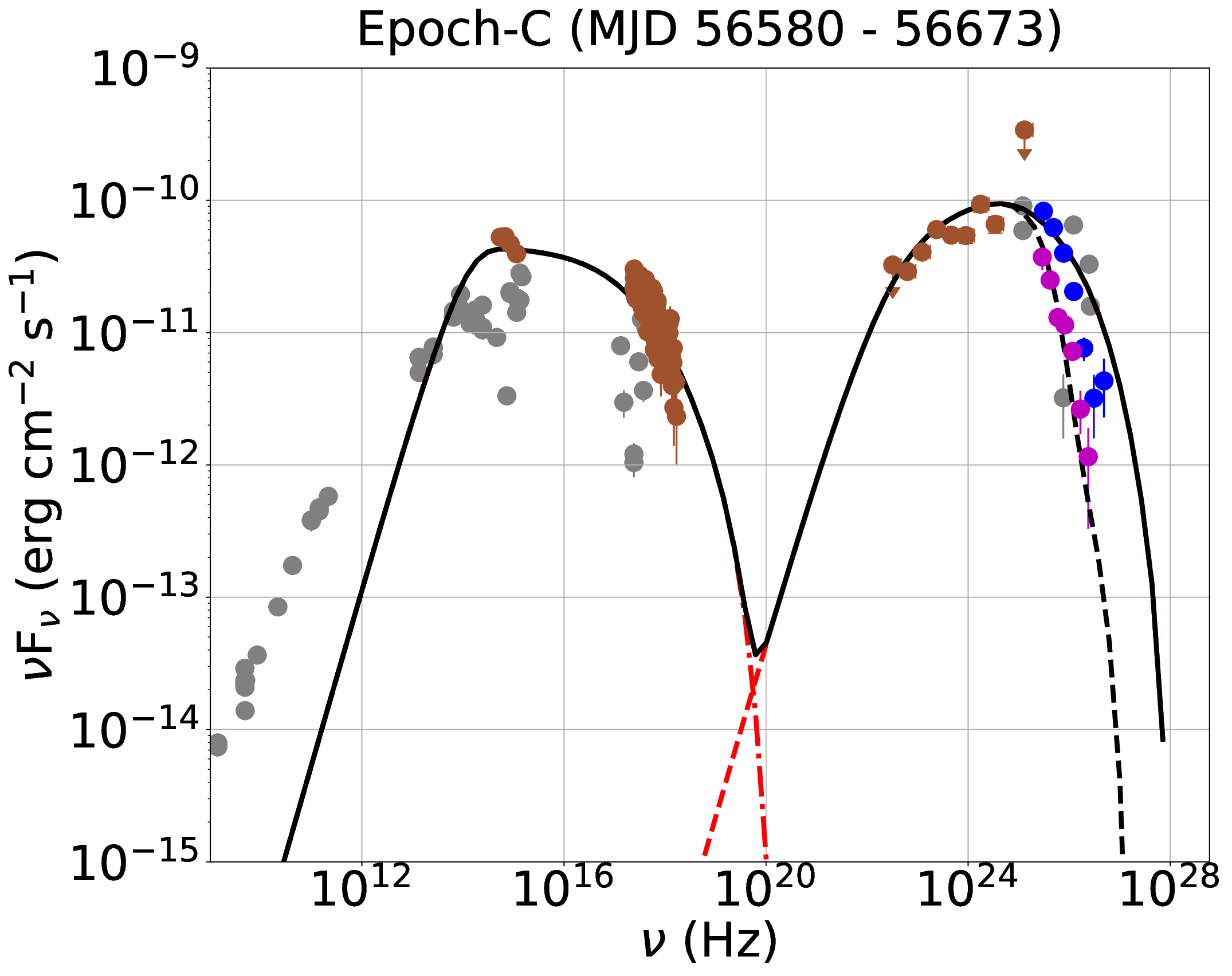}
\includegraphics[height=2.2in,width=3.0in]{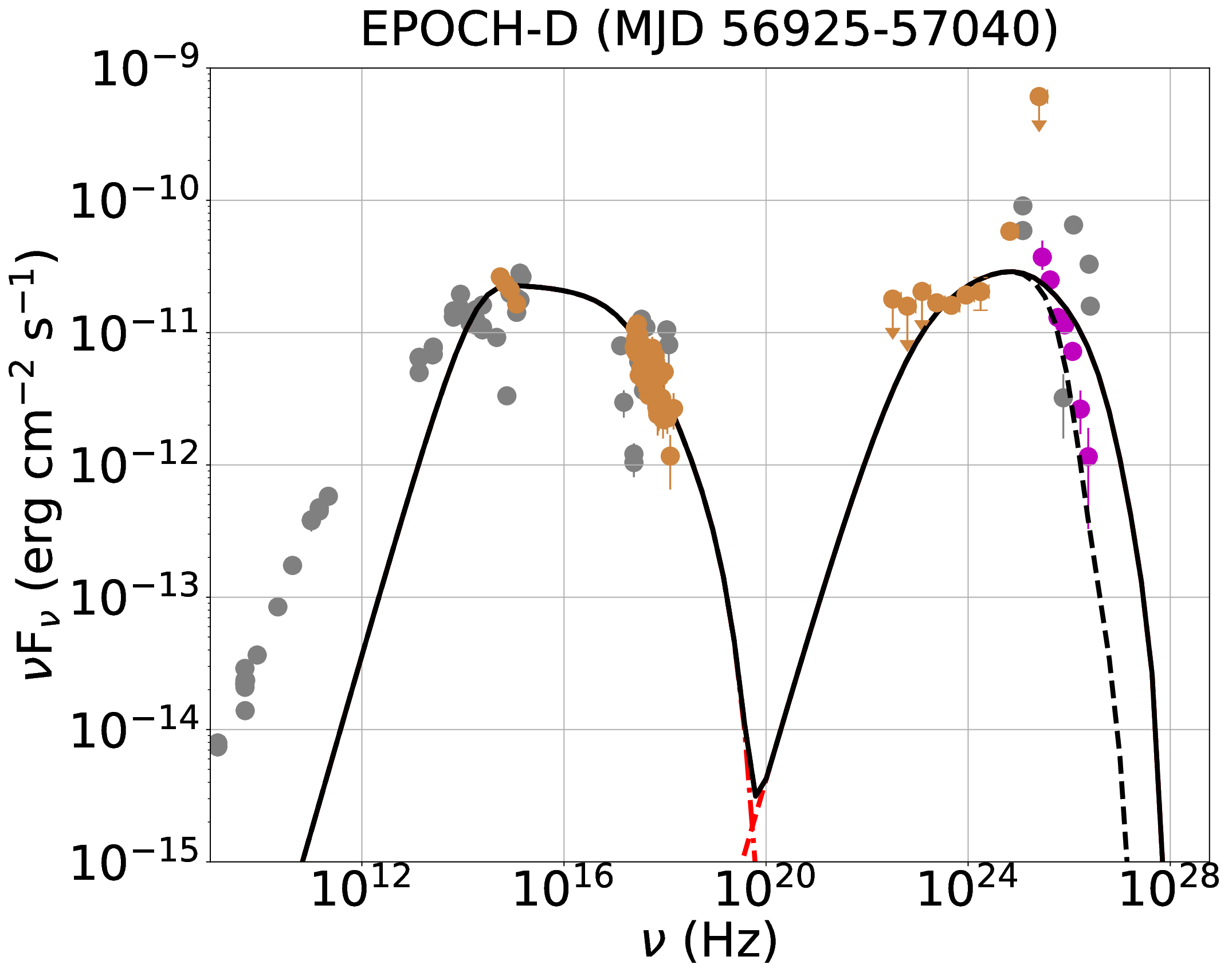}
\includegraphics[height=2.2in,width=3.0in]{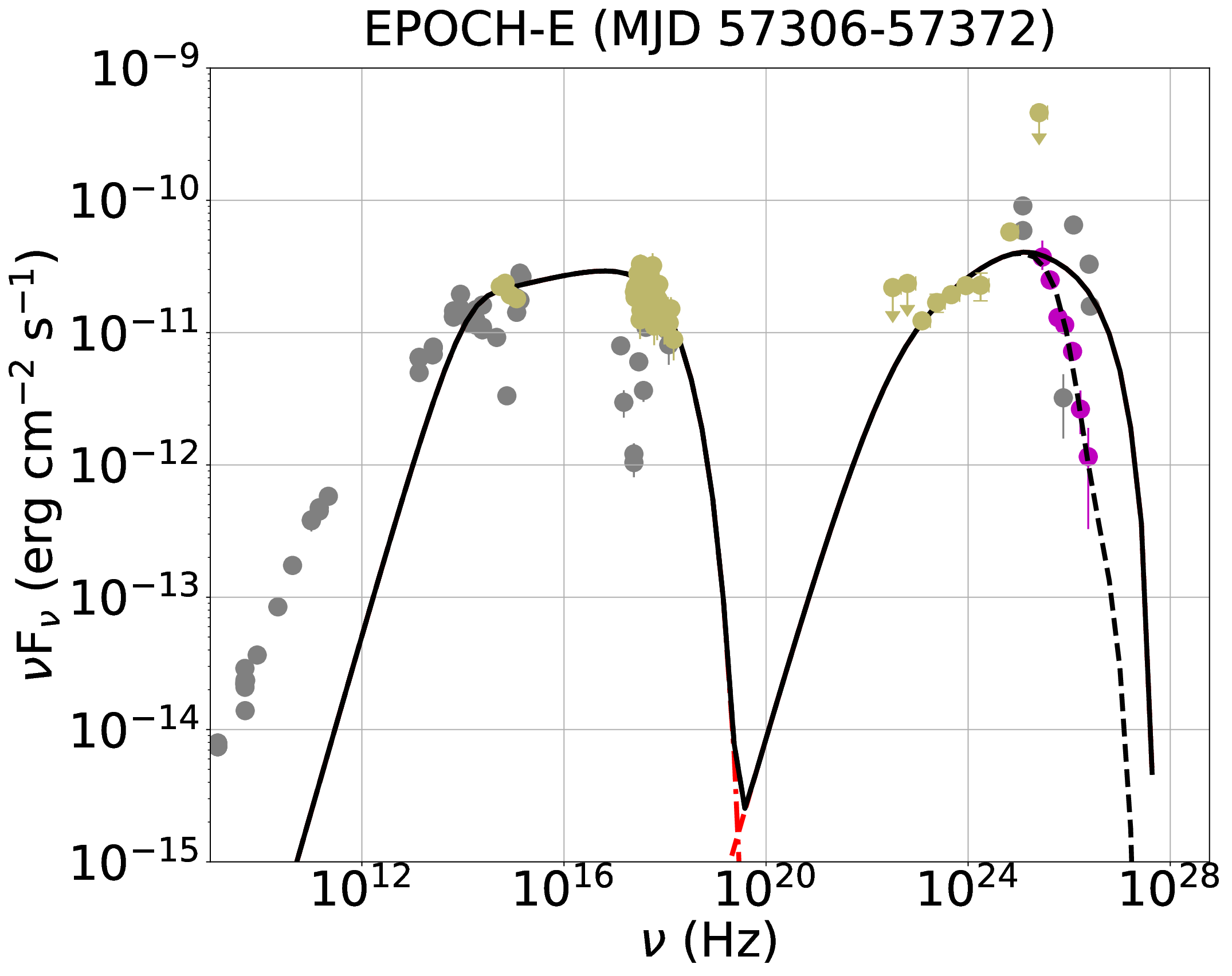}
\includegraphics[height=2.2in,width=3.0in]{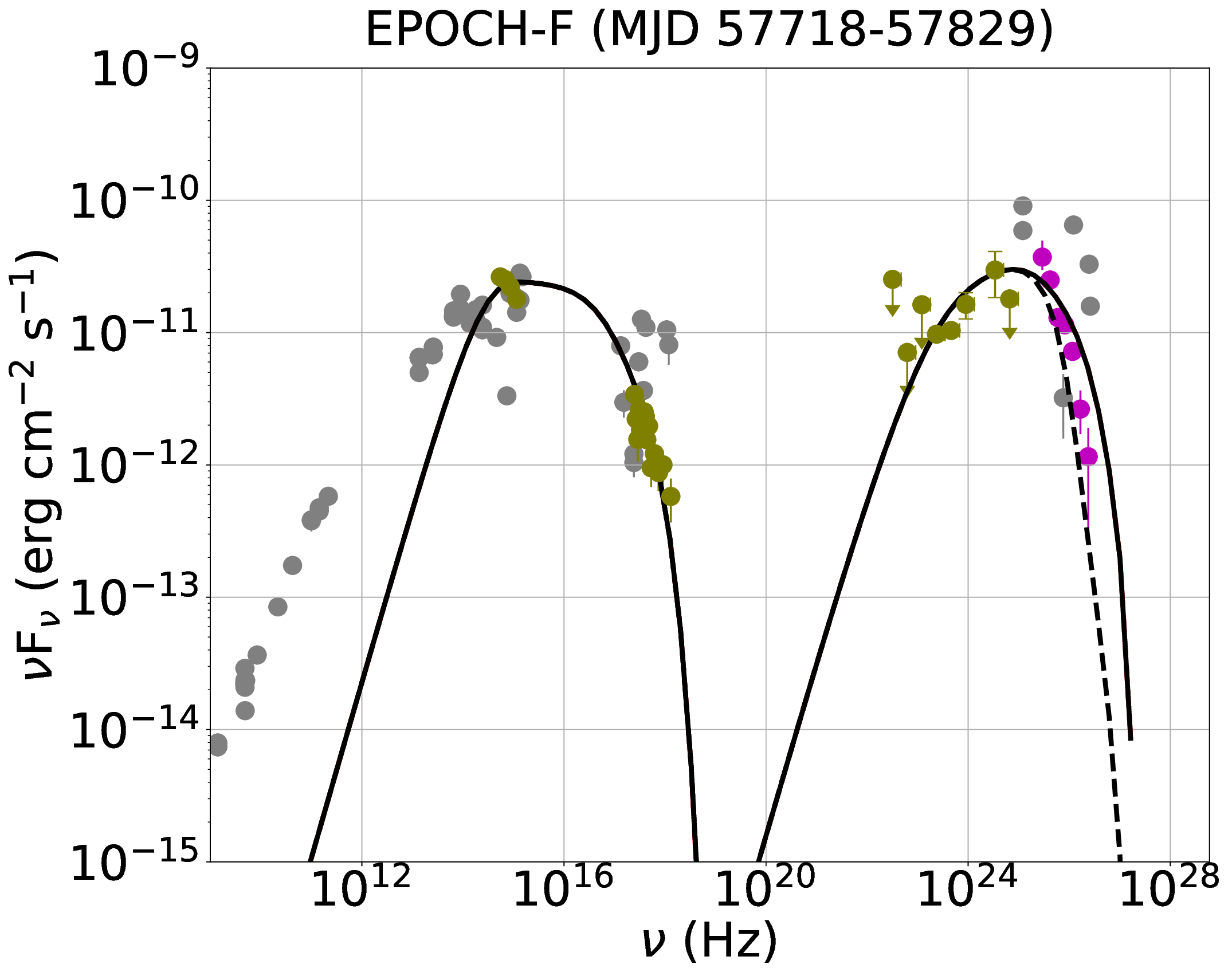}
\includegraphics[height=2.2in,width=3.0in]{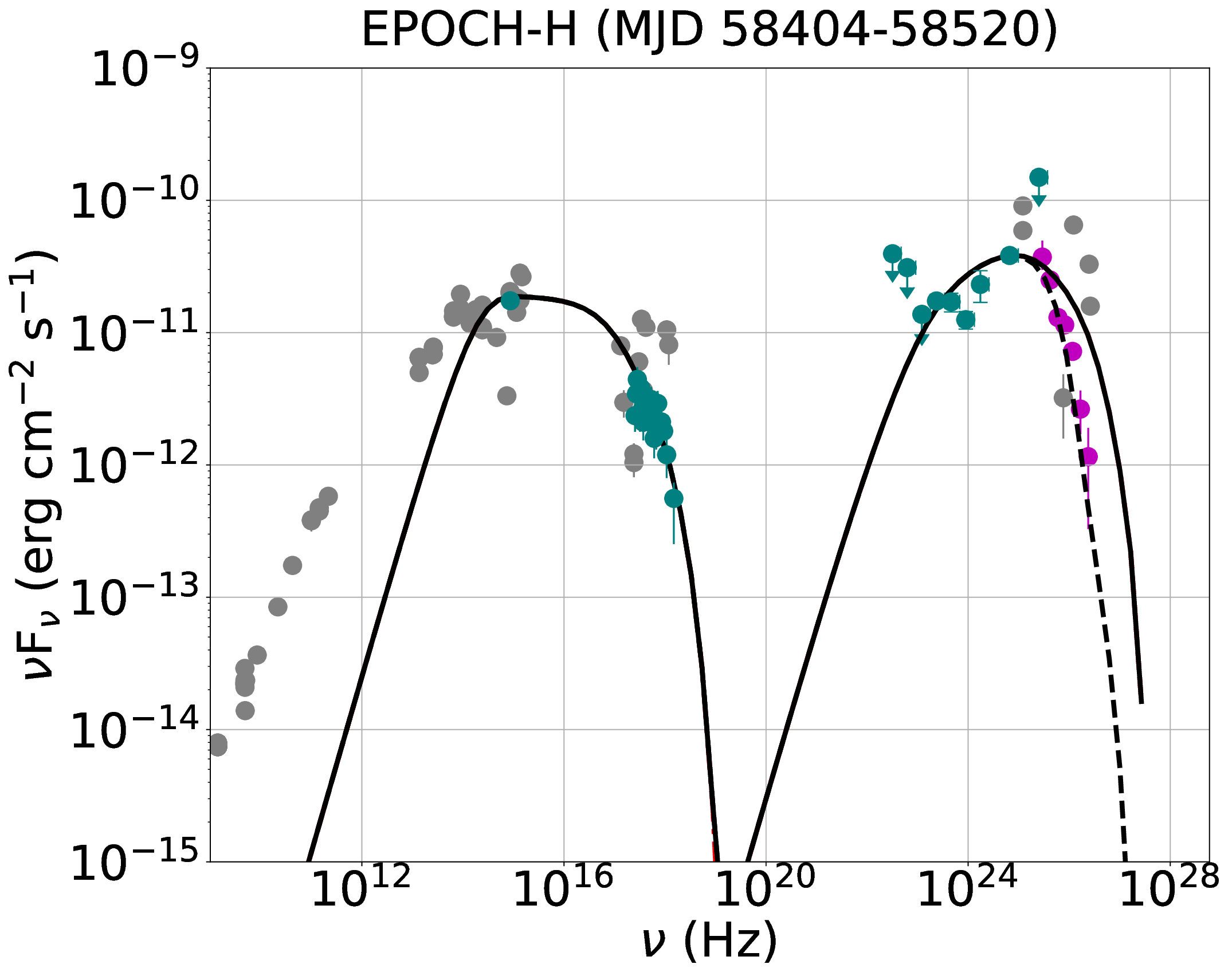}
\includegraphics[height=2.2in,width=3.0in]{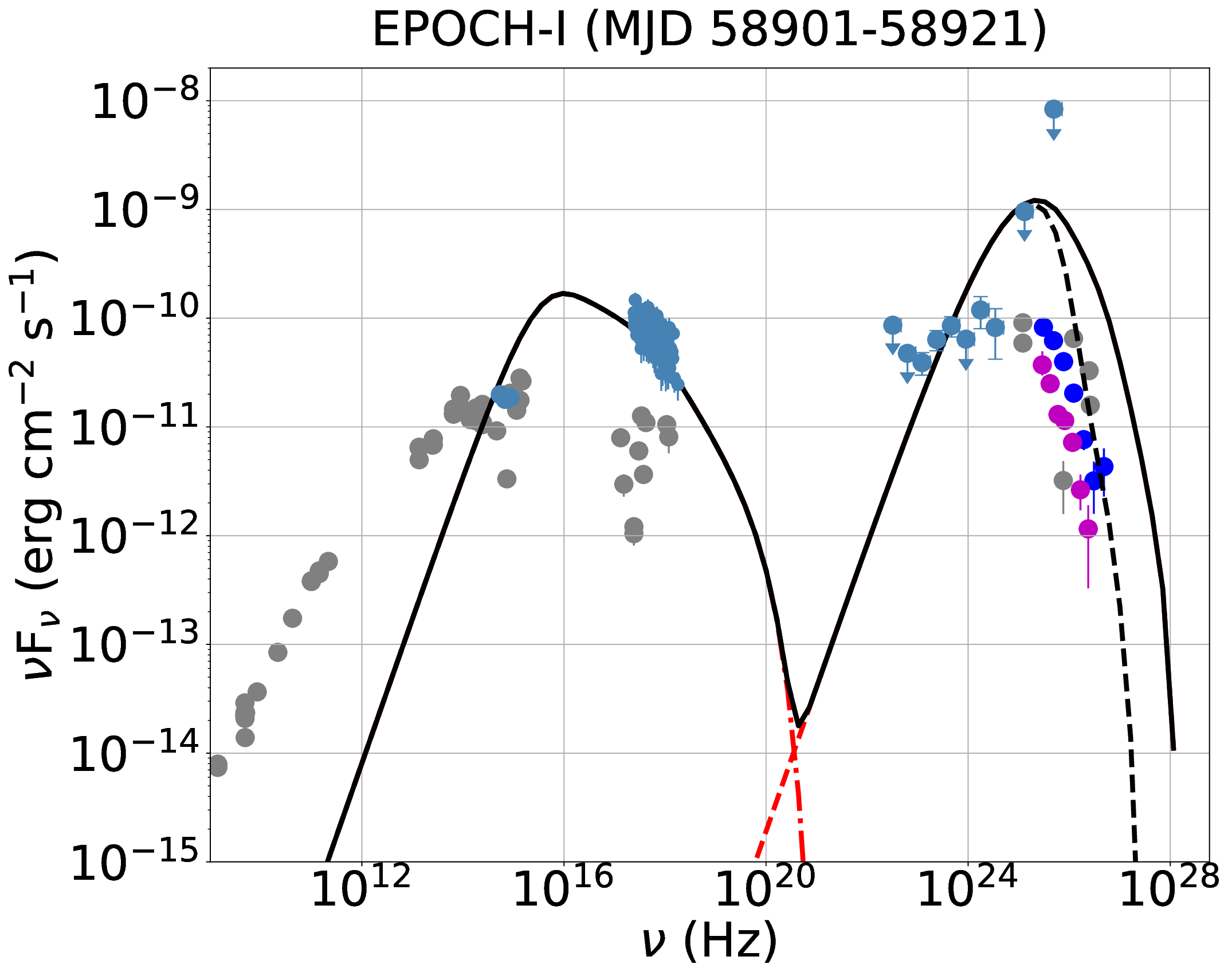}
\includegraphics[height=2.2in,width=3.0in]{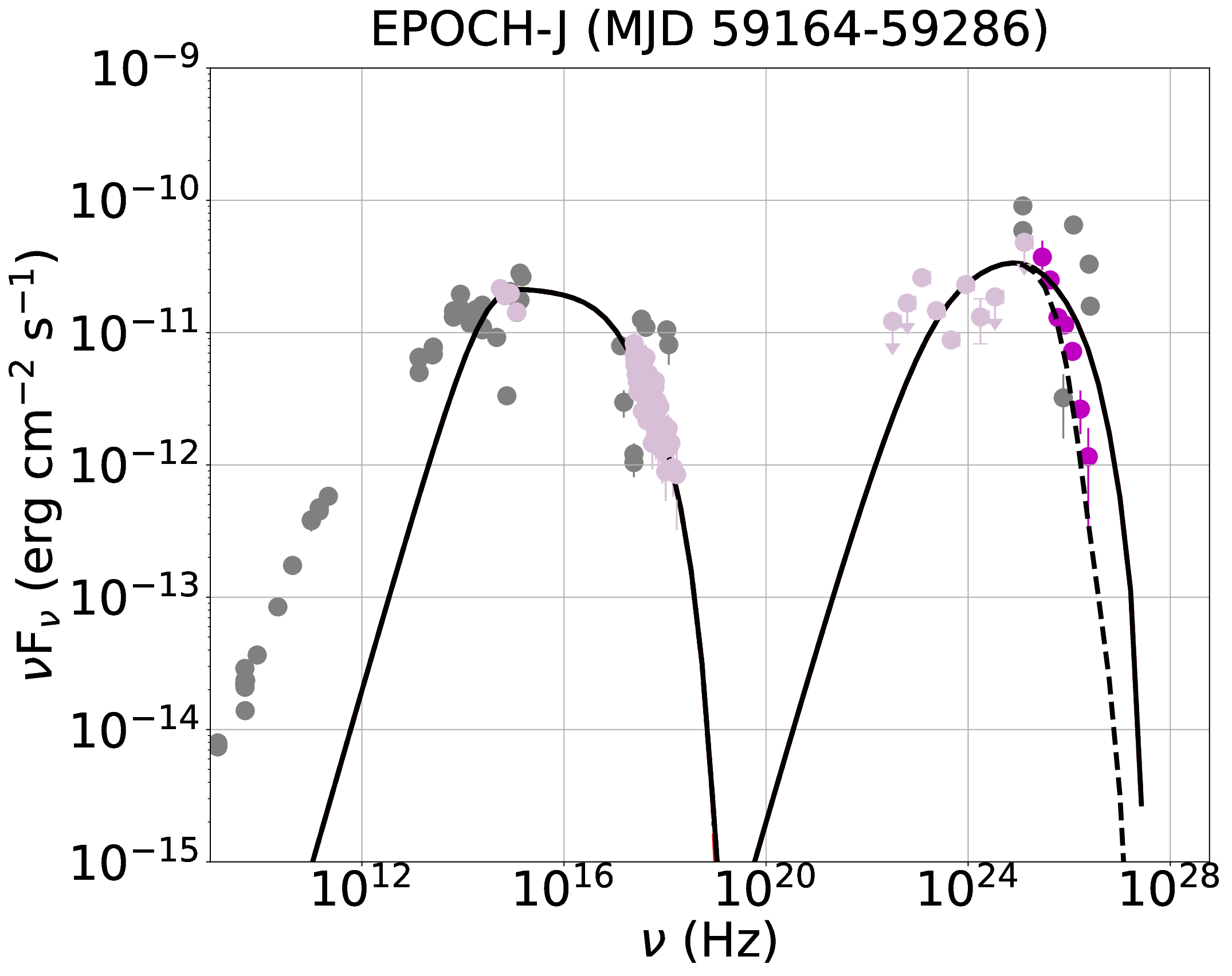}
\caption{One-zone leptonic model curve of the broad-band (optical to high-energy gamma-ray) SEDs for different epochs. Non-simultaneous archival data are depicted by grey colored points in each plot. The magenta- and blue-colored VHE data taken from \citet{Adams2022Jun} partially overlap with the time interval of Epoch C and are therefore non-simultaneous to the SEDs of the other epochs. The magenta data represent a low-VHE flux state (MJD 56632.5-56689.0) while the blue data correspond to a high-VHE flux state (MJD 56628.5-56632.5). The time intervals of these states are shown in Figures \ref{fig:1} and \ref{fig:B1}. The black dashed curve represents the EBL-corrected model.}
\label{fig:6}
\end{figure*}

In one-zone SSC model, the low-energy hump (radio to X-ray) of SED is described by synchrotron emission of relativistic electrons present in the jet, whereas the high-energy hump (gamma-ray to VHE gamma-ray) is described by SSC emission of the same electron population. This model has been extensively used in the literature for the SED modeling of BL Lac sources and in many cases, it successfully explained the emission spectra (e.g., \citealt{Sahakyan2020Aug, Mondal2021Nov, Prince2022Sep}, and reference therein). We assumed particle injection spectra governed by broken power-law given as:
\begin{equation} \label{eq:7}
    N(\gamma^{\prime}) = \begin{cases}    
    N_{0} \times \gamma^{\prime -\alpha_{1}}, & \gamma^{\prime}_{min} \leq \gamma^{\prime} \leq \gamma^{\prime}_{br} \\
    N_{0} \times \gamma_{br}^{\prime -(\alpha_{1} - \alpha_{2})} \times \gamma^{\prime -\alpha_{2}}, & \gamma^{\prime}_{br} \leq \gamma^{\prime} \leq \gamma^{\prime}_{max}
                \end{cases} 
\end{equation}

where, primed quantities represent parameters in the co-moving frame. To fit the broadband SED of each epoch we varied several parameters e.g., spectral index of the injection spectrum ($\alpha_{1}$), minimum and maximum energy of electrons ($\gamma_{min}$ and $\gamma_{max}$), break energy ($\gamma_{br}$), magnetic field (B), and radius (R) of the emission region. We assume the Doppler factor ($\delta$) to be 26 for all epochs, with $\Gamma \sim \delta$  \citep{Adams2022Jun}. It should be noted that the jet opening angle is not measured for this source \citep{2020A&A...640A.132M}, therefore $\delta$ is not well constrained. The energy density of the electrons/positrons is given as :
\begin{equation}
        U^{\prime}_e=\frac{3}{4 \pi R^{3}} \int_{\gamma^{\prime}_{min}}^{\gamma^{\prime}_{max}} EQ(E)dE 
\end{equation}
and the energy density in the magnetic field is expressed as:
\begin{equation}
        U^{\prime}_{B^{\prime}}= \frac{B^{\prime 2}}{8 \pi}
\end{equation}
The total absolute jet power required to model broadband SED is given by:
\begin{equation}
        P_{tot}= P_{e}+P_{B}+P_{p} = \pi R^{\prime 2} \Gamma^{2} c (U^{\prime}_{e}+U^{\prime}_{B}+U^{\prime}_{p})
\end{equation}
We assumed charge neutrality scenario and ratio of electron-positron pairs to cold protons of 10:1 in the emission region to calculate the energy density of cold protons ($U_{p}$) and, subsequently, the cold proton power in the jet. We also incorporated the correction due to extra-galactic background light (EBL) absorption using the model described in \cite{2012MNRAS.422.3189G}. The one-zone leptonic model fits for all epochs (except Epoch-B, Epoch-G, and Epoch-K) are shown in 
Figure~\ref{fig:6}. The values of all the model parameters required for SED modeling are listed in Table~\ref{tab:3}. 
\\
In case of TXS 0518+211, we found that the one-zone model cannot fully explain the broad-band SED. It is unable to account for the observed optical–UV slope and gamma-ray fluxes. Therefore, we also attempt to fit the SEDs 
using two-zone SSC model. 

\begin{figure*}
\centering
\includegraphics[height=2.2in,width=3.0in]{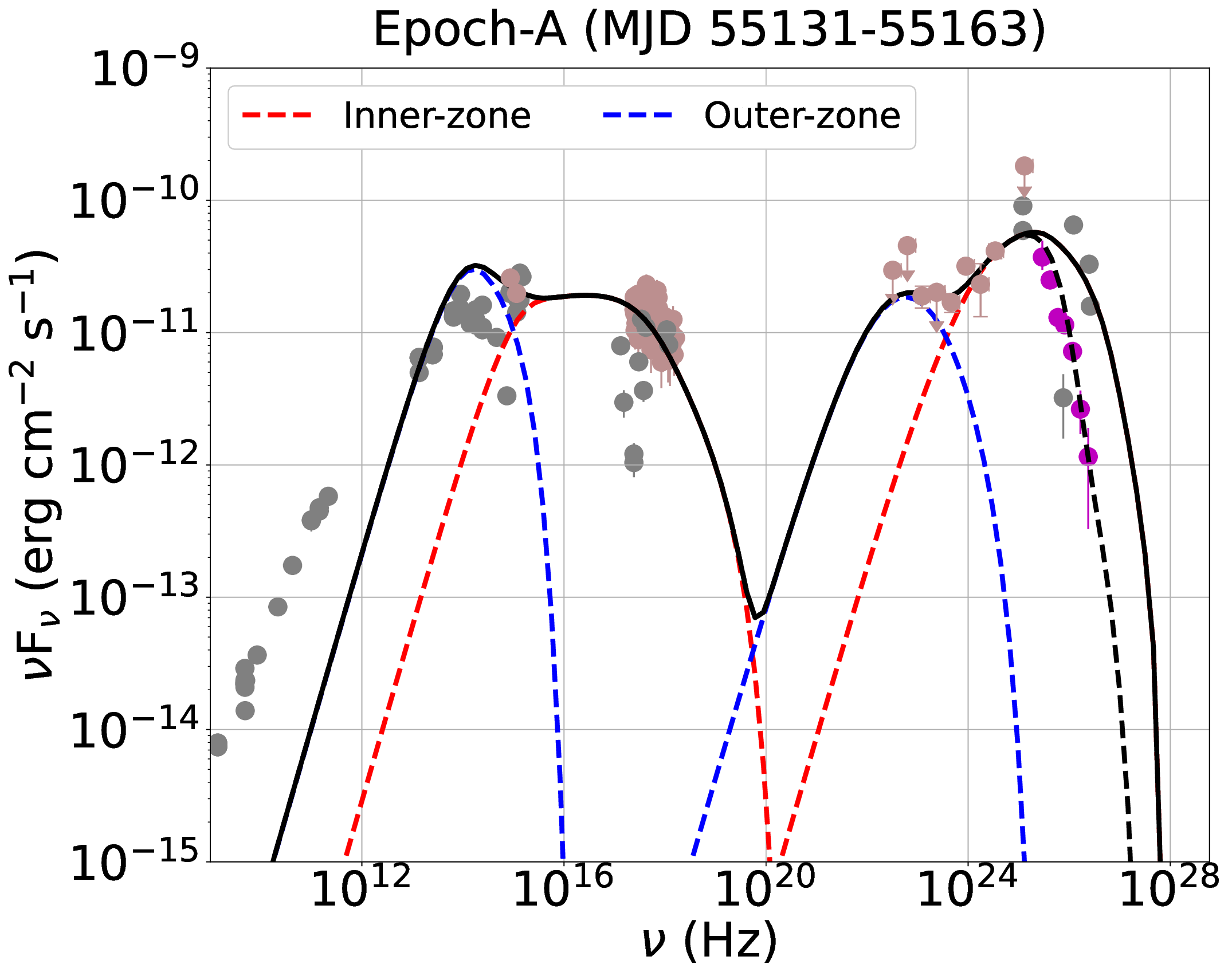}
\includegraphics[height=2.2in,width=3.0in]{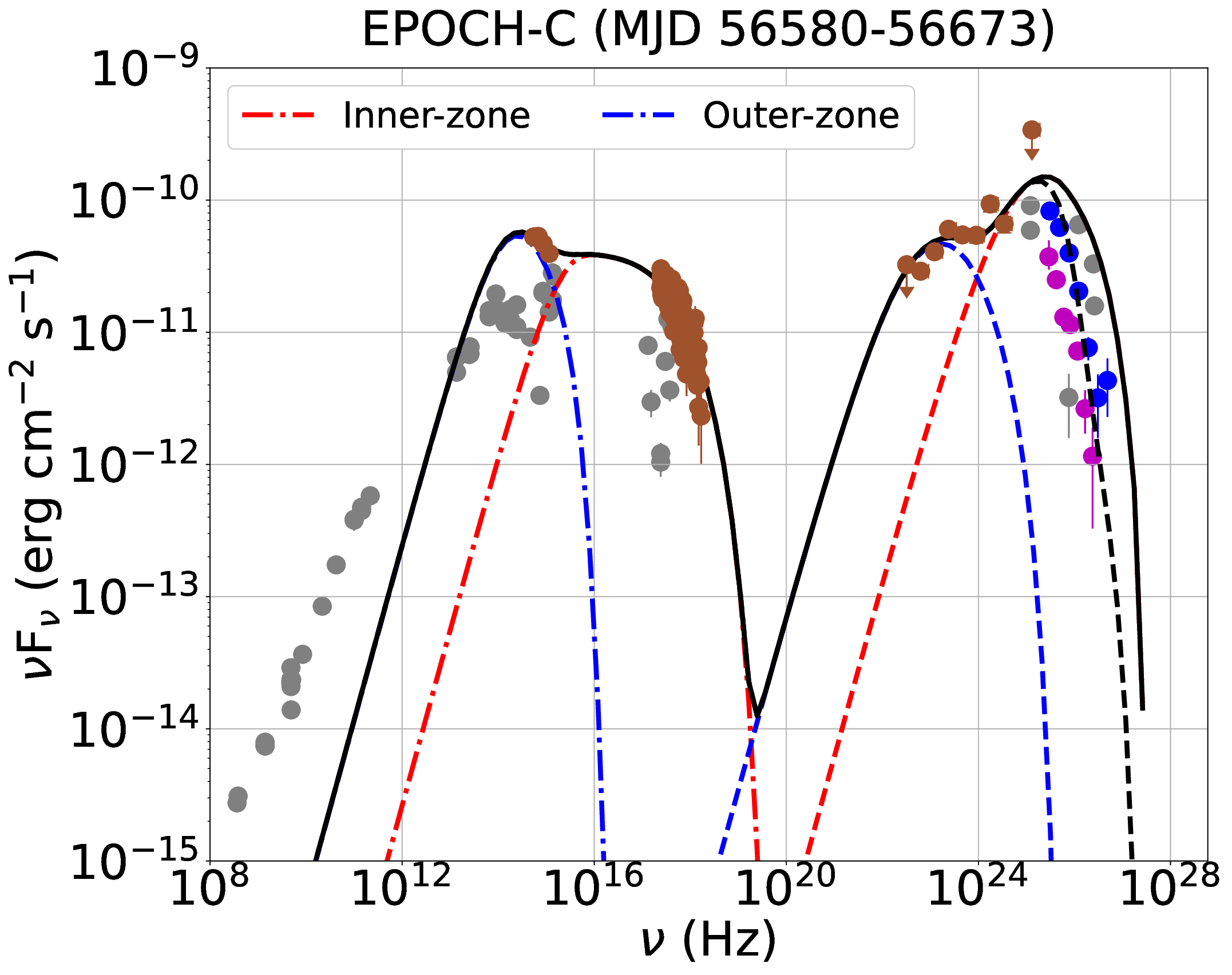}
\includegraphics[height=2.2in,width=3.0in]{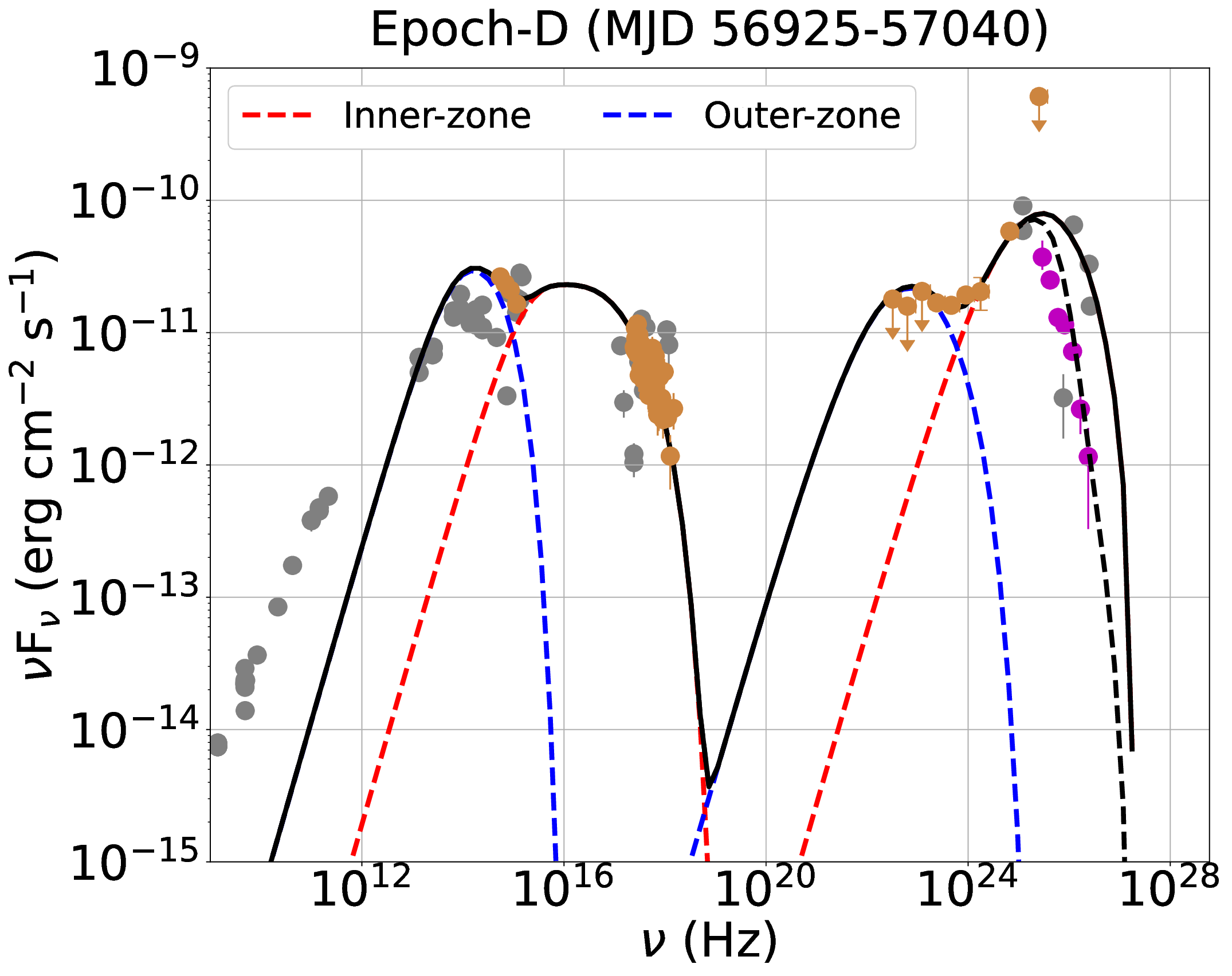}
\includegraphics[height=2.2in,width=3.0in]{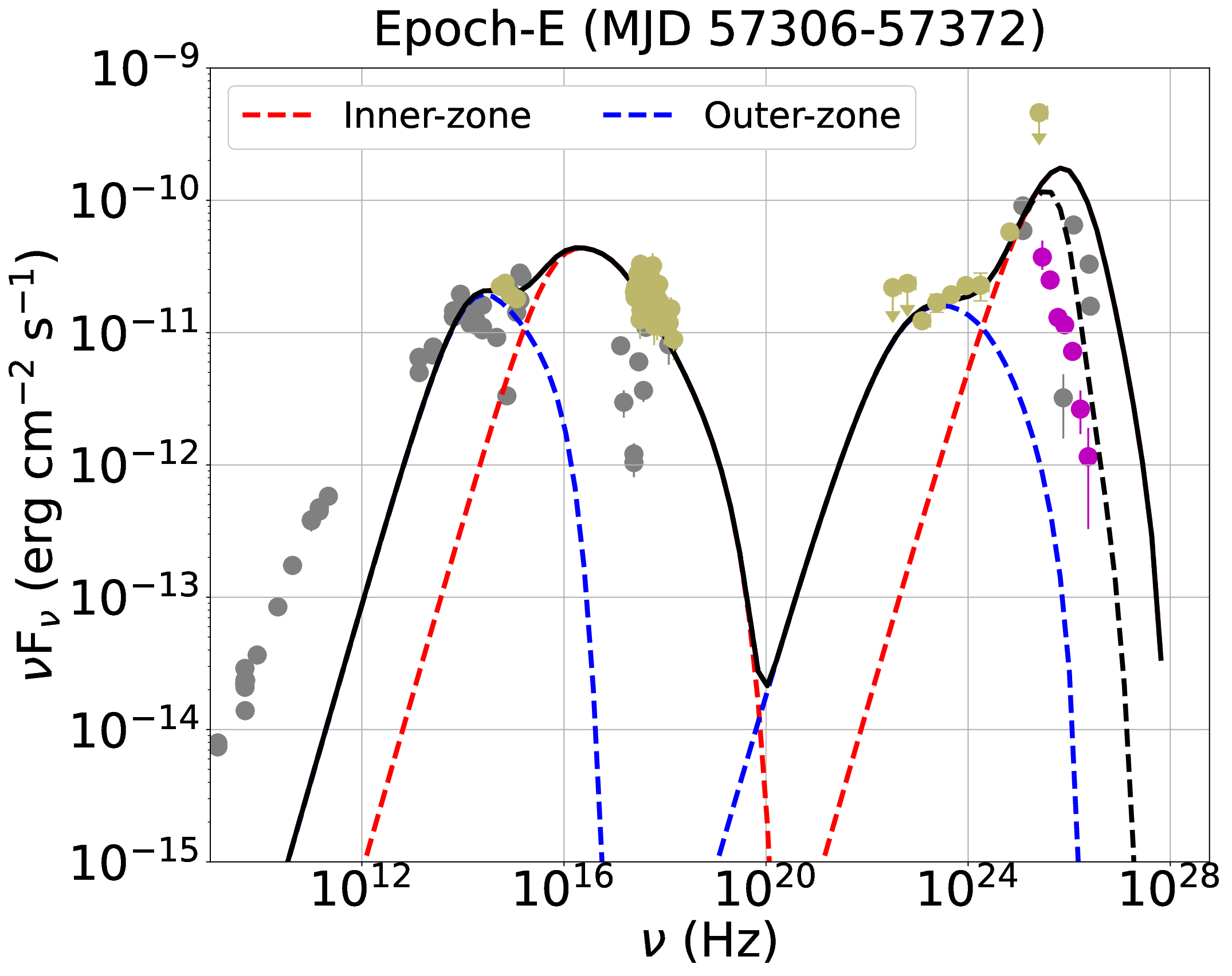}
\includegraphics[height=2.2in,width=3.0in]{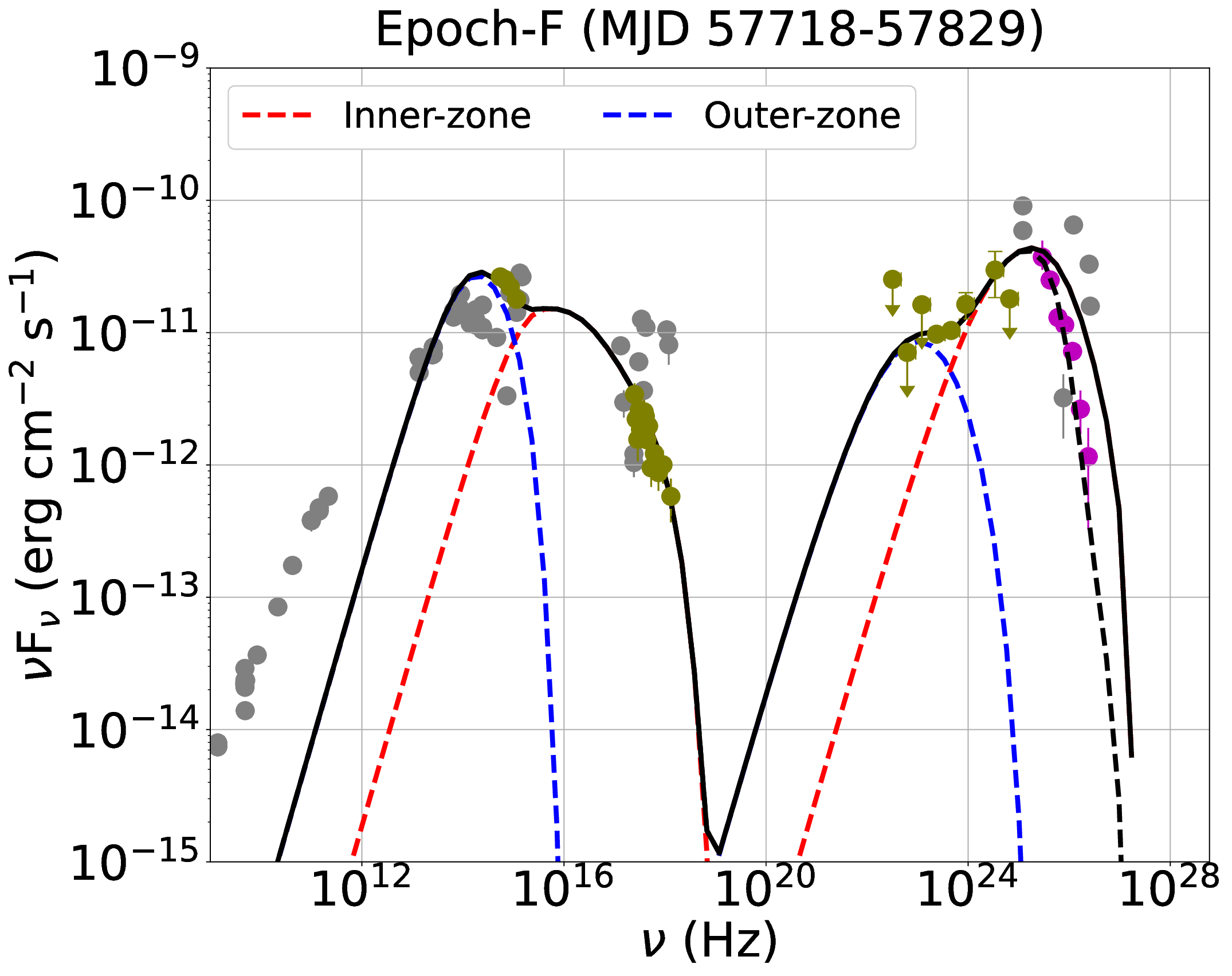}
\includegraphics[height=2.2in,width=3.0in]{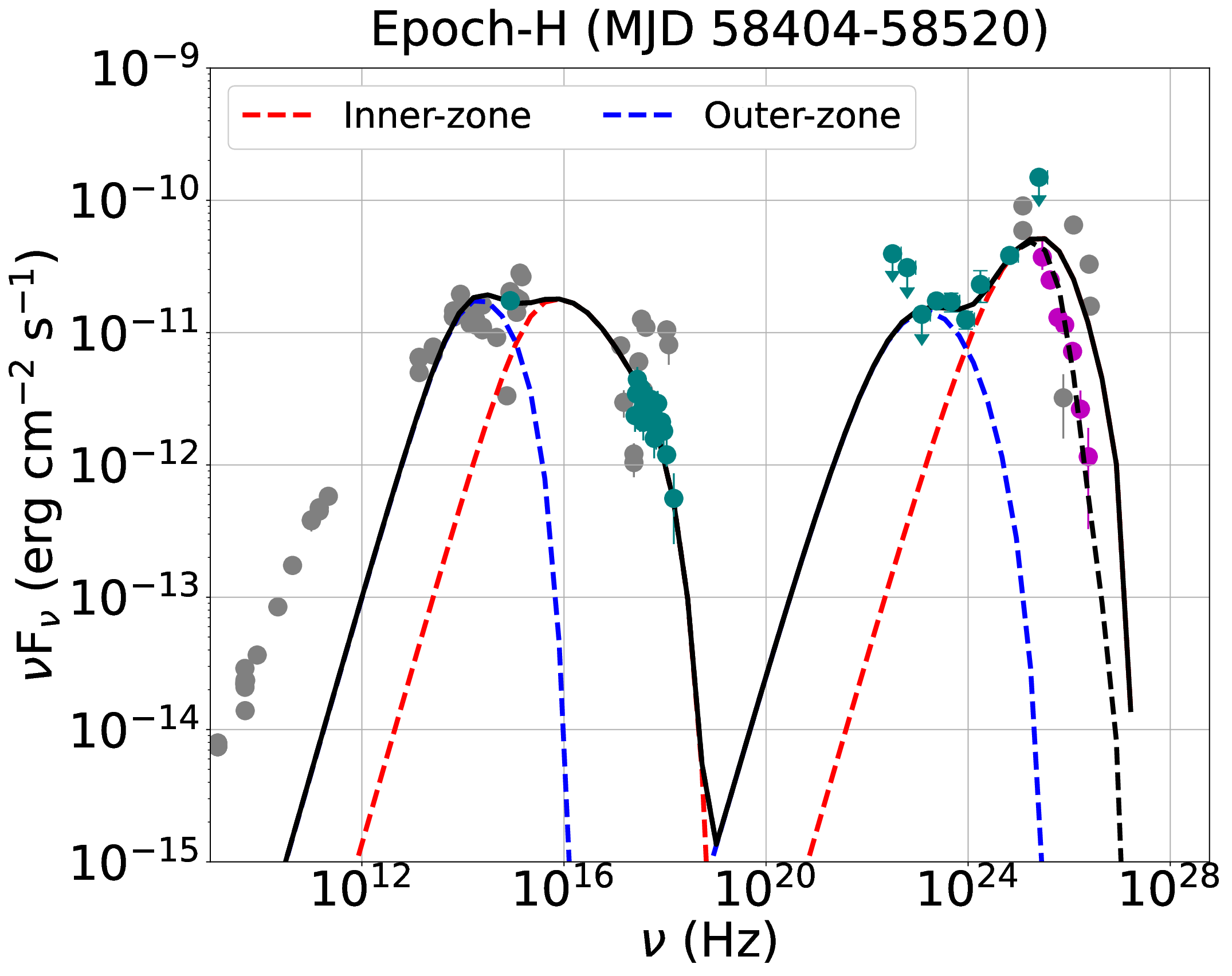}
\includegraphics[height=2.2in,width=3.0in]{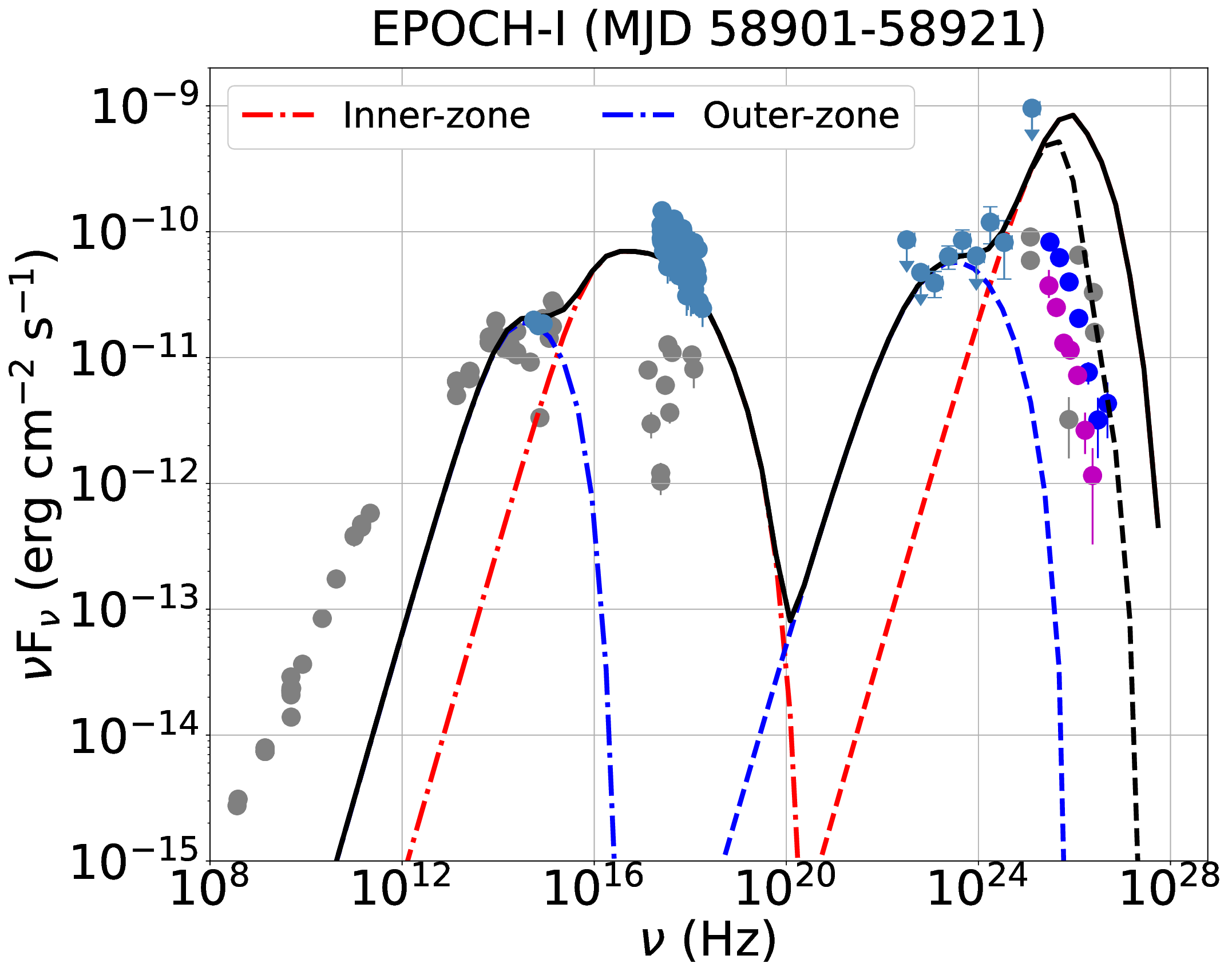}
\includegraphics[height=2.2in,width=3.0in]{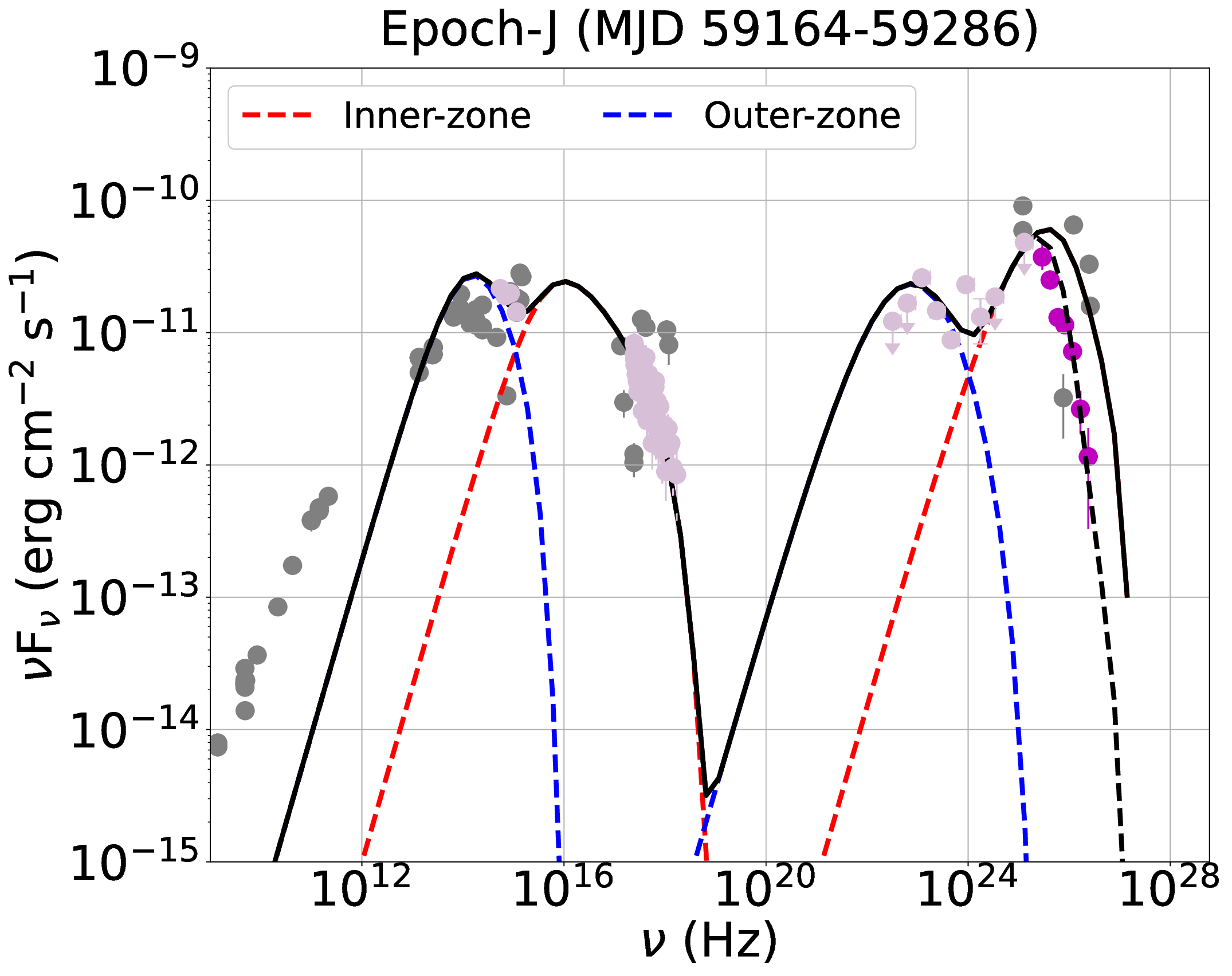}
\caption{Two-zone leptonic model curve of the broadband SEDs (optical to high-energy gamma-ray) for different epochs. The data and details are the same as in Figure-\ref{fig:6}.}
\label{fig:7}
\end{figure*}

\begin{table*}
\caption{Results of multi-wavelength SED modeling (One-zone). The time duration of different phases is given in the last column. The second to eleventh columns represent the value of various parameters used in the one-zone modeling. Here, $\alpha_{1}$, $\alpha_{2}$ = Spectral indices of injected electron spectrum; $\gamma_{min}$, $\gamma_{max}$, $\gamma_{br}$ = Minimum, maximum and break Lorentz factor of injected electron spectrum; R = Size of the emission region; $P_e$ = Power in the injected spectrum; $P_B$ = Power in the magnetic field; $P_{tot}$ = Total required jet power. Several parameters are kept fixed at a specific value during modeling/fitting including $\delta$ = 26, $\Gamma$ = 25 (see text for more details).}
\label{tab:3}
\hskip-3.5cm
\centering
\scalebox{0.90}{
\begin{tabular}{cccccccc rrrr}   
\hline\hline
Activity Epochs & $\alpha_{1}$ & $\alpha_{2}$ & $\gamma^{\prime}_{min}$ & $\gamma^{\prime}_{max}$ & $\gamma^{\prime}_{br}$ & B$^{\prime}$ & R$^{\prime}$ & $P_e$ & $P_B$ & $P_{tot}$ & Time duration  \\
& & & & & & (G) &(cm.) & (erg/sec) & (erg/sec) & (erg/sec) & (days) \\
(1) & (2) & (3) & (4) & (5) & (6) & (7) & (8) & (9) & (10) & (11) & (12) \\
\hline\hline
Epoch-A & 2.95 & 3.95 & 8.50$\times10^{3}$  & 2.30$\times10^{6}$ & 5.00$\times10^{5}$ & 0.01 & 1.40$\times10^{17}$ & 2.44$\times10^{45}$ & 4.59$\times10^{42}$  & 2.45$\times10^{45}$ & 32 \\
\hline
Epoch-C & 2.98 & 3.98 & 9.00$\times10^{3}$ & 4.00$\times10^{6}$ & 9.00$\times10^{5}$ & 0.017 & 1.60$\times10^{17}$ & 2.53$\times10^{45}$  & 1.18$\times10^{43}$ & 2.55$\times10^{45}$ & 93 \\
\hline
Epoch-D & 2.98 & 3.98 & 1.20$\times10^{4}$ & 3.60$\times10^{6}$ & 5.00$\times10^{5}$ & 0.012 & 1.80$\times10^{17}$ & 1.36$\times10^{45}$  & 1.00$\times10^{43}$ & 1.38$\times10^{45}$ & 115  \\
\hline
Epoch-E & 2.80 & 3.80 & 9.30$\times10^{3}$ & 2.00$\times10^{6}$ & 9.00$\times10^{5}$ & 0.012 & 1.10$\times10^{17}$ & 2.26$\times10^{45}$  & 4.08$\times10^{42}$  & 2.27$\times10^{45}$ & 66 \\
\hline
Epoch-F & 2.98 & 3.98 & 1.55$\times10^{4}$ & 8.50$\times10^{5}$& 2.00$\times10^{5}$ & 0.011 & 1.90$\times10^{17}$ & 1.16$\times10^{45}$  & 9.32$\times10^{42}$ & 1.17$\times10^{45}$  & 111  \\
\hline
Epoch-H & 2.98 & 3.98 & 1.45$\times10^{4}$ & 1.50$\times10^{6}$ & 3.00$\times10^{5}$ & 0.009 & 1.60$\times10^{17}$ & 1.80$\times10^{45}$  & 4.86$\times10^{42}$ & 1.81$\times10^{45}$  & 116  \\
\hline
Epoch-I & 2.98 & 3.98 & 3.50$\times10^{4}$ & 5.50$\times10^{6}$ & 4.50$\times10^{5}$ & 0.038 & 4.00$\times10^{16}$ & 2.50$\times10^{45}$  & 5.41$\times10^{42}$ & 2.51$\times10^{45}$  & 20  \\
\hline
Epoch-J & 2.98 & 3.98 & 1.45$\times10^{4}$ & 1.30$\times10^{6}$ & 3.00$\times10^{5}$ & 0.012 & 1.40$\times10^{17}$ & 1.41$\times10^{45}$  & 6.61$\times10^{42}$ & 1.42$\times10^{45}$  & 107  \\
\hline\hline
\end{tabular}
}
\end{table*}

\begin{table*}
\caption{Results of two-zone multi-wavelength SED modeling. The third to thirteenth columns represent the value of various parameters used in the one-zone modeling (see text for more details).}
\label{tab:4}
\hskip-3.5cm
\centering
\scalebox{0.85}{
\begin{tabular}{ccccccccc rrrr}   
\hline\hline
Activity & Emission & $\alpha_{1}$ & $\alpha_{2}$ & $\gamma^{\prime}_{min}$ & $\gamma^{\prime}_{max}$ & $\gamma^{\prime}_{br}$ & B$^{\prime}$ & R$^{\prime}$ & $P_e$ & $P_B$ & $P_{tot}$ & Time duration \\
Epochs & region & & & & & & (G) & (cm.) & (erg/sec) & (erg/sec) & (erg/sec) & (days) \\
(1) & (2) & (3) & (4) & (5) & (6) & (7) & (8) & (9) & (10) & (11) & (12) & (13) \\
\hline\hline
Epoch-A & Inner-zone  & 2.84 & 3.84 & 1.70$\times10^{4}$  & 3.00$\times10^{6}$ & 4.00$\times10^{5}$ & 0.03 & 2.50$\times10^{16}$ & 1.32$\times10^{45}$ & 1.32$\times10^{42}$  & 1.33$\times10^{45}$  & 32 \\
& Outer-zone & 2.85 & 3.85 & 7.90$\times10^{3}$  & 3.20$\times10^{4}$ & 1.00$\times10^{4}$ & 0.011 & 1.30$\times10^{17}$ & 2.52$\times10^{45}$ & 4.84$\times10^{42}$  & 2.54$\times10^{45}$  & \\ 
\hline
Epoch-C & Inner-zone & 2.95 & 3.95 & 2.50$\times10^{4}$ & 1.20$\times10^{6}$ & 4.00$\times10^{5}$ & 0.03 & 3.30$\times10^{16}$ & 3.29$\times10^{45}$  & 2.30$\times10^{42}$ & 3.30$\times10^{45}$  & 93 \\
& Outer-zone & 2.86 & 3.86 & 9.50$\times10^{3}$ & 4.30$\times10^{4}$ & 2.00$\times10^{4}$ & 0.01 & 1.90$\times10^{17}$ & 4.21$\times10^{45}$  & 7.79$\times10^{42}$ & 4.25$\times10^{45}$ & \\
\hline
Epoch-D & Inner-zone  & 2.84 & 3.84 & 2.80$\times10^{4}$ & 7.70$\times10^{5}$ & 2.00$\times10^{5}$ & 0.02 & 4.20$\times10^{16}$ & 3.00$\times10^{45}$  & 1.49$\times10^{42}$ & 3.01$\times10^{45}$ & 115 \\
& Outer-zone & 2.84 & 3.84 & 8.00$\times10^{3}$ & 3.05$\times10^{4}$ & 2.00$\times10^{4}$ & 0.008 & 1.80$\times10^{17}$ & 4.21$\times10^{45}$  & 5.48$\times10^{42}$ & 4.24$\times10^{45}$ & \\
\hline
Epoch-E & Inner-zone & 2.75 & 3.75 & 4.80$\times10^{4}$ & 3.50$\times10^{6}$ & 2.00$\times10^{5}$ & 0.02 & 4.30$\times10^{16}$ & 3.13$\times10^{45}$  & 1.73$\times10^{42}$  & 3.13$\times10^{45}$ & 66 \\
& Outer-zone & 2.84 & 3.84 & 1.00$\times10^{4}$ & 9.50$\times10^{4}$ & 2.00$\times10^{4}$ & 0.008 & 1.80$\times10^{17}$ & 3.23$\times10^{45}$  & 5.35$\times10^{42}$ & 3.25$\times10^{45}$ & \\
\hline
Epoch-F & Inner-zone & 2.84 & 3.84 & 2.35$\times10^{4}$ & 8.00$\times10^{5}$& 1.00$\times10^{5}$ & 0.02 & 3.80$\times10^{16}$ & 2.50$\times10^{45}$ & 1.35$\times10^{42}$ & 2.50$\times10^{45}$  & 111 \\
& Outer-zone & 2.84 & 3.84 & 9.50$\times10^{3}$ & 3.30$\times10^{4}$& 2.00$\times10^{4}$ & 0.008 & 2.50$\times10^{17}$ & 2.14$\times10^{45}$  & 9.75$\times10^{42}$ & 2.16$\times10^{45}$ & \\
\hline
Epoch-H & Inner-zone & 2.80 & 3.80 & 2.90$\times10^{4}$ & 7.80$\times10^{5}$ & 1.00$\times10^{5}$ & 0.02 & 3.80$\times10^{16}$ & 2.52$\times10^{45}$  & 1.35$\times10^{42}$ & 2.53$\times10^{45}$  & 116 \\
& Outer-zone & 2.84 & 3.84 & 9.00$\times10^{3}$ & 4.50$\times10^{4}$ & 2.00$\times10^{4}$ & 0.008 & 1.50$\times10^{17}$ & 3.56$\times10^{45}$  & 3.37$\times10^{42}$ & 3.58$\times10^{45}$ & \\
\hline
Epoch-I & Inner-zone & 2.88 & 3.88 & 4.65$\times10^{4}$ & 3.00$\times10^{6}$ & 5.00$\times10^{5}$ & 0.035 & 1.20$\times10^{16}$ & 3.75$\times10^{45}$  & 4.13$\times10^{41}$ & 3.75$\times10^{45}$ & 20 \\
& Outer-zone & 2.87 & 3.87 & 9.50$\times10^{3}$ & 5.50$\times10^{4}$ & 2.80$\times10^{4}$ & 0.011 & 5.50$\times10^{16}$ & 3.49$\times10^{45}$  & 8.26$\times10^{41}$ & 3.51$\times10^{45}$ & \\
\hline
Epoch-J & Inner-zone & 2.80 & 3.80 & 3.80$\times10^{4}$ & 7.50$\times10^{5}$ & 1.00$\times10^{5}$ & 0.02 & 4.20$\times10^{16}$ & 2.32$\times10^{45}$  & 1.65$\times10^{42}$ & 2.32$\times10^{45}$  & 122 \\
& Outer-zone & 2.84 & 3.84 & 9.50$\times10^{3}$ & 3.50$\times10^{4}$ & 1.00$\times10^{4}$ & 0.008 & 1.70$\times10^{17}$ & 4.56$\times10^{45}$  & 4.33$\times10^{42}$ & 4.60$\times10^{45}$ & \\
\hline\hline
\end{tabular}
}
\end{table*}

\subsubsection{Two-zone SSC model:} \label{sec:4.1.2}
In the literature it has been shown that the simple one-zone SSC models are often insufficient to explain the broad-band emissions from VHE detected BL Lac type blazars (\citealt{2017A&A...603A..31A, 2020A&A...640A.132M}). 
In two-zone model we consider two emission regions, namely inner-zone and outer-zone, which are spatially separated from each other in the jet \citep[see][]{2021ApJ...920..117D}. Both the zones emit radiation via synchrotron and SSC processes. The light curves in the X-ray and VHE wavebands are strongly correlated in VHE-detected blazars \citep{2020A&A...640A.132M}. The inner zone, located closer to the central engine, primarily explains the X-ray and VHE components of the SED, whereas the outer-zone explains the optical-UV and partially the high-energy gamma-ray components. We assume that the shape of the injected spectrum is the same for both zones (described by Equation-\ref{eq:7}). 
\par 
Two-zone model requires more number of parameters than simple one-zone model. To use two-zone model we placed constraints in some parameters that include Doppler factor of the two emission regions assumed to be equal and same (i.e., $\delta$ = 26 and $\delta \sim \Gamma$) as for the one-zone emission model. 
The radius (R) of the two emission regions estimated from the variability timescales given by X-ray ($t_{var} \sim$ 1 day) and optical/UV ($t_{var} \sim$ 2.5 days) light curves. Using causality condition (R$^{\prime}$ $\leq$ $\frac{c \delta t_{var}}{1+z}$), we obtain inner and outer-zone's radii to be $\sim$ 5.4$\times$10$^{16}$ cm and $\sim$ 1.4$\times$10$^{17}$ cm, respectively. However, we note that these values are only approximate and several effects can introduce large uncertainties (see \citet{2002PASA...19..486P}). The variability timescales derived from individual epochs imply, in most cases, upper limits on the radii of the inner and outer emission regions of the order of $\sim$ 10$^{18}$ cm (assuming $\delta$ = 26), which appear too large and do not provide a reasonably good fit. Previous works using two-zone models to fit the broadband SEDs of other BL Lac–type sources have found that the inner- and outer-zone radii lie in the ranges 10$^{14}$–10$^{16}$ cm and 10$^{16}$–10$^{17}$ cm, respectively (\citealt{2015ApJ...798....2S, 2018A&A...611A..44P, 2022MNRAS.512.1557A}). For our work, we set $R^{\prime}$ value to be the order of $\sim$ 10$^{16}$ cm and $\sim$ 10$^{17}$ cm for the two emission zones.
The magnetic field strength (B) is typically assumed to decrease with distance from the central engine (Z), i.e., B $\propto Z^{-1}$ \citep{2009MNRAS.400...26O}. The Z value can be estimated from the variability timescale (Z = $\frac{2 \Gamma^{2} c t_{var}}{1+z}$), assuming the conical jet approximation. This implies $\frac{B^{\prime}_{outer}}{B^{\prime}_{inner}}$ = $\frac{t_{var, inner}}{t_{var, outer}}$ or $B^{\prime}_{outer} \sim \frac{B^{\prime}_{inner}}{3}$ G. 
We assumed high-energy electron index for each emission zone as $\alpha_{2} = \alpha_{1} + 1$, which is expected for the canonical cooling break in a homogeneous model. Further, to compute power carried by the cold protons, we assumed the charge neutrality condition and the ratio of electron–positron pairs to cold protons same as in the one-zone model.
\par    
After constraining the above parameters, we performed the two-zone model fit. The total jet power is computed by summing the contributions from each emission zone:
\begin{equation} \label{eq:11}
        P_{tot}= \sum_{i=1}^{2} P_{e,i}+P_{B,i}+P_{p,i} = \sum_{i=1}^{2} \pi R^{\prime 2} \Gamma^{2} c (U^{\prime}_{e,i}+U^{\prime}_{B,i}+U^{\prime}_{p,i})
\end{equation}

The two-zone leptonic model fit for all epochs (except Epoch-B, Epoch-G, and Epoch-K) are shown in Figure-\ref{fig:7}. In our two-zone model fit, we found that the spectral index ($\alpha_{1}$) lies in the range 2.75–2.95 for both emission zones. The magnetic field (B$^{\prime}$) of the inner emission zone lies in the range 0.02–0.035 G. For the high-flux state, a relatively stronger magnetic field is required to model the broadband SEDs. We found that the minimum ($\gamma^{\prime}_{min}$) and maximum ($\gamma^{\prime}_{max}$) Lorentz factors of electrons in the inner zone are relatively higher than those in the outer emission zone. All the model parameters values required for two-zone model fit are listed in Table-\ref{tab:4}.

\section{DISCUSSIONS} 
\label{sec:5}
Since the beginning of the \textit{Fermi}-LAT monitoring program (August 2008), TXS 0518+211 has been found to be in active-state on several occasions in optical to high-energy gamma-ray bands. This source was first reported and studied in the TeV band by \cite{Archambault2013Sep}. They explored the observational properties (i.e., spectral indices, optical polarization, Compton dominance, etc.) in radio, optical, X-ray, and TeV bands and argued that the source resembles an IBL-type BL Lac during low flux phase. However, during high flux state on 27 November 2009, the source exhibited elevated X-ray level by a factor of $\sim$15 with significant spectral hardening behavior ($\Gamma_{X}$ = 2.00$\pm$0.10) and the synchrotron component showed HBL-like properties. Later, \cite{Adams2022Jun} studied this source (using simultaneous optical to TeV data) from October 2013 to February 2014 during both low and high flux states. They did not observe any significant spectral hardening with increasing X-ray flux levels, as reported during the previous November 2009 high flux state, and the broadband SED in this time span exhibited similar properties to IBLs in all flux states. \\
\\
\subsection{Low and high flux states characteristics from the multi-wavelength data} \label{sec:5.1}
In this study, we have divided the entire multi-wavelength light curve into 11 different Epochs (Epoch-A to Epoch-K). We computed the fractional variability ($F_{var}$) and reduced chi-square value ($\chi^{2}_{red}$) for each epoch and the entire light curve to characterize the long-term variability across optical, UV, X-ray, and gamma-ray wavebands. The value of $F_{var}$ and $\chi^{2}_{red}$ (ref Table-\ref{tab:1}) reveal that the light curve in X-ray band shows more prominent variability than optical and gamma-ray light curves. However, we caution that for low number of data points, $\chi^{2}_{red}$ values are not reliable as the standard deviation of this quantity follows $\sqrt{\frac{2}{df}}$ (where, df = degrees of freedom). In the X-ray band, the largest $F_{var}$ is observed for the entire light curve, with a value of $1.10 \pm 0.03$, followed by Epoch-A ($0.96 \pm 0.04$) and Epoch-I ($0.71 \pm 0.07$). A similar result was also reported by \cite{Adams2022Jun}, based on $\sim$ 1 year (2013 - 2014) of near-simultaneous optical to VHE data. Similarly, in the optical-UV bands, the largest $F_{var}$ is for the entire light curve, but with comparatively lower values ranging from 0.35 to 0.50. Whereas, in the case of gamma-ray, it is observed for the Epoch-H with a value of $0.89 \pm 0.15$. During Epoch-H, flaring behavior was observed over a short duration (MJD 58402 to 58417) in the gamma-ray band with an average flux of $0.47 \pm 0.10$, resulting in a higher fractional variability. Corresponding optical U-band and X-ray light curves showed comparatively low flux values ($1.73 \pm 0.08$ and $1.04^{+0.20}_{-0.17}$, respectively during MJD 58402 - 58417), which are similar to the average flux of Epoch-H. However, the lack of high-cadence data in these bands (i.e., optical and X-ray) during this short period makes it difficult to draw any decisive conclusion about the flux state (i.e, possibility of orphan gamma-ray flare). During the total observing period, the source remains in either a low- or intermediate-activity phase in the X-ray band for most of the time. As the source evolves from the low- to the high-flux phase, we observe a decrease in the X-ray spectral index (i.e., spectral hardening trend), with the highest value of $\Gamma_{X}$ in Epoch-K ($2.74^{+0.23}_{-0.23}$) and the lowest in Epoch-I ($2.18^{+0.04}_{-0.04}$). A similar trend is also observed in the optical–UV band, although the variation of spectral indices ($\Gamma_{UVOT}$) between the low- and high-flux phases is relatively smaller and within the error bars. This result is consistent with the index–flux correlation study discussed in section~\ref{sec:3.4} (see Figure~\ref{fig:3}).
\par
The INOV study of this source reveals that none of the sessions exhibit strong variability on minute-to-hour timescales. It is worth pointing out that \cite{2023MNRAS.524L..66N} conducted INOV study on six TeV-detected HBLs, and detected variability in only one out of 24 intra-night sessions. The low INOV duty cycle is comparable to IBL sub-class but much less compared to LBL sub-class \citep{2017ApJ...844...32P}, Despite both sub-classes exhibit fairly high degree of optical polarization. HBLs/TeV-HBLs generally exhibit a parsec-scale jet with either subluminal or mildly superluminal radio knots. It has been argued that the presence of super-luminal/fast-moving radio knots could be a key diagnostic for INOV detection \citep{2023MNRAS.524L..66N}. According to \cite{2019ApJ...874...43L}, TeV-detected TXS 0518+211 showed a combination of quasi-stationary and mildly super-luminal radio knots. Hence, our INOV results suggest that the source is consistent with either belonging to IBL or HBL subclass. The INOV study alone can not identify its specific subclass. A detailed comparable INOV study between TeV-detected IBL and HBL sources could be helpful in the future to resolve this issue. 
\par
\begin{figure}
\centering
\includegraphics[height=2.5in,width=3.6in]{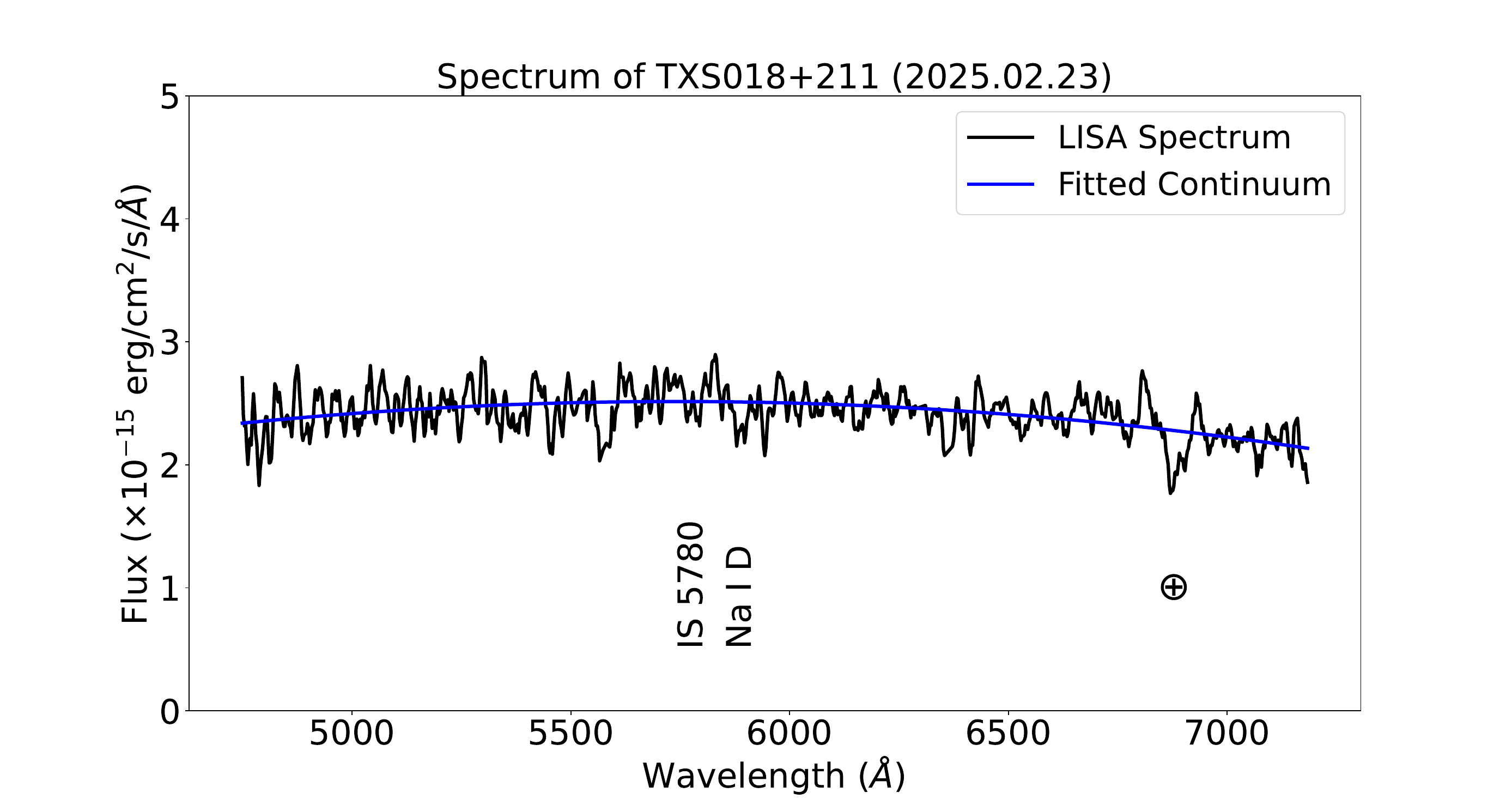}
\caption{Spectroscopic result (de-reddened) of TXS 0518+211, observed on 23rd February, 2025. The spectrum is de-reddened (corrected for interstellar reddening) using R = 3.1 and E(B-V) = 0.5875 \citep{Schlafly2011Aug}. The telluric line is indicated by $\oplus$ sign. Absorption lines due to the interstellar medium of our Galaxy are visible but not prominent.}
\label{fig:8}
\end{figure}

As discussed earlier, the redshift of this source is still uncertain, and only lower and upper limits are available in the literature. To estimate the redshift and flux state, we conducted spectroscopic observations of this source with an exposure time of $\sim$ 50 minutes using the low-resolution spectrograph LISA ($R \sim 850$) mounted on the PRL 1.2 m telescope, Mount Abu, India. The data analysis was performed using standard \textsc{Python} scripts and the IRAF package\footnote{\url{https://iraf-community.github.io/install.html}}. 
The details of the instrument and the standard reduction procedures, including bias subtraction, flat-fielding, cosmic-ray correction, one-dimensional spectrum extraction, and wavelength and flux calibrations, are presented in \citet{2016MNRAS.463.2137V} and \citet{10.1093/mnras/staf1528}. 
Due to the limited sensitivity of the instrument and average weather condition, we obtained a low signal-to-noise ratio (S/N$<$6). The optical spectrum is shown in Figure-\ref{fig:8}. It is evident that no clear emission lines are seen in the spectrum. The continuum emission (shown by the solid blue line) during this period is relatively bluer than that observed in earlier spectra taken in October and December 2009 by \cite{Archambault2013Sep}. 
\\

\subsection{Emission mechanisms during low and high flux states} \label{sec:5.2}
The broadband temporal as well as spectral studies (specifically during Epoch-I, Epoch-C, and Epoch-K discussed in Section~\ref{sec:3.4}), suggest that the source exhibits complex emission mechanisms. We used simple one-zone and two-zone (non-interacting) leptonic models to investigate the emission processes across different epochs. In the one-zone scenario, we found that the emission region is particle-dominated for all epochs, with a maximum total required jet power ($P_{tot}$) of 3.11 × 10$^{45}$ erg/sec during Epoch-I. The values of various model parameters, including the magnetic field (B) and minimum, maximum, break Lorentz factors ($\gamma^{'}_{min}$, $\gamma^{'}_{max}$ and $\gamma^{'}_{br}$) of the injected electron spectrum, and radius of the emission region($R^{'}$), determined in this work (for one-zone scenario) are comparable to those reported in \cite{Adams2022Jun}. However, we noticed that none of the optical–UV slopes can be reproduced with the simple one-zone scenario. Furthermore, we notice a clear plateau in the low-energy gamma-ray range ($\lesssim$3.0 GeV) during several epochs, 
such as Epoch-C, Epoch-D, Epoch-E, and Epoch-H. Such a trend has previously been observed in other BL Lac-type objects (e.g., 1ES 1959+650; \citealt{2018A&A...611A..44P}) and could indicate that emission from two or more zones contributes to the observed SED. 
\par

In the two-zone scenario, we considered two emission region sites with similar Doppler factor ($\delta$ = 26) that are non-interacting and spatially separated from each other. The inner-zone predominantly explains the X-ray and high-energy to VHE gamma-ray ($\gtrsim$ 3 GeV) emission component, while the outer emission-zone explains the optical-UV and low-energy gamma-ray (0.1 to $\sim$3 GeV) emission component of the broadband SED. According to the flux levels of different epochs in the X-ray band, Epoch-F and Epoch-H can be considered as quiescent states. From the SED modeling, we found that the change in total electron number density (greater than 80\%) in both emission zones was greater than the change in the magnetic field during the high-flux states of the source (e.g, Epoch-C and Epoch-I). The injection of fresh electrons and/or the re-acceleration of particles could increase the electron energy density in the emission region, 
which might be responsible for the enhanced or high-state activities of this source. In our two-zone modeling, we require a 
relatively smaller emission region radius (R$'_{\text{outer}}$ = 6.0 × 10$^{16}$ cm, which implies a denser seed photon field) 
for Epoch-I to explain the stronger SSC component from the outer zone. 
According to our SED modeling, during Epoch-E and Epoch-I, the source exhibits an HBL-like property ($\nu_{\text{peak}} > 10^{15}$ Hz). 
In both cases, we noticed a spectral hardening trend in the X-ray band ($\Gamma_{X} = 2.32^{+0.09}_{-0.09}$ for Epoch-E 
and $\Gamma_{X} = 2.18^{+0.04}_{-0.05}$ for Epoch-I). The intrinsic bolometric synchrotron emission (and the corresponding SSC emission) from the inner zone dominates the emission from the outer zone ($\sim$ 3 times higher for Epoch-E and $\sim$ 6 times higher for Epoch-I) for these two epochs, whereas in all other epochs it is nearly comparable (0.8–1.3 times higher) to the outer zone emission. The maximum total required jet power (computed using Equation \ref{eq:11}) in the two-zone scenario is found to be 8.06 × 10$^{45}$ erg/s for Epoch-I, followed by Epoch-C. 
\\
From visual inspection, it is clear that the two-zone model better describes the contemporaneous Fermi observations (energy range $\sim$ 0.1–15 GeV) of Epoch-I ($\chi^{2}_{red}$ = 3.10 for one-zone model and $\chi^{2}_{red}$ = 1.19 for two-zone model). However, both models fail to explain the high-energy – VHE ($>$ 100 GeV - TeV range) part of the broadband SED of this epoch. The data points corresponding to this energy range (i.e., $>$ 100 GeV–TeV), shown in blue and magenta colors in Figures \ref{fig:6} and \ref{fig:7}, are archival and do not overlap with the time interval of Epoch I (MJD 58901–58921), but instead partially overlap with the time interval of Epoch C (MJD 56580–56673; see Figure-\ref{fig:1} and Figure-\ref{fig:B1}). It has been observed that, for this source, the correlation between the X-ray and TeV gamma-ray fluxes is stronger than between the optical and TeV gamma-ray fluxes \citep{Adams2022Jun}. Hence, simultaneous GeV–TeV observations during such exceptional orphan X-ray flare epochs are required to test one-zone and two-zone models. We observed a significant dip between the low-energy (synchrotron) hump and high-energy (SSC) hump in our two-zone model fit for all epochs (see Figure-\ref{fig:7}). The depth of this dip decreases as the source evolves from a quiescent state to a high-flux state. Hence, observations in wavebands that cover the frequency range between X-rays and gamma-rays (e.g., NuSTAR observations) 
would help better constrain the model parameters in future, especially the minimum Lorentz factor ($\gamma'_{\text{min}}$) of the outer zone and maximum Lorentz factor ($\gamma^{'}_{max}$) of the inner-zone. Recently, \cite{2025A&A...694A.308M} studied this source during the activity period of Epoch-I over a short timescale duration and found that, with a one-zone lepto-hadronic model fit, the dip between the two emission humps almost disappears. Therefore, panchromatic simultaneous observations, along with NuSTAR data, could also be helpful in understanding the emission mechanisms in more detail.
\section{CONCLUSIONS} \label{sec:6}
In this work, we have performed a broad-band temporal and spectral study of TeV-detected BL Lac TXS 0518+211 using long-term 
(MJD 54682 - 60670; 16 years) optical, UV, X-ray (from \textit{Swift} UVOT/XRT) and gamma-ray (from \textit{Fermi}-LAT) data. Based on the flux level of Bayesian block in the entire X-ray light curve, we identified  11 different epochs (Epoch-A to Epoch-K) and investigated the emission properties of the source in all the epochs. All data points in the other wavebands (optical–UV and gamma-ray) within the time interval of a given epoch are considered to be part of that epoch. In the following, we highlight some important findings of our work:
\begin{itemize}
\item Fractional variability reveals that, in all the epochs, the source exhibits higher variability in the X-ray band compared to the optical/UV and gamma-ray bands. This is also evident from the average fractional variability parameter of long-term light curves spanning across full 16 years, wherein fractional variability in X-ray band is 1.10$\pm$0.03, while it ranges between 0.35 to 0.45 in the other bands.

\item We also probed variability on short timescales by conducting three intra-night photometric monitoring sessions in optical $R$-band during January and March 2025. Notably, none of DLCs show any strong variability signature. The INOV behavior of TXS 0518+211 is consistent with a typical INOV trend seen among IBL/HBL-type sources.

\item Correlation study shows a weak to moderate correlation across different wavebands, with the Spearman correlation coefficient ranging from 0.29 to 0.58 in the long-term light curve.  

\item We identified two high-flux states in the long-term X-ray light curve using a Bayesian block representation, namely Epoch-C (MJD 56580 - 56673) and Epoch-I (MJD 58895 - 58922). During Epoch-C, the source showed simultaneous variability across all three wavebands, whereas during Epoch-I, a significant increase in the X-ray flux level was observed without a corresponding counterpart in the optical/UV bands. Another recent flux state, during Epoch-K (MJD 60611 - 60670), we noticed a significant decrease in X-ray flux, but the optical/UV and gamma-ray flux levels remained nearly steady. These behaviors in the multi-wavelength light curve suggest complex emission region structure in the jet.

\item We used both one-zone and two-zone (non-interacting) leptonic models to explain the broadband SED of all epochs. The two-zone model provides a better explanation of the optical to high-energy gamma-ray SED. The maximum total jet power required in the two-zone model is found to be 8.06$\times$10$^{45}$ erg/s for Epoch-I. The high X-ray flux observed during this epoch could be explained by the increased activity in the inner emission region.
    
\end{itemize}


\section*{Data Availability}
This work has made use of publicly available \textit{Fermi}-LAT data obtained from FSSC's website data server and provided by NASA Goddard Space Flight Center. This work has also made use of data, software/tools obtained from NASA High Energy Astrophysics Science Archive Research Center (HEASARC), developed by the Smithsonian Astrophysical Observatory (SAO), and the XRT Data Analysis Software (XRTDAS) developed by ASI Science Data Center, Italy. We acknowledge the use of data collected from the CMOS imager on the 1.2 m telescope at Mt. Abu Observatory.

\begin{acknowledgments}
We acknowledge the use of public data from the \textit{Swift} data archive. The data used in this work were obtained with the Fermi Gamma-ray Space Telescope, a NASA mission. We thank the \textit{Fermi}-LAT Collaboration for the use of their data. The research work carried out at the Physical Research Laboratory (PRL) is funded by the Department of Space, Government of India. A.K.D. thanks PRL for the fellowship. We thank Prof. Shashikiran Ganesh, and local support staff at Mt. Abu observatory for their help during observations. P.K. acknowledges support from the Department of Science and Technology (DST), Government of India, through the DST-INSPIRE faculty grant (DST/INSPIRE/04/2020/002586). 
\end{acknowledgments}

\facilities{\textit{Fermi}-LAT, \textit{Swift} (XRT and UVOT), Mt Abu observatory: 1.2m}

\software{Fermitools \citep{2019ascl.soft05011F}, HEASARC \citep{2014ascl.soft08004N}, XSPEC \citep{1996ASPC..101...17A}, AstroimageJ \citep{2017AJ....153...77C}, IRAF \citep{Tody86}, GAMERA \citep{Hahn:2016CO}, astropy \citep{2013A&A...558A..33A,2018AJ....156..123A,2022ApJ...935..167A}}

\bibliography{bibliography}{}
\bibliographystyle{aasjournalv7}

\appendix

\section{The duration of optical, UV and X-ray observations in different epochs} \label{app:A} Here, we list the corresponding observation time duration for each instrument for each epoch (ref Table-\ref{tab:A1}). We only show time information of only those epochs which are used in the broadband SED modeling. According to our definition, Epoch-F, Epoch-H, and Epoch-J can be considered low or quiescent flux states, as the Bayesian block flux levels in the X-ray light curve are below the average flux of the entire light curve for these epochs. The time durations of all these quiescent epochs are very wide. We could further divide these epochs into different sub-epochs by applying the same Bayesian block criteria. However, it is noted that the short durations of such sub-epochs used for the spectral study would result in upper limits on the gamma-ray flux for most data points. 

\setcounter{table}{0}
\renewcommand{\thetable}{A\arabic{table}}

\begin{table*}[b]
\caption{The duration of optical, UV and X-ray observations in different epochs. The number of observation IDs for the Swift-UVOT and XRT instruments for a given epoch are given in brackets.}
\label{tab:A1}
\hskip-3.5cm
\centering
\begin{tabular}{ccccccc rrrr}   
\hline\hline
Activity Epochs & UVOT-V & UVOT-B & UVOT-U & UVOT-W1 & \textit{Swift}-XRT & \textit{Fermi}-LAT \\
\hline\hline
Epoch-A & - & - & 55132-55160 (2) & 55131-55159 (2) & 55131-55162 (7) & 55131-55162  \\
\hline
Epoch-C & 56621-56673 (17) & 56621-56673 (17) & 56580-56673 (20) & 56583-56673 (16) & 56580-56673 (27) & 56580-56673 \\
\hline
Epoch-D & 56925-57040 (9) & 56925-57040 (9) & 56925-57040 (9) & 56925-57040 (8) & 56925-57040 (9) & 56925-57040   \\
\hline
Epoch-E & 57306-57372 (5) & 57306-57372 (5) & 57306-57372 (5) & 57306-57372 (5) & 57306-57372 (7) & 57306-57372 \\
\hline
Epoch-F & 57728-57808 (6) & 57728-57808 (6) & 57728-57808 (6) & 57728-57808 (4) & 57718-57808 (6) & 57718-57808 \\
\hline
Epoch-H & - & - & 58404-58520 (7) & - & 58404-58520 (8) & 58404-58520  \\
\hline
Epoch-I & 58908-58921 (2) & 58907-58921 (5) & 58904-58921 (6) & 58908-58921 (4) & 58901-58921 (8) & 58901-58921  \\
\hline
Epoch-J & 59164-59227 (8) & 59164-59227 (9) & 59164-59252 (13) & 59164-59231 (9) & 59164-59286 (22) & 59164-59286  \\
\hline\hline
\end{tabular}
\end{table*}

\section{Broadband SED Modeling of sub-epochs} \label{app:B}
Blazar light curves generally show rapid short-term variability over minute to day time scale across multi-wavelength regime, especially during the interval of high activity episodes. In our work, we applied the Bayesian block method to the long-term X-ray light curve and if the flux levels of consecutive Bayesian blocks are higher than the average flux of the entire X-ray light curve ($\sim$ 16 years), that time duration is defined as a high-activity state. This approach to defining high-activity states may result in wider epoch time intervals and could overlook short-term variability features within a given epoch. For example, in our study, Epoch-C (MJD 56580-56673) is considered as a high flux state and time duration of this epoch is very wide (93 days). We further divided this epoch into three different sub-epochs, namely Epoch-C1 (MJD 56580-56590), Epoch-C2 (MJD 56622-56629), and Epoch-C3 (MJD 56650-56657) using the same procedure (flux level of Bayesian block in the X-ray light curve) as discussed above (shown in Figure-\ref{fig:B1}). We then performed two-zone pure leptonic modeling for each sub-epoch, adopting the same constraints on the parameters as for the other epochs (ref section-\ref{sec:4.1.2}). This approach also mitigates the effect of using non-simultaneous data in the SED modeling. The two-zone leptonic model fits for all sub-epochs are shown in Figure \ref{fig:B2}. The values of all the model parameters required for SED modeling are listed in Table-\ref{tab:B1}.

\setcounter{figure}{0}
\renewcommand{\thefigure}{B\arabic{figure}}
\setcounter{table}{0}
\renewcommand{\thetable}{B\arabic{table}}

\begin{figure*}[h] 
\centering
\includegraphics[height=3.8in,width=7.0in]{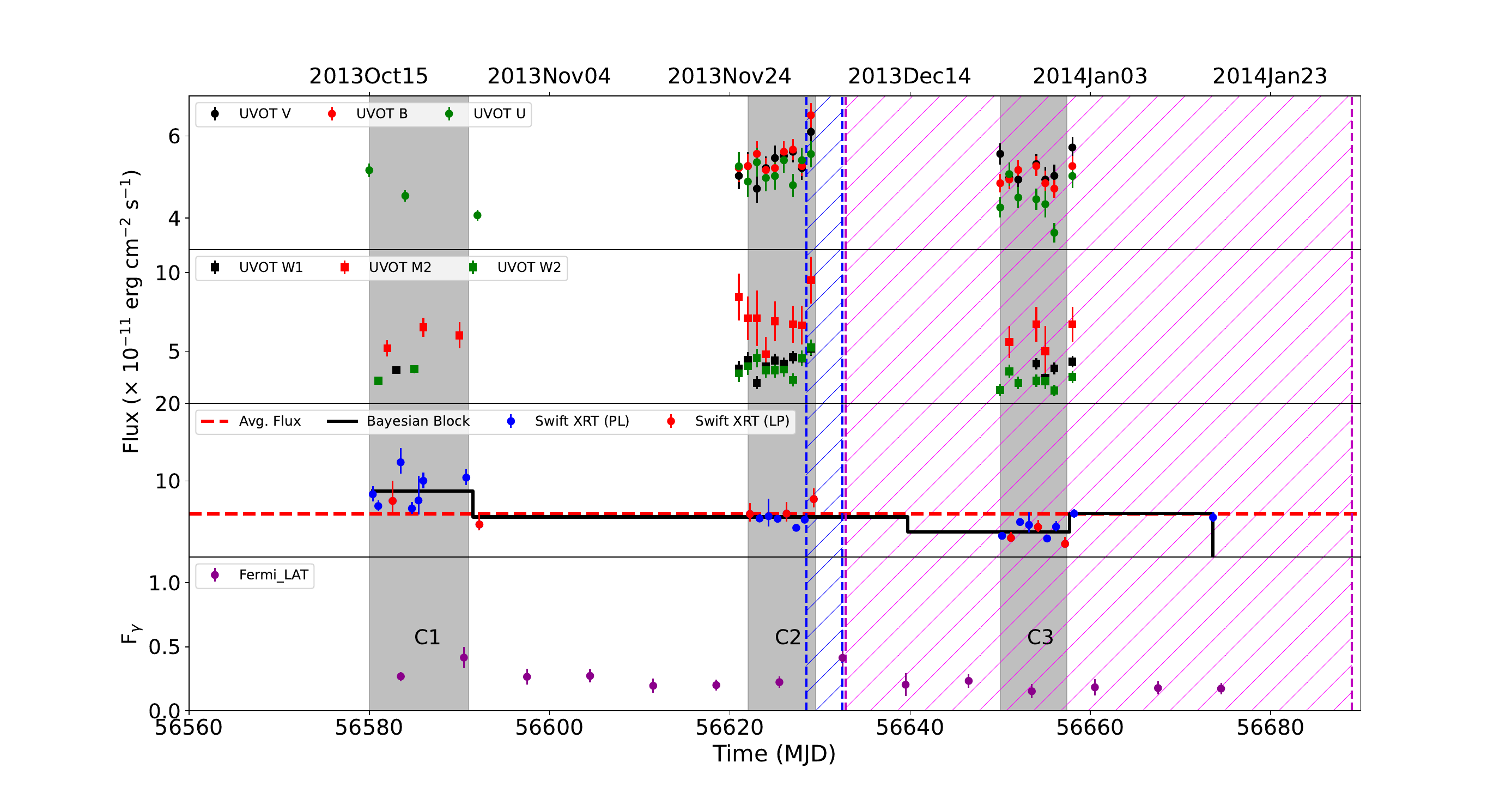}
\caption{The zoomed-in view of the multi-wavelength light curves of Epoch-C (MJD 56580 $-$ 56673). From top to bottom panels represent optical ($V$, $B$, $U$ bands), UV ($UVW1$, $UVM2$, $UVW2$ bands), X-ray, and gamma-ray light curves, respectively. The three different sub-epochs (Epoch-C1, Epoch-C2 and Epoch-C2) identified based on the flux level of Bayesian block in the X-ray band are shown by grey shaded vertical strips. The blue (MJD 56628.5–56632.5) and magenta (MJD 56632.5–56689.0) hatched regions indicate the time intervals of the VERITAS observations (also shown in Figure \ref{fig:1}).}
\label{fig:B1}
\end{figure*}

\begin{figure}
\centering
\includegraphics[height=2.2in,width=3.0in]{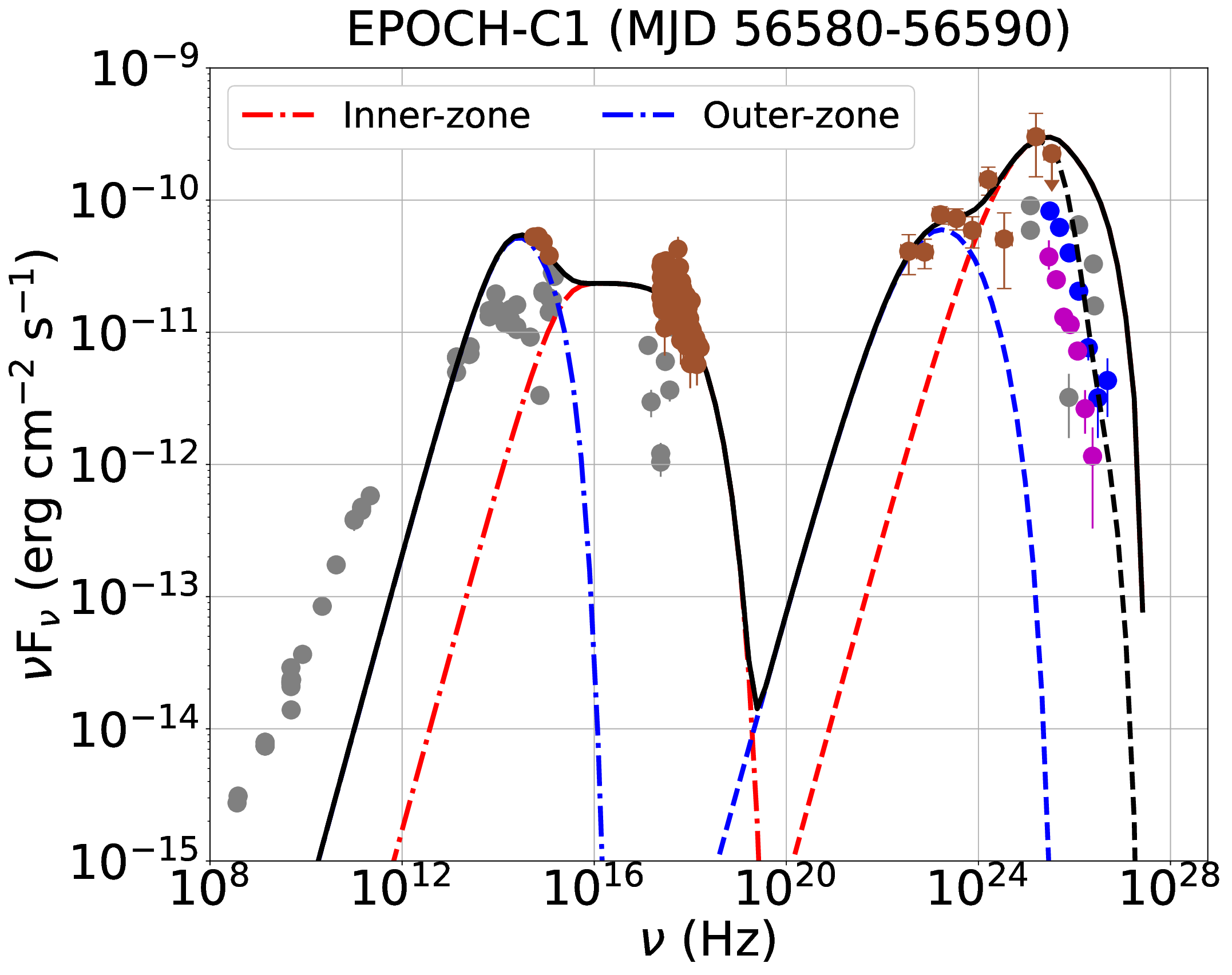}
\includegraphics[height=2.2in,width=3.0in]{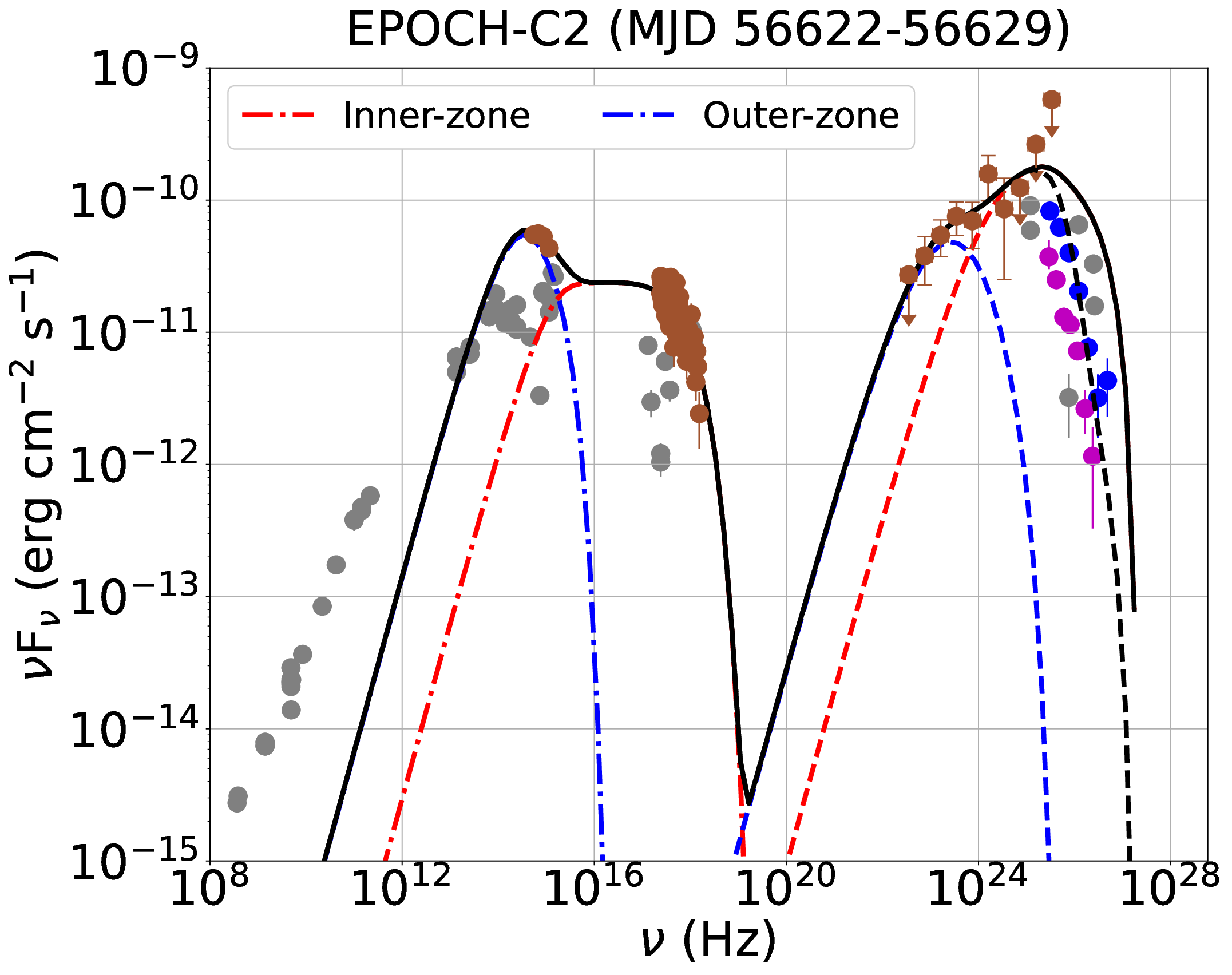}
\includegraphics[height=2.2in,width=3.0in]{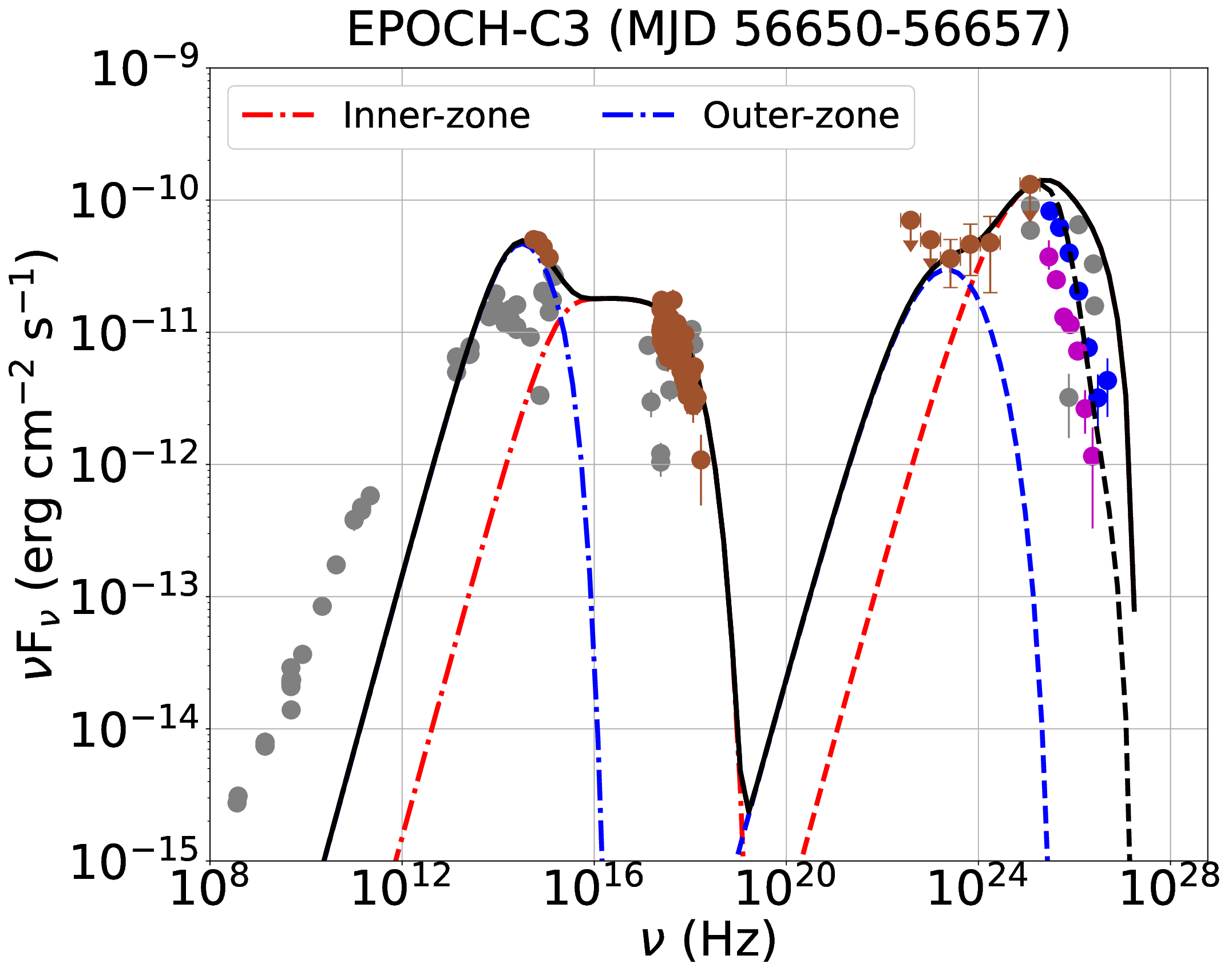}
\caption{The two-zone leptonic model (shown by solid curves) fitted to the broadband SEDs (from optical to high-energy gamma-rays) for different sub-epochs of Epoch-C. Non-simultaneous archival are represented by grey colored data points in each figure. Non-simultaneous VHE data taken from \citet{Adams2022Jun} are shown by magenta and blue colored points corresponding to low-VHE flux state (MJD 56632.5-56689.0) and high-VHE flux state (MJD 56628.5-56632.5), respectively. The black dashed curve represents the EBL-corrected model.}
\label{fig:B2}
\end{figure}

\begin{table*}
\caption{Results of two-zone multi-wavelength SED modeling. The third to thirteenth columns represent the value of various parameters used in the one-zone modeling (see Table-\ref{tab:4} for more details).}
\label{tab:B1}
\hskip-3.5cm
\centering
\scalebox{0.85}{
\begin{tabular}{ccccccccc rrrr}   
\hline\hline
Activity & Emission & $\alpha_{1}$ & $\alpha_{2}$ & $\gamma^{\prime}_{min}$ & $\gamma^{\prime}_{max}$ & $\gamma^{\prime}_{br}$ & B$^{\prime}$ & R$^{\prime}$ & $P_e$ & $P_B$ & $P_{tot}$ & Time duration \\
Epochs & region & & & & & & (G) & (cm.) & (erg/sec) & (erg/sec) & (erg/sec) & (days) \\
(1) & (2) & (3) & (4) & (5) & (6) & (7) & (8) & (9) & (10) & (11) & (12) & (13) \\
\hline\hline
Epoch-C1 & Inner-zone  & 2.95 & 3.95 & 2.35$\times10^{4}$ & 1.20$\times10^{6}$ & 4.00$\times10^{5}$ & 0.03 & 1.10$\times10^{16}$ & 2.08$\times10^{45}$ & 2.55$\times10^{41}$  & 2.08$\times10^{45}$ & 10 \\
& Outer-zone & 2.86 & 3.86 & 9.80$\times10^{3}$  & 4.00$\times10^{4}$ & 2.00$\times10^{4}$ & 0.01 & 1.30$\times10^{17}$ & 3.30$\times10^{45}$ & 3.96$\times10^{42}$  & 3.32$\times10^{45}$  & \\ 
\hline
Epoch-C2 & Inner-zone & 2.94 & 3.94 & 2.00$\times10^{4}$ & 8.00$\times10^{5}$ & 4.00$\times10^{5}$ & 0.03 & 1.40$\times10^{16}$ & 1.38$\times10^{45}$  & 4.13$\times10^{41}$ & 1.39$\times10^{45}$  & 7 \\
& Outer-zone & 2.86 & 3.86 & 1.20$\times10^{4}$ & 4.00$\times10^{4}$ & 2.00$\times10^{4}$ & 0.01 & 1.45$\times10^{17}$ & 2.80$\times10^{45}$  & 3.99$\times10^{42}$ & 2.82$\times10^{45}$ & \\
\hline
Epoch-C3 & Inner-zone  & 2.94 & 3.94 & 2.20$\times10^{4}$ & 8.00$\times10^{5}$ & 4.00$\times10^{5}$ & 0.03 & 1.15$\times10^{16}$ & 1.22$\times10^{45}$  & 2.79$\times10^{41}$ & 1.23$\times10^{45}$ & 7 \\
& Outer-zone & 2.86 & 3.86 & 1.10$\times10^{4}$ & 4.00$\times10^{4}$ & 2.00$\times10^{4}$ & 0.01 & 1.60$\times10^{17}$ & 2.08$\times10^{45}$  & 6.00$\times10^{42}$ & 2.10$\times10^{45}$ & \\
\hline\hline
\end{tabular}
}
\end{table*}

\end{document}